\documentclass{jfm}
\usepackage{graphicx}
\usepackage{epstopdf, epsfig}
\usepackage{subfig} 
\usepackage{amsmath}

\usepackage{multirow}
\usepackage{lscape}
\usepackage{rotating}

\usepackage{array} 
\newcolumntype{L}[1]{>{\raggedright\let\newline\\\arraybackslash\hspace{0pt}}m{#1}}
\newcolumntype{C}[1]{>{\centering\let\newline\\\arraybackslash\hspace{0pt}}m{#1}}
\newcolumntype{R}[1]{>{\raggedleft\let\newline\\\arraybackslash\hspace{0pt}}m{#1}}

\shorttitle{Modelling and Simulation of Turbulent Flows with Coherent Structures}
\shortauthor{F.S. Pereira, L. Eca, G. Vaz, and S.S. Girimaji}

\title{Challenges in Modelling and Simulation of Turbulent Flows with Spatially-Developing Coherent Structures}

\author{F.S. Pereira\aff{1}\aff{2}
  \corresp{\email{filipemsoares@ist.utl.pt}},
L. E\c{c}a\aff{2},
G. Vaz\aff{3},
 \and  S.S. Girimaji\aff{1} }
\affiliation{\aff{1}Department of Ocean Engineering, Texas A\&M University,
College Station, United States of America
\aff{2}Department of Mechanical Engineering, Instituto Superior T\'ecnico, Lisbon, Portugal
\aff{3}Maritime Institute Research Netherlands, Wageningen, The Netherlands}
\begin{document}

\maketitle

\begin{abstract}
The objective of this work is to investigate the fundamental challenges encountered in modelling and simulation of turbulent flows driven by spatially-developing coherent structures. Scale-Resolving Simulations (SRS's) of practical interest are expressly intended for efficiently computing such flows by resolving only the most important features of the coherent structures and modelling the remainder as stochastic field. With reference to a typical Large-Eddy Simulation (LES), practical SRS methods seek to resolve a considerably narrower range of scales (physical resolution) to achieve an adequate degree of accuracy at reasonable computational effort. The success of SRS methods depends upon three important factors: $i)$ ability to identify key flow mechanisms responsible for the generation of coherent structures; $ii)$ determine the {optimum range of resolution} required to adequately capture key elements of coherent structures; and $iii)$ ensure that the modelled part is comprised nearly exclusively of fully-developed stochastic turbulence. This study considers the canonical case of the flow around a circular cylinder to address the aforementioned three key issues. It is first demonstrated using experimental evidence that the vortex-shedding instability and flow-structure development involves four important stages. The inherent limitations of the Reynolds-Averaged Navier-Stokes (RANS) approach in addressing the flow physics in these four stages of development are explained. A series of SRS computations of progressively increasing resolution (decreasing cut-off length) are performed. An {\em a priori} basis for locating the origin of the coherent structures development is proposed and examined. The criterion is based on the fact that the coherent structures are generated by the Kelvin-Helmholtz (KH) instability. The most important finding is that the key aspects of coherent structures can be resolved only if the effective computational Reynolds number (based on total viscosity)  exceeds the critical value of the KH instability in laminar flows. Finally, a quantitative criterion assessing the nature of the unresolved field based on the strain-rate ratio of mean and unresolved fields is examined. Based on the findings, quantitative guidelines for determining the optimal degree of physical resolution in flows of practical interest are proposed.
\end{abstract}

\begin{keywords}
Scale-Resolving Simulation; Coherent Structures; Physical Resolution; Free Shear-Layer; Circular Cylinder.
\end{keywords}

\section{Introduction}
\label{sec:1}

Turbulent flows with spatially-developing coherent structures constitute an important classification of interest in many natural phenomena and engineering applications. Coherent structures drive the flow dynamics in wakes, jets, mixing layers and internal flows. For example, the vortex-shedding in many wake flows is preceded by Kelvin-Helmholtz rollers that develop in the free shear-layers emerging from the surface of the body. A typical bluff-body wake in such regime consists of four important flow regions: upstream low intensity turbulence; initiation and development of coherent structures in the free shear-layer; subsequent vortex-shedding; and high-intensity stochastic turbulence in the far-wake. 

The velocity or pressure field, $\Phi$, in flows with coherent structures is most conveniently described with a triple decomposition \citep{REYNOLDS_JFM_1970,SCHIESTEL_PF_1987}:
\begin{equation}
\label{eq:1_1}
\Phi=\overline{\Phi}+\tilde{\phi}+\phi'\; .
\end{equation}
Here $\overline{\Phi}$ is the appropriately averaged flow field, $\tilde{\phi}$ is the velocity field associated with the coherent structures, and $\phi'$ represents the stochastic background turbulence. The rationale of the triple decomposition can be explained as follows. The coherent velocity field encapsulates the space and time correlations of the large-scale structures that are specific to the flow conditions and geometry. The background velocity field represents more universal stochastic features of a fully-developed turbulent field which is characterized by shorter spatio-temporal correlation distances. In a typical wake flow, the upstream and far-wake velocity field may be reasonably represented as fully-developed stochastic turbulence. In the near-wake, the coherent structures clearly dominate the flow dynamics.

Predictive computations of flows with spatially-developing coherent structures is rife with challenges. While Direct Numerical Simulations (DNS) and Large-Eddy Simulations (LES) are suitable for capturing the coherent field physics, they are computationally prohibitive for many applications. On the other hand, the Reynolds-Averaged Navier-Stokes (RANS) approach is inherently deficient as it does not distinguish between the coherent and stochastic parts of the velocity field. These closure models cannot account for two or multi-point coherence in the velocity field and hence are best suited only for the stochastic flow field

Over the last two decades Scale-Resolving Simulations (SRS's) of practical interest have emerged expressly to address flows with large-scale coherent structures. The objective is to resolve a small but sufficient range of scales to capture the main aspects of the coherent structures and model the remainder of the flow field with an appropriate stochastic closure. These SRS strategies are intended to render a reasonable representation of the coherent flow field while resolving a substantially smaller range of scales than in a typical LES. Whereas the SRS rationale is meritorious, its utilization is difficult as there are no clear guidelines on the precise range of coherent scales that must be resolved. Undoubtedly, the success of the SRS approach depends upon \textit{i)} the ability to identify the flow physics underlying the generation of coherent structures; \textit{ii)} designating the optimum range of scales to be resolved and \textit{iii)} ensuring that only the stochastic field is modelled.

The SRS approaches of practical interest available in literature can be broadly classified into two types: hybrid \citep{SPALART_AFOSRIC_1997, KOK_AIAA42_2004, SHUR_IJHFF_2008, KOK_FTC_2017} and bridging \citep{SPEZIALE_ICNMFD_1996, FASEL_JFE_2002, GIRIMAJI_JAM_2005, GIRIMAJI_JAM2_2005, SCHIESTEL_TCFD_2005, CHAOUAT_PF_2005} methods. The SRS strategy used in this work is the bridging Partially-Averaged Navier-Stokes (PANS) technique of  \cite{GIRIMAJI_JAM_2005}. In this method, the range of resolved scales is controlled by specifying the fraction of turbulence kinetic energy to be modelled. The Sub-Filter Stresses (SFS) tensor is modelled using a Boussinesq-type constitutive relation and evolution equations are solved for the unresolved turbulence kinetic energy and specific dissipation.

In general, the SRS approach is amenable to the paradigm that such a simulation is equivalent to a DNS of a variable viscosity non-Newtonian fluid \citep{RIVLIN_QAM_1957,MUSCHINSKI_JFM_1996,DASIA_PF_2014}. This permits the consideration of the spatially-developing coherent structures in a SRS computation as the onset and development of instabilities in a non-Newtonian fluid. The instability under investigation here is the Kelvin-Helmholtz rollers that develop in a thin free shear-layer around the cylinder.

The goals of the study are therefore two-fold: \textit{i)} identify the key flow physics underlying the coherent structures development in a prototypical flow; and \textit{ii)} develop criteria for optimum flow resolution. An optimal SRS entails two important factors: $i)$ the resolution should be adequate, but not excessive, to directly compute the key features of the coherent structures; and $ii)$ ensure that the unresolved part of the velocity field is exclusively fully-developed stochastic turbulence. The flow past a circular cylinder, an archetypal flow with wake vortices, is employed in the study to address these questions. 

The remainder of the paper is arranged as follows. In Section 2, the key elements of the Von-K\'arm\'an vortex-street development in the flow past a circular cylinder are identified. Experimentally measured ranges of important flow features are established. The PANS method is described in Section 3. The computational setup is explained in Section 4. PANS circular cylinder simulations are performed with different ranges of resolved scales (degrees of physical resolution). In Section 5, PANS flow fields are analysed using the non-Newtonian turbulence paradigm and compared against experimental data. Criteria for optimal degree of coherent structures resolution are developed and demonstrated. The paper concludes in Section 6 with a summary of the major findings.

\section{Flow Around Circular Cylinder}
\label{sec:2}
The flow around circular cylinders can be classified into distinct regimes based on the Reynolds number,
\begin{equation}
\label{eq:2_1}
 Re\equiv \frac{\overline{V}_\infty D}{\nu}\; ,
\end{equation}
where $\overline{V}_\infty$ is the incoming time-averaged stream-wise velocity, $D$ is the cylinder diameter, and $\nu$ is the fluid kinematic viscosity. According to \cite{ZDRAVKOVICH_BOOK_1997}, it is possible to distinguish five different regimes depending on $Re$ and on the location of turbulent transition: fully laminar (L, $Re<180-200$); transition in the wake (TW, $Re<350-400$); transition in the free shear-layer (TSL, $Re<1.0\times 10^5-2.0\times 10^5$); transition in the boundary-layer (TBL); and fully turbulent boundary-layer (T). All these categories contain sub-regimes with well-defined features. 

In addition to $Re$, there are other parameters that may influence the limits of such regimes. The most relevant and common influencing parameters \citep{ZDRAVKOVICH_BOOK_1997} are the turbulence intensity \citep{FAGE_ARC_1929,SADEH_NASA_1982,NORBERG_JFS_1987}; aspect ratio \citep{WEST_JFM_1982,SZEPESSY_JFM_1992,NORBERG_JFM_1994}; wall-blockage \citep{WEST_JFM_1982}; surface roughness \citep{FAGE_ARC_1929, ACHENBACH_JFM_1971, GUVEN_JFM_1980}; end effects \citep{GERICH_JFM_1982,PRASAD_JFM_1997B}; and cylinder oscillations \citep{KOOPMANN_JFM_1967,TOKUMARU_JFM_1991}. The importance of the turbulence intensity, $I$, and aspect ratio, $\Lambda=L/D$, is demonstrated in table \ref{tab:2_1_1}. For example, the data show that the time-averaged vortex-shedding formation length, $\overline{L}_F$, may vary up to $11\%$ with $I$.

\begin{table}
  \begin{center}
\def~{\hphantom{0}}
\begin{tabular}{C{1.3cm}C{1.3cm}C{1.3cm}C{2.0cm}C{2.0cm}C{2.0cm}C{2.0cm}}
$Re$	&	$I(\%)$	& $\Lambda$ & $St$ &  $\overline{C}_{D}$ &  $\overline{C}_{pb}$ &    $\overline{L}_F/D$ \\[3pt]\hline
\multirow{1}{*}{3000} 	& 0.1-1.4	&80& 0.213-0.209 & 0.98-1.03 & 0.84-0.89	& 1.44-1.65\\
\multirow{1}{*}{8000} 	& 0.1-1.4	&80& 0.204-0.199 & 1.13-1.20 & 1.05-1.12 & 0.90-0.99\\
\multirow{1}{*}{$3900^*$} 	& $0.06$	&5-50& 0.194-0.209 & - & 0.68-0.87 & -\\ \hline
\end{tabular}
\caption{Experimental measurements of the Strouhal number, $St$, time-averaged drag coefficient, $\overline{C}_D$, base pressure coefficient, $\overline{C}_{pb}$, and formation length, $\overline{L}_F$, in terms of free-stream turbulence intensity, $I$, and aspect ratio, $\Lambda$. Data from \cite{NORBERG_TREP_1987} and \cite{NORBERG_JFM_1994}. $^*$ denotes interpolated values using data at $Re=3800$ and $4400$.}
\label{tab:2_1_1}
  \end{center}
\end{table}

Each of the flow regimes poses a different set of challenges to modelling and simulation. The focus of this study is the TSL regime which is crucially dependent upon the development of coherent structures in the free shear-layer coming off the cylinder.  The details of this regime are summarized in Section \ref{sec:2_3}. A thorough overview of all flow regimes is given in \cite{WILLIAMSON_ARFM_1996} and \cite{ZDRAVKOVICH_BOOK_1997}.
%
%
\subsection{Transition in the Free Shear-Layer Regime}
\label{sec:2_3}
\begin{figure*}
\centering
\subfloat[Cross-section view.]{\label{fig:2.1_2a}
\includegraphics[scale=0.25]{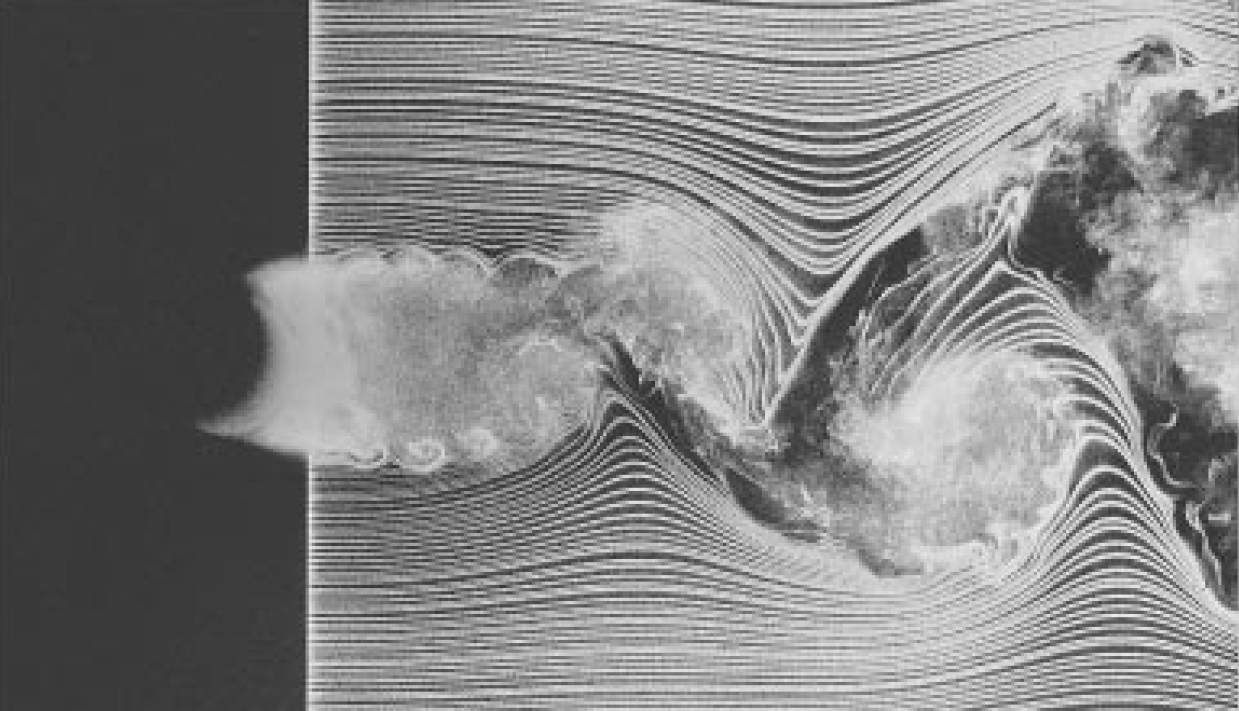}}
\\
\subfloat[$V_1(t)$.]{\label{fig:2.1_2b}
\includegraphics[scale=0.22]{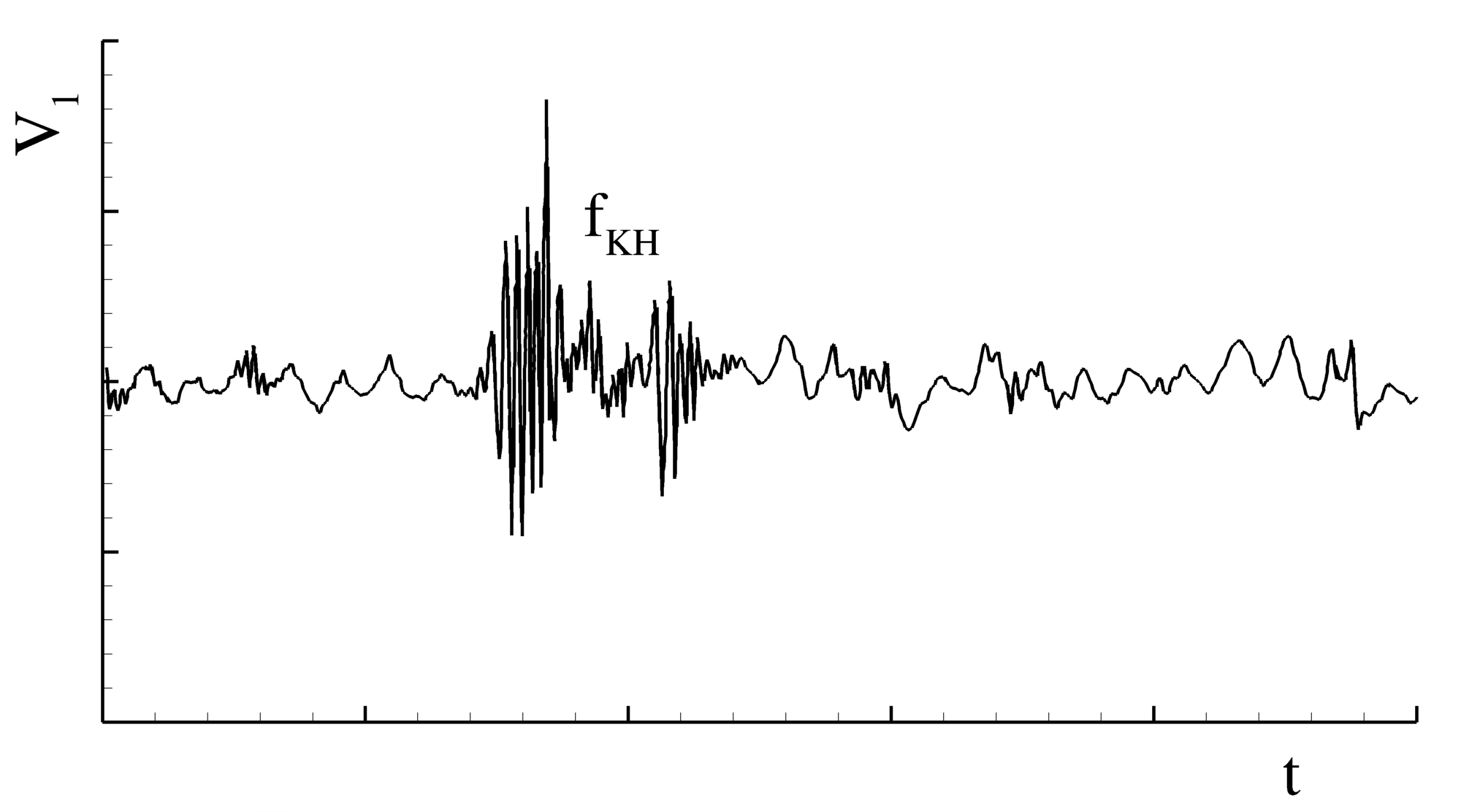}}
\\
\subfloat[$E(V_1)$.]{\label{fig:2.1_2c}
\includegraphics[scale=0.22]{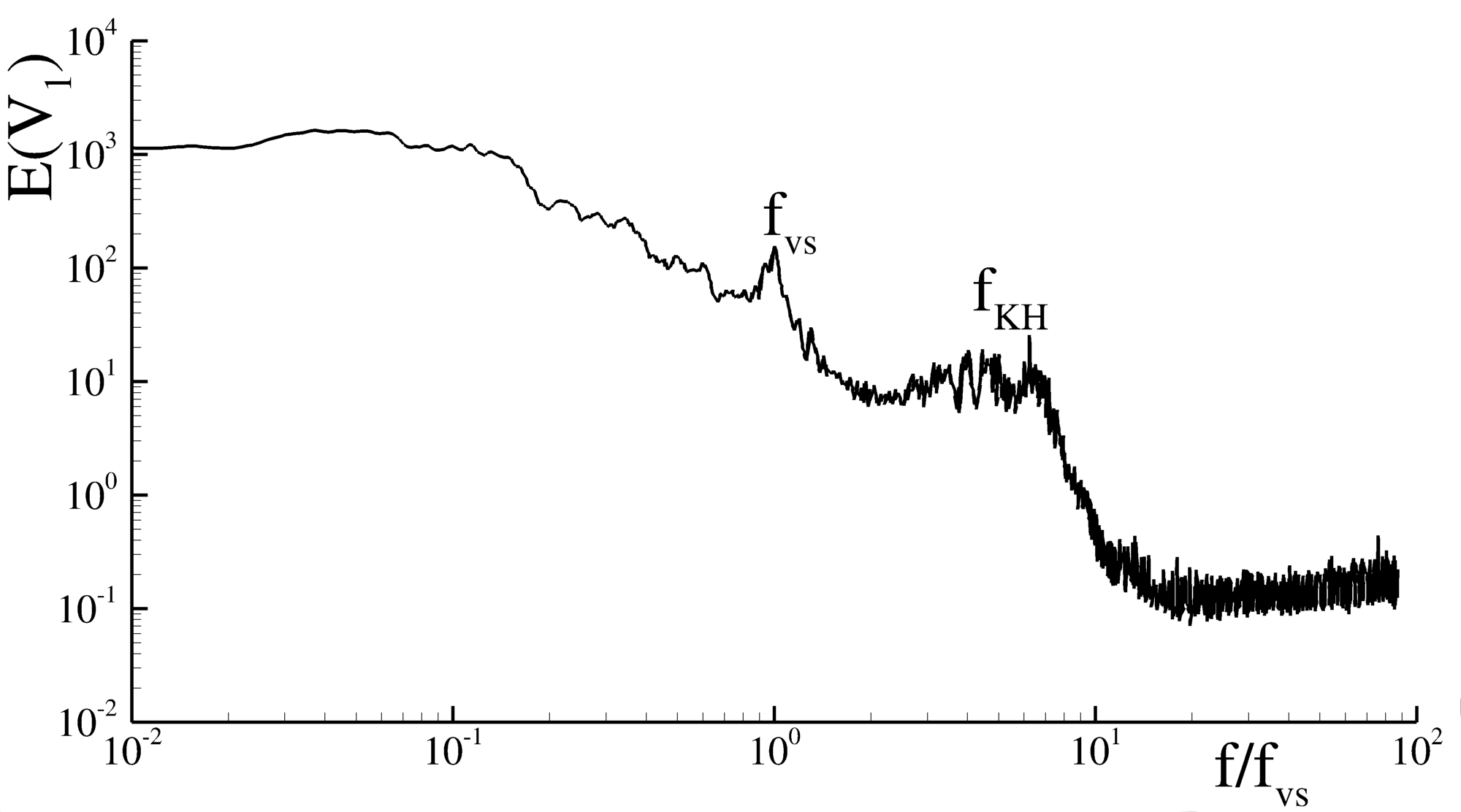}}
\caption{Cross-section view of the flow around a circular cylinder in the near-wake ($Re=10,000$), and stream-wise velocity, $V_1$, time trace ($Re=4000$) and spectrum ($Re=3700$) in the free shear-layer. Figures taken from \cite{PRADAD_JFM_1997A} and \cite{RAJAGOPALAN_EF_2005}.}
\label{fig:2.1_2}
\end{figure*}

The flow in the TSL regime is characterized by three shear-layers - boundary-layer, free shear-layer, and wake. In this range of $Re$, two boundary-layers detach from the cylinder's surface creating a recirculation region and two free shear-layers. Kelvin-Helmholtz (KH) rollers are then generated in these shear-layers. Their features govern the flow dynamics in the free shear-layer. The importance of this coherent structure has motivated multiple investigations in the past decades \citep{ROSHKO_TECH_1954,BLOOR_JFM_1964,PRADAD_JFM_1997A,RAJAGOPALAN_EF_2005}. Figure \ref{fig:2.1_2a} shows a flow visualization in this regime from \cite{PRADAD_JFM_1997A}.

The Kelvin-Helmholtz instability manifests as intermittent high frequency bursts in the velocity field with a characteristic frequency $f_{KH}$ - figures \ref{fig:2.1_2b} and \ref{fig:2.1_2c}. The irregular behaviour of this instability increases with $Re$. Although Reynolds number values ranging from 350 to 2600 are also reported in the literature \citep{RAJAGOPALAN_EF_2005, PRADAD_JFM_1997A}, there is a general consensus that this instability can only be reliably observed beyond $Re>1200$ \citep{PRADAD_JFM_1997A}. This phenomenon is also critically dependent on influencing parameters such as ending conditions \citep{UNAL_JFM_1988, PRADAD_JFM_1997A}. These small coherent structures, called transition waves in \cite{BLOOR_JFM_1964}, are responsible for the turbulence onset/transition.

After the onset of turbulence, the two free shear-layers roll-up and generate the primary wake instability characterized by a periodic transverse motion with frequency $f_{vs}$. Such instability is variously called vortex-shedding, K\'arm\'an-B\'ernard or simply Von-K\'arm\'an vortex-street. Throughout this manuscript this instability will be referred as primary or vortex-shedding instability. Figure \ref{fig:2.1_2c} depicts a representative spectra of a stream-wise velocity signal in the free shear-layer. It shows that while $f_{vs}$ has a sharp peak, $f_{KH}$ presents a relatively broad band resulting from the intermittency of this instability. 

Literature reports the existence of more instabilities in this regime. \cite{MANSY_JFM_1994} and \cite{LIN_JFS_1995} observed the existence of mode B instabilities up to $Re\approx 10,000$. Instabilities resulting from the pairing and breakdown of the vortex-shedding and Kelvin-Helmholtz vortices have also been reported \citep{RAJAGOPALAN_EF_2005}. These are understood to be sub-harmonics and harmonics of $f_{vs}$ and $f_{KH}$. \cite{LEHMKUHL_PF_2013} observed the existence of a frequency significantly lower than $f_{vs}$. Apart from some harmonics, none of these instabilities is expected to have higher frequencies than $f_{KH}$.

In summary, the development of a flow in the TSL regime entails four important steps:
\begin{enumerate}
\item onset of the Kelvin-Helmholtz instability in the free shear-layer;
\item spatial development of the Kelvin-Helmholtz rollers;
\item breakdown to high intensity turbulence;
\item turbulent shear-layer roll-up leading to vortex-shedding.
\end{enumerate}
It is evident that these flow features must be accurately simulated in order to reasonably predict the flow statistics for practical applications. We will now develop the criteria for optimal simulation of these structures and flow features.

\section{Closure Modelling and Simulation}
\label{sec:3}

As mentioned in the introduction, fluid motions containing coherent structures can be best described with a triple decomposition \citep{REYNOLDS_JFM_1970, SCHIESTEL_PF_1987}. Hence, any dependent flow variable $\Phi$ is seen as a combination of an appropriate averaged ($\overline{\Phi}$), coherent ($\tilde{\phi}$), and stochastic background turbulence ($\phi'$) parts,
\begin{equation}
\label{eq:3_0}
\Phi=\overline{\Phi}+\tilde{\phi}+\phi'\; .
\end{equation}
The coherent field is typically dominated by strong flow-dependent spatio-temporal correlations. For this reason, the simulation and modelling of this component is particularly complex. The stochastic turbulent field, on the other hand, exhibits more universal features with shorter spatio-temporal correlation distances. Although DNS and LES are capable of accurately computing such flows, their numerical demands are prohibitively expensive in many cases. A computationally viable approach is therefore often necessary.

\subsection{Reynolds-Averaged Navier-Stokes Closures}
\label{sec:3.2}
By definition, current one-point turbulence closures account for the effect of the turbulent field on the mean flow equations by employing either an algebraic constitutive relation or evolution equations for the Reynolds stresses. One-point closures employing algebraic constitutive relations obtain length and time scales by solving zero, one, and two extra transport equations. The algebraic constitutive relationships are typically based on one of the following three tenets:  molecular analogy (Boussinessq approximation), weak-equilibrium assumption (algebraic Reynolds stress models - \cite{POPE_JFM_1975}; \cite{RODI_ZAMM_1976}; \cite{GATSKI_JFM_1993} and \cite{GIRIMAJI_TCFD_1996}) or extended thermodynamic concepts \citep{HUANG_TCFD_1996}. The more advanced algebraic models can also account for anisotropic turbulent viscosity and non-linear constitutive relation effects. However, all these closures assume turbulence in a fully-developed near-equilibrium state (wherein this is a balance between linear, non-linear and viscous mechanisms). \cite{GIRIMAJI_TCFD_2001} has demonstrated how some transient effects can be included in an algebraic closure. All these closures preclude turbulence far from equilibrium and exclude the presence of any organized motion or coherent structures in the modelled flow field.

Reynolds stress transport models \citep{LAUNDER_JFM_1975,SPEZIALE_JFM_1991} can potentially incorporate more non-local effects and render a more accurate account of the energy transfer amongst  different normal components. The pressure-strain correlation closure proposed by \cite{GIRIMAJI_JFM_2000} goes further by including a reasonable transition from the rapid distortion limit to a fully-developed state of turbulence. Yet, as pointed out in \cite{MISHRA_FTC_2010, MISHRA_JFM_2013, MISHRA_JFM_2017}, non-local pressure effects present a formidable challenge even in homogeneous flows. Much of the physics of instabilities is incumbent in the pressure-strain correlation term \citep{MISHRA_JFM_2017}. Most closures of this term are based on the assumption of homogeneous turbulence. Hence, such models are inherently incapable of capturing instabilities that arise due to inhomogeneity effects. Thus, the organized motion arising out of mean flow inhomogeneity and flow instabilities cannot still be accurately represented by general Reynolds stress transport models.

One-point closure models account for flow inhomogeneities only via turbulent transport closures. Typically, these are modelled using the gradient transport hypothesis which also excludes any instabilities. Overall, the main shortcoming of one-point closures are: $i)$ the inability to accurately model the physical features of flow instabilities in the unresolved field; and $ii)$ the inadequacy in accounting for transient turbulence effects. These effects are critical in flows with spatially-developing coherent structures.

Multi-point closures can, in principle, account for some degree of spatial coherence in the flow field. However, at the current time, neither suitable closure models nor viable numerical approaches have been developed for computing inhomogeneous turbulent flows with coherent structures.

\subsection{Scale-Resolving Simulation Approaches}
\label{sec:3.3}
\begin{figure}
\centering
\includegraphics[scale=0.18]{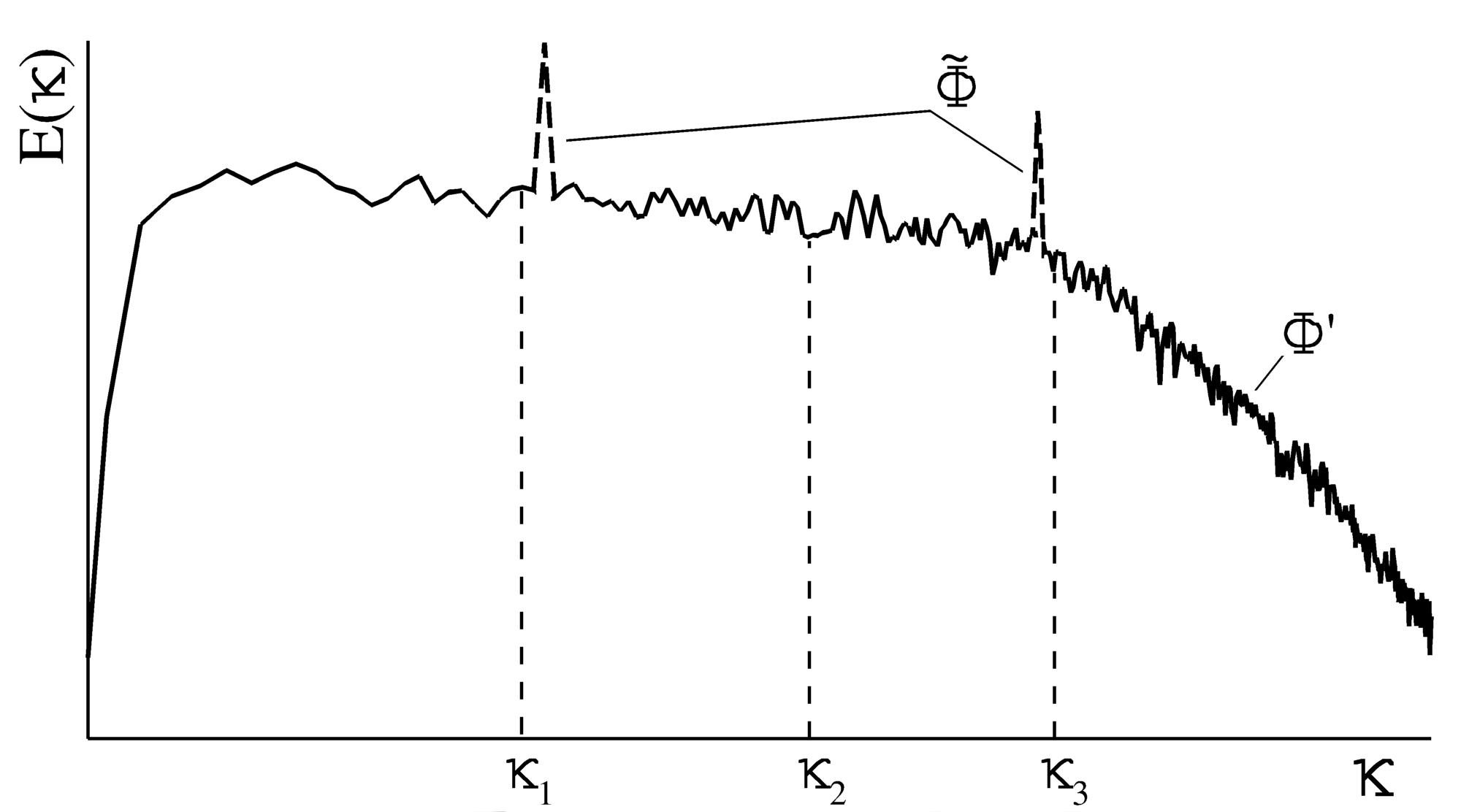}
\caption{Typical energy spectrum containing coherent and turbulent instabilities.}
\label{fig:3_1}
\end{figure}

Over the last two decades, hybrid and bridging SRS formulations have emerged for the express purpose of simulating practical flows containing coherent structures. The aim of any SRS formulation of practical interest is to resolve only the coherent component of the flow field and model the stochastic part with a suitable closure - figure \ref{fig:3_1}. The resolution of the coherent field obviates the need for accurate closure modelling. Therefore, the extent of resolution is dictated by the complexity of the coherent flow field and the required degree of accuracy. The concept of accuracy-on-demand is embedded in these methods. 

The SRS governing equations are based upon the scale invariance property of the filtered Navier-Stokes equations \citep{GERMANO_JFM_1992}. Let us assume a general filtering operator, constant preserving, and commuting with spatial and temporal differentiation. This filter decomposes any dependent quantity $\Phi$ into a filtered/resolved, $\langle \Phi \rangle$, and a unresolved/modelled, $\phi$, component,
\begin{equation}
\label{eq:3_1}
 \Phi \equiv \langle \Phi \rangle+ \phi\; .
\end{equation}
Ideally, the resolved flow field $\langle \Phi \rangle$ comprises both mean and coherent parts. The unresolved component $\phi$ is then modelled as stochastic field. The challenge of SRS lies in guaranteeing that an optimal portion the coherent field is resolved.

The application of the aforementioned operator to the incompressible Navier-Stokes equations leads to its filtered form \citep{GERMANO_JFM_1992},
\begin{equation}
\label{eq:3_2}
\frac{\partial \langle V_i \rangle}{\partial x_i} = 0\; ,
\end{equation}
\begin{equation}
\label{eq:3_3}
\frac{\partial \langle V_i \rangle}{\partial t} + \langle V_j\rangle \frac{\partial \langle V_i \rangle}{\partial x_j} = -\frac{1}{\rho}\frac{\partial \langle P\rangle}{\partial x_i} + \nu \frac{\partial^2 \langle V_i \rangle}{\partial x_j \partial x_j} +\frac{1}{\rho}\frac{\partial \tau \left(V_i,V_j\right)}{\partial x_j}\; ,
\end{equation}
where $V_i$ are the Cartesian velocity components, $P$ is the pressure, $\rho$ is the fluid density, and $\tau(V_i,V_j)$ is the generalized central second moment or Sub-Grid Stresses (SGS) tensor which accounts for the effects of the modelled velocity field on the resolved field. Clearly, a closure model for the sub-grid stresses tensor is needed. In LES, Smagorinsky-type closures are often used. Other SRS formulations employ $n$-equations closures. The Detached Eddy Simulations of \cite{SPALART_AFOSRIC_1997}, and the PANS of \cite{GIRIMAJI_JAM_2005} and \cite{BASARA_AIAA_2011} are cases of such formulations. The use of second moment closure models is also possible as demonstrated in \cite{CHAOUAT_PF_2005} where such type of closure is adapted to the Partially-Integrated Transport Model \citep{CHAOUAT_PF_2005,SCHIESTEL_TCFD_2005}. PANS is the SRS approach employed in this investigation.

\subsection{Partially-Averaged Navier-Stokes Equations}
\label{sec:3.1}

The quality of a SRS is clearly dependent on the range of resolved scales (physical resolution or cut-off length-scale), and on the fidelity of the closure model used to represent the stochastic field. It is expected that the accuracy of a SRS computation will improve with an increase in the range of resolved scales as the closure model contribution progressively diminishes. 

Although one-point closures are inadequate to represent the physics of a coherent velocity field, they are reasonably accurate for modelling fully-developed stochastic turbulence. Since the intent of PANS is to resolve the coherent structures, it can be argued that the residual unresolved field can be adequately represented by a simple one-point closure such as the Boussinesq relation:
\begin{equation}
\label{eq:3_9}
\frac{\tau_{ij}(V_i,V_j)}{\rho}=2\nu_u \langle S_{ij}\rangle - \frac{2}{3}k_u\delta_{ij}\;,
\end{equation}
where $\langle S_{ij}\rangle$ is the resolved strain-rate tensor,
\begin{equation}
\label{eq:3_9}
\langle S_{ij} \rangle = \frac{1}{2} \left( \frac{\partial \langle V_i\rangle}{\partial x_j} +\frac{\partial \langle V_j\rangle}{\partial x_i} \right) \;,
\end{equation}
$\nu_u$ is the turbulent or eddy viscosity of the unresolved field, $k_u$ is the modelled turbulence kinetic energy, and $\delta_{ij}$ is the Kronecker delta.  For the sake of clarity we use the subscript $u$ to denote the unresolved component of equation \ref{eq:3_1} (instead of just lower-case and capital symbols). The turbulent viscosity and kinetic energy are then calculated through a PANS closure-equations. 

The range of resolved scales in PANS closures is controlled by two parameters,
\begin{equation}
\label{eq:3_4}
f_k\equiv \frac{k_u}{k} \;, \hspace{.5cm} f_\epsilon\equiv \frac{\epsilon_u}{\epsilon} \;,
\end{equation}
which characterize the fraction of turbulence kinetic energy and dissipation rate ($\epsilon$) being modelled. Alternatively, the decomposition can also be effected in terms of $f_k$ and $f_\omega$:
\begin{equation}
\label{eq:3_5}
f_\omega \equiv \frac{\omega_u}{\omega} \equiv  \frac{f_\epsilon}{f_k} \;,
\end{equation}
where $\omega$ is the specific dissipation.

The $k-\omega$ PANS \citep{LAKSHMIPATHY_AIAA_2006} and the $k-\omega$ Shear-Stress Transport (SST) PANS \citep{PEREIRA_THMT_2015} are the closures employed  in this investigation:
\begin{equation}
\label{eq:3_6}
\frac{Dk_u}{Dt}=\tau_{ij}\frac{\partial \langle V_i\rangle}{\partial x_j}-\beta^*k_u\omega_u +\frac{\partial}{\partial x_j}\left[\left( \nu+\nu_u \sigma_k \frac{f_\omega}{f_k} \right)\frac{\partial k_u}{\partial x_j}\right]\;,
\end{equation}
\begin{equation}
\label{eq:3_7}
\frac{D\omega_u}{Dt}=\frac{\alpha}{\nu_u}\tau_{ij} \frac{\partial \langle V_i\rangle}{\partial x_j} - \left( \alpha \beta^* -\frac{\alpha \beta^*}{f_\omega} + \frac{\beta}{f_\omega} \right)\omega_u^2 + \frac{\partial}{\partial x_j} \left[\left(\nu+\nu_u\sigma_\omega\frac{f_\omega}{f_k}\right)\frac{\partial \omega_u}{\partial x_j}\right] +D_{c}\;.
\end{equation}
In equation \ref{eq:3_7}, $D_c$ is the cross-diffusion term,
\begin{equation}
\label{eq:3_8}
D_c= \left\{
\begin{array}{ll}
0\; ,  &(k-\omega)\\ 
&\\
2 \frac{\sigma_{\omega_2}}{\omega} \frac{f_\omega}{f_k}\left(1-F_1\right) \frac{\partial k_u}{\partial x_j} \frac{\partial \omega_u}{\partial x_j} \; , &(\text{SST})
\end{array}
\right.
\; ,
\end{equation}
and the turbulent viscosity is defined as
\begin{equation}
\label{eq:3_10}
\nu_u= \left\{
\begin{array}{ll}
\frac{k_u}{\omega_u}\; ,  &(k-\omega)\\ 
&\\
\frac{a_1k_u}{\max\left\{a_1\omega_u; \langle S \rangle F_2\right\}} \; , &(\text{SST})
\end{array}
\right.
\; .
\end{equation}
The remaining coefficients and auxiliary functions are as given in \cite{WILCOX_AIAA_1988} and \cite{MENTER_THMT_2003}.

The turbulent viscosity of the unresolved field in a PANS simulation can be approximated as
\begin{equation}
\label{eq:3_15}
\nu_u \approx \frac{f_k}{f_\omega} \frac{k}{\omega} \approx \frac{f_k^2}{f_\epsilon} C_\mu \frac{k^2}{\epsilon}\; ,
\end{equation}
for one-point closures based on $k-\omega$ and $k-\epsilon$ formulations. In high Reynolds number flows, all of the dissipation is expected to occur at the unresolved scales, leading to $f_\epsilon=1.0$. Then, we can write,
\begin{equation}
\label{eq:3_16}
\nu_u \approx f_k^2 \frac{k}{\omega} \approx f_k^2 C_\mu \frac{k^2}{\epsilon} = f_k^2 \nu_t \; ,
\end{equation}
where $\nu_t$ is the corresponding turbulent viscosity of modelling the entire turbulent field (RANS). Clearly, the turbulent viscosity vanishes as $f_k$ tends to zero. At the other extreme, the turbulent viscosity of the unresolved fields goes to the RANS values ($\nu_t$) when $f_k$ approaches unity. \cite{GIRIMAJI_AIAA43_2005} have proposed the following relation to estimate the maximum physical resolution $f_k$ that a given spatial grid resolution can support,
\begin{equation}
\label{eq:3_17}
f_k \ge \frac{1}{\sqrt{C_\mu}} \left( \frac{\Delta}{L} \right)^{2/3} \; .
\end{equation}
In equation \ref{eq:3_17}, $\Delta$ is an appropriate measure of the grid spatial resolution, and $L$ is the characteristic length of the largest turbulent scales - $L=k^{3/2}/\epsilon$. From this relation, it is possible to infer that as $f_k$ decreases, the computational requirements increase.

The flow field of PANS and other SRS methods can be analysed most conveniently as the direct numerical simulation of a non-Newtonian fluid at a lower effective computational Reynolds number $Re_e$ \citep{DASIA_PF_2014},
\begin{equation}
\label{eq:3_13}
Re_e \equiv \frac{\overline{V}_\infty D}{\nu+\nu_u}\; .
\end{equation}
The consequence of this decrease of Reynolds number is the reduction of the numerical effort. In order to illustrate this, assume a high Reynolds number flow so that all dissipation occurs entirely in the unresolved scales - $f_\epsilon=1.00$. Then, it is possible to estimate the range of scales being resolved for a given $f_k$ \citep{DASIA_PF_2014},
\begin{equation}
\label{eq:3_14}
\frac{L}{\eta_u} \sim  Re_e^{3/4} =  \left( \frac{\overline{V}_\infty D}{\nu + \nu_t f_k^2} \right)^{3/4}\; .
\end{equation}
Here, $\eta_u$ is the computational Kolmogorov length scale (smallest scales being resolved) given by
\begin{equation}
\label{eq:3_31}
 \eta_u=\left(\frac{\left(\nu+\nu_u\right)^3}{\epsilon}\right)^{1/4}\; .
\end{equation}
Note that the length of $\eta_u$ should be of the size of the spatial grid resolution. From equations \ref{eq:3_13} and \ref{eq:3_14}, it is possible to infer that as $f_k$ tends to unity, the simulation approaches a RANS computation. At the other extreme, as $f_k$ vanishes the effective computational Reynolds number approaches the flow Reynolds number, leading to a DNS. Between these extremes of $f_k$, the effective Reynolds number takes intermediate values.
%

\section{Problem Setup}
\label{sec:4}

The selected test-case is the flow around a circular cylinder at $Re=3900$. This flow is representative of the regime described in Section \ref{sec:2_3} where  the vortex-shedding and Kelvin-Helmholtz instabilities are the dominant spatially-developing coherent structures. It is therefore expected that the findings from this study can be used to establish general guidelines for the minimum resolution necessary for flows with coherent structures.

%
\subsection{Numerical Details}
\label{sec:4_2}

The numerical simulations are carried out with the finite volume solver \cite{ReFRESCO} wherein all terms of the closed set of equations are discretized with second-order accurate schemes. The values prescribed in PANS simulations for the fraction of the turbulence kinetic energy being modelled, $f_k$, are $0.25$, $0.50$, $0.75$ and $1.00$ (0.15 is also used for the SST closure). It is assumed that dissipation, $\epsilon$, occurs entirely in the unresolved scales so that $f_\epsilon=1.00$ ($f_\omega=f_\epsilon/f_k$). The computational domain is a rectangular prism centred at the cylinder's axis with stream-wise, $x_1$, transverse, $x_2$, and span-wise, $x_3$, lengths of $50D$, $22D$, and $3D$. The boundary conditions are similar to those employed in \cite{PEREIRA_IJHFF_2017}: velocity and turbulent quantities are set constant at the inlet boundary, $x_1/D=-10$, whereas the pressure is extrapolated from the interior of the domain. The turbulent quantities $k_u$ and $\omega_u$ result from setting the turbulent intensity as $I=0.2\%$ \citep{PARNAUDEAU_PF_2008}, and $\nu_t/\nu=10^{-3}$; at the outlet, $x_1/D=40$, the stream-wise derivatives of all dependent variables are set equal to zero; at the top and bottom boundaries, $x_2/D=\pm 12$, the transverse derivatives are zero and the pressure is imposed; and symmetry boundary conditions are employed in the transverse direction \citep{PEREIRA_IJHFF_2017}. All calculations are conducted on a multi-block structured grid with $4,546,800$ volumes and a dimensionless time-step, $\Delta t V_\infty /D$, of $5.209\times 10^{-3}$. The selection of this spatial-temporal resolution is consequence of the verification studies executed in \cite{PEREIRA_IJHFF_2017}. In order to minimize round-off and iterative errors, the calculations run on double precision and the maximum norm of the normalized residual of all dependent quantities is equal to $10^{-5}$ at each time-step. The simulated time is 500 dimensionless time-units. The flow statistics are calculated with a minimum of 350 time-units. All quantities shown in this study are normalized by $V_\infty$, $D$, and $\rho$ as reference quantities.

%
\subsection{Experimental Measurements}
\label{sec:4_3}

The experimental measurements of \cite{PARNAUDEAU_PF_2008} are used as reference in this study. These experiments were carried out in a wind tunnel at $Re=3900$. The experimental facility has a square section with length $23.3D$, and the incoming flow has a turbulence intensity, $I$, lower than $0.20\%$. The quantities of interest taken from this study are the Strouhal number, St, and the time-averaged velocity magnitude, $\langle\overline{V_i}\rangle$, variance and covariance, $\overline{v_iv_j}$, fields. These fields were measured using Particle Image Velocity (PIV), and the dimensionless time, $\Delta T V_\infty/D$, used to converge the flow statistics is approximately $2.08\times 10^5$ time-units. Although at a slightly different Reynolds number, $Re=4000$, the experiments of \cite{NORBERG_BBVIV3_2002} are also used as reference for the pressure distribution on the cylinder's surface, $\overline{C}_p(\theta)$. The data were obtained on a wind tunnel with a $215.5D$ ($L_2$) by $80.0D$ ($L_3$) section and $I<0.06\%$. The measurements were carried out through pressure taps during $2.10\times 10^4$ time-units. The root-mean-square lift coefficient $C_L'$ measurements of \cite{NORBERG_JFS_2003} are also considered. The main difference to the experimental setup of \cite{NORBERG_BBVIV3_2002} are the $Re$, $L_2$, $L_3$ and $\Delta T V_\infty/D$. These are, respectively, $4400$, $314.1D$, $105.0D$ and $3.20\times 10^4$. The results shown in table \ref{tab:2_1_1} are also used in this study to demonstrate the relevance of influencing parameters. 

\section{Results and Discussion}
\label{sec:5}

This section starts by analysing the influence of the physical resolution (range of resolved scales) on the fidelity of the simulations. The results obtained at different resolutions are compared with experimental measurements in Section \ref{sec:5.1}. We seek to establish general convergence of SRS results towards a range of observations over a collection of similar experiments. Thereafter, the physical rationale behind the behaviour of the simulated results is investigated in Section \ref{sec:5.2}. Finally, in Section \ref{sec:5.3} we develop a set of guidelines for optimal SRS computations of flows with coherent structures.

%
\subsection{Effect of Physical Resolution}
\label{sec:5.1}

\begin{figure}
\centering
\subfloat[$f_k=1.00$.]{\label{fig:5.1_1a}
\includegraphics[scale=0.16,trim=0 0 0 0,clip]{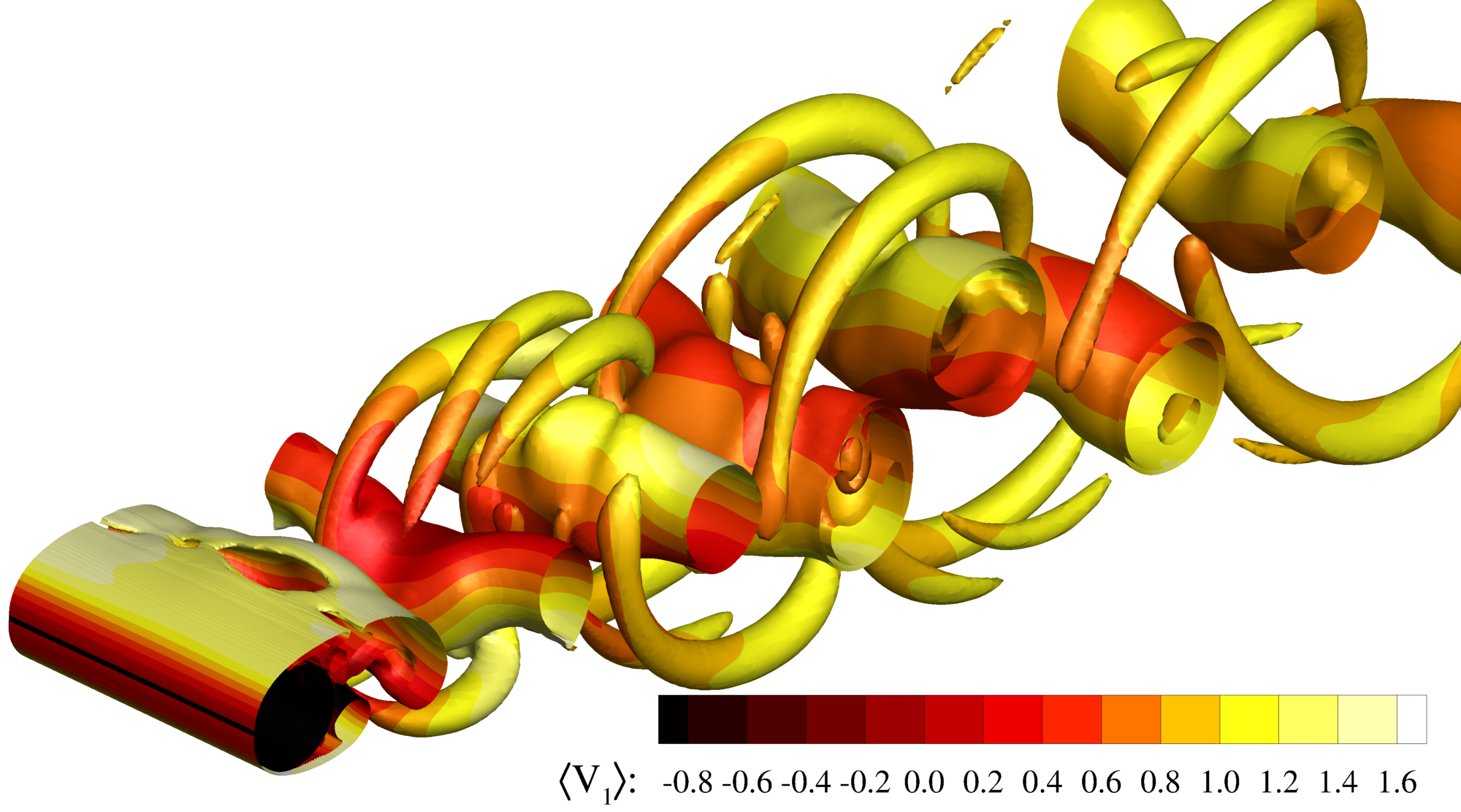}}
~
\subfloat[$f_k=0.75$.]{\label{fig:5.1_1a}
\includegraphics[scale=0.16,trim=0 0 0 0,clip]{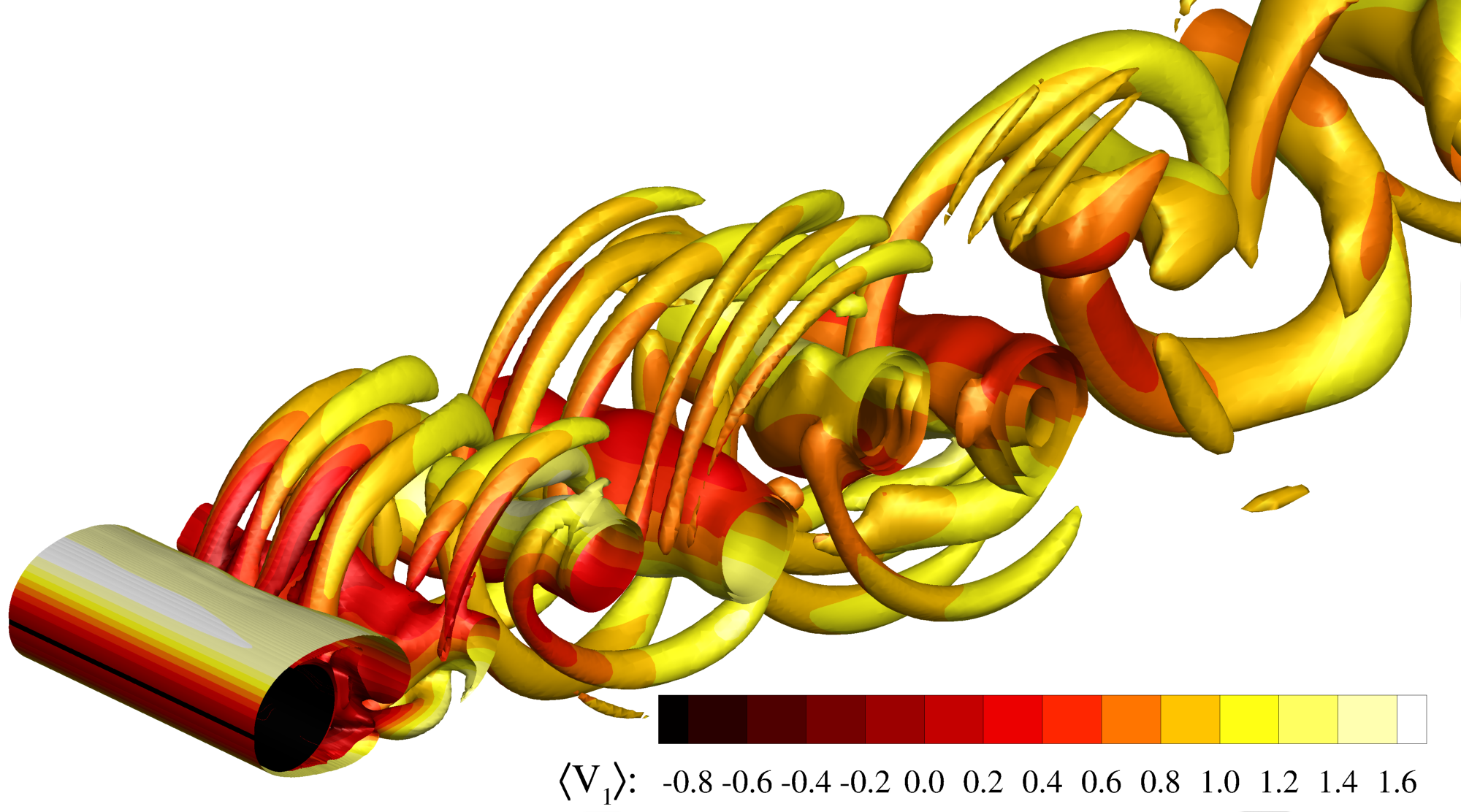}}
\\
\subfloat[$f_k=0.50$.]{\label{fig:5.1_1c}
\includegraphics[scale=0.16,trim=0 0 0 0,clip]{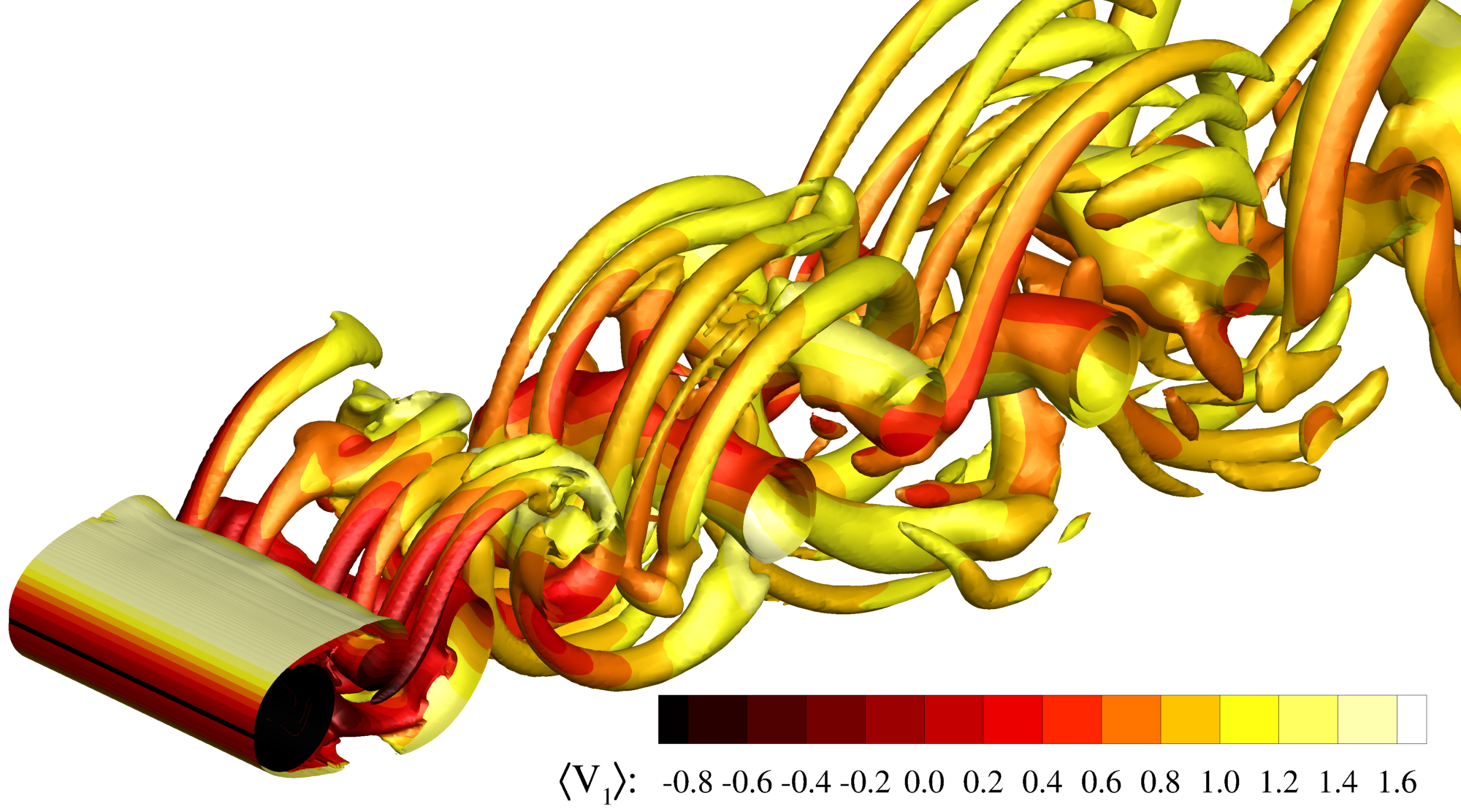}}
~
\subfloat[$f_k=0.25$.]{\label{fig:5.1_1d}
\includegraphics[scale=0.16,trim=0 0 0 0,clip]{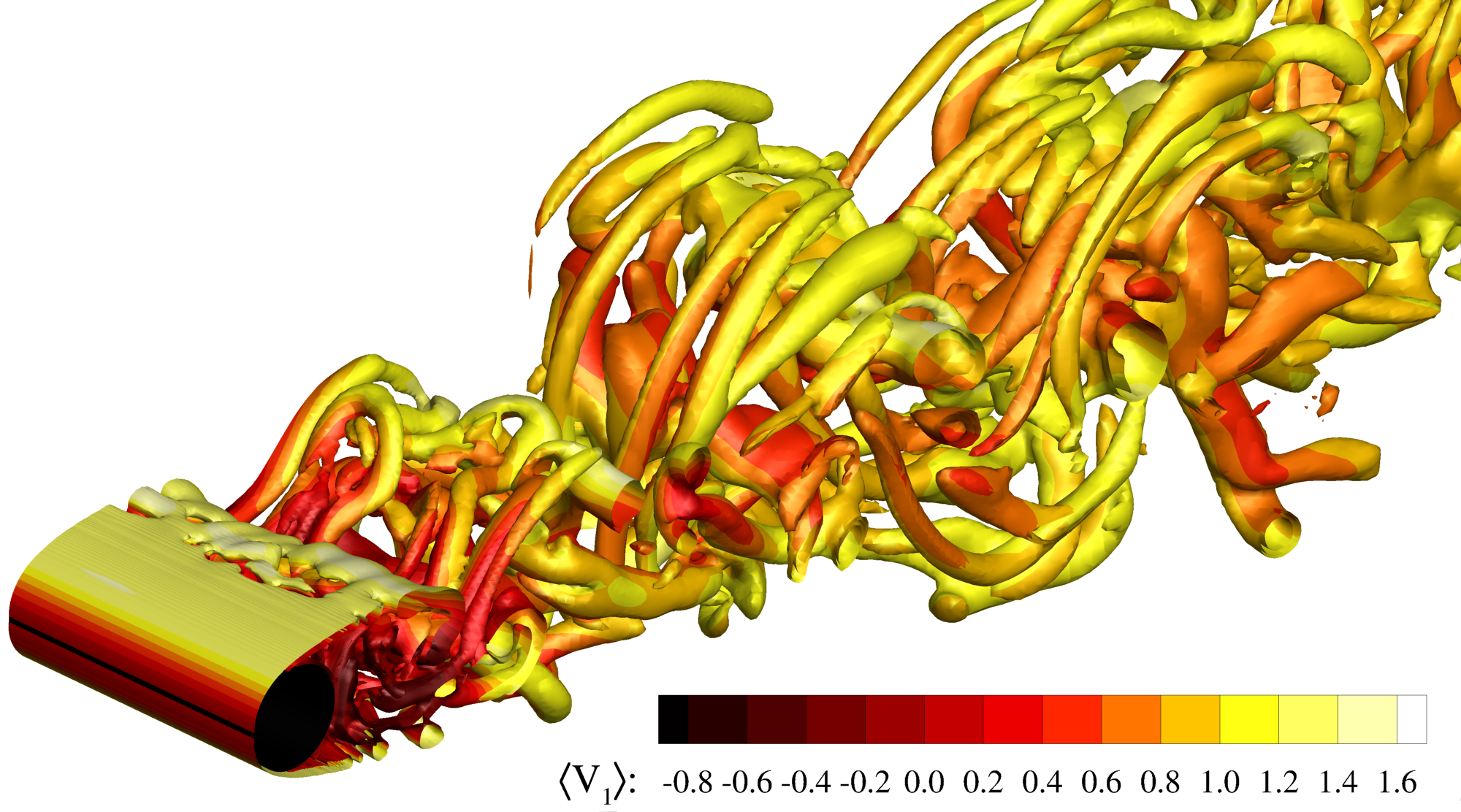}}
\caption{Instantaneous $Q$ factor iso-surfaces ($Q=0.1$, $0.2$ and $0.5$) coloured with the stream-wise velocity magnitude, $\langle V_1 \rangle$, for different physical resolutions. Results for SST PANS closure.}
\label{fig:5.1_1}
\end{figure}

\begin{figure}
\centering
\subfloat[ $C_L'$.]{\label{fig:5.1_2a}
\includegraphics[scale=0.16,trim=0 0 0 0,clip]{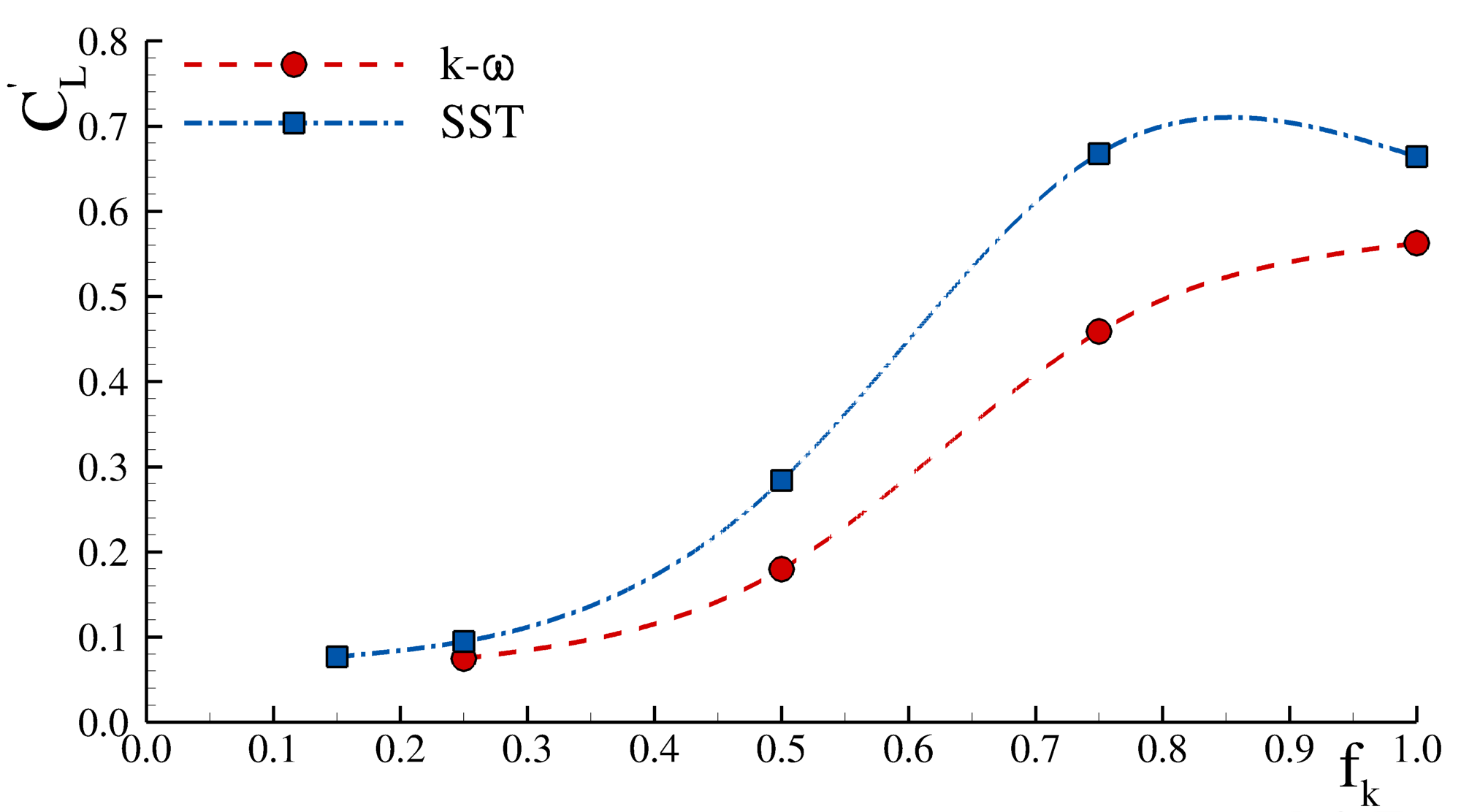}}
~
\subfloat[$\overline{C}_D$.]{\label{fig:5.1_2b}
\includegraphics[scale=0.16,trim=0 0 0 0,clip]{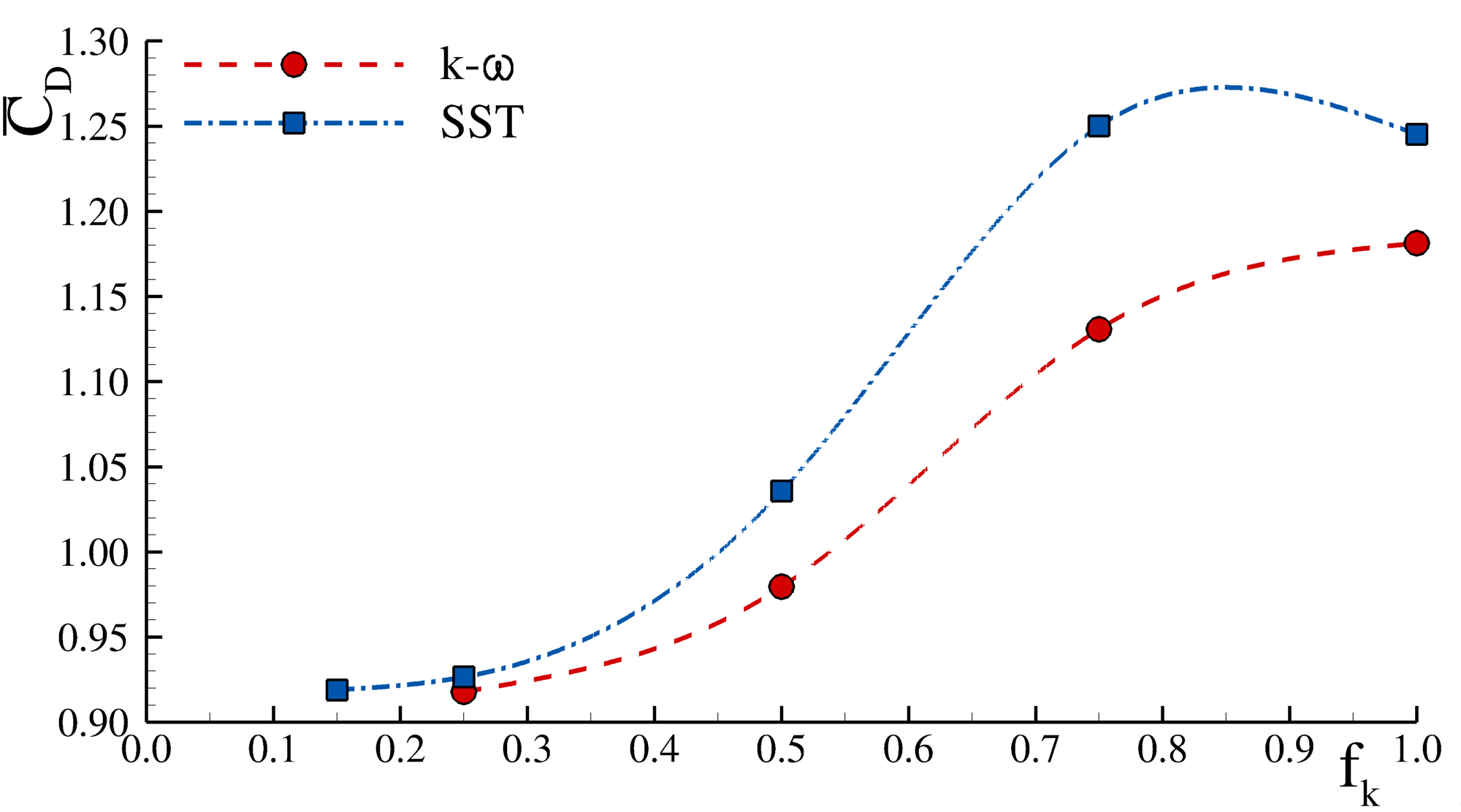}}
\\
\subfloat[ $\overline{C}_{pb}$.]{\label{fig:5.1_2c}
\includegraphics[scale=0.16,trim=0 0 0 0,clip]{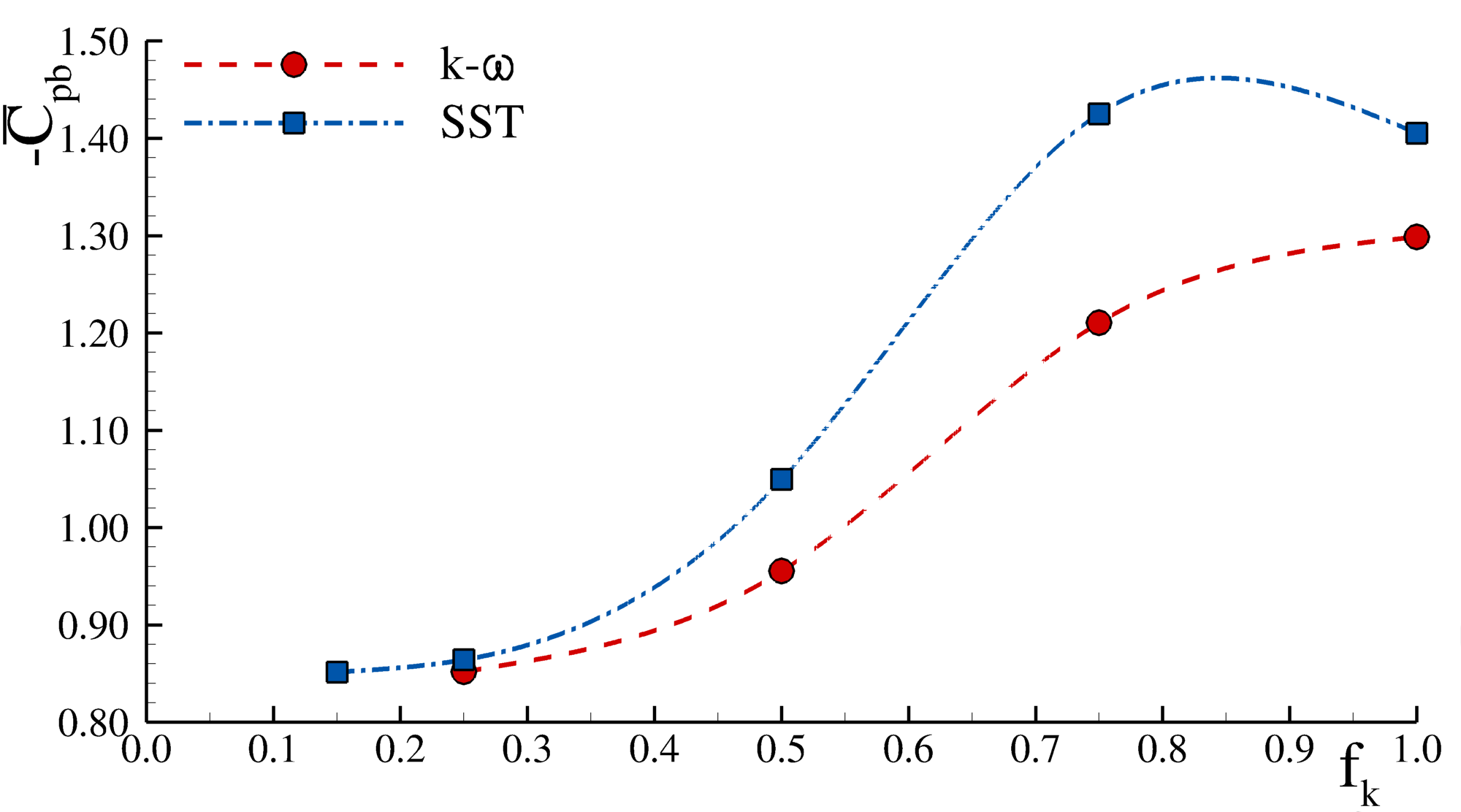}}
~
\subfloat[$\overline{L}_r$.]{\label{fig:5.1_2d}
\includegraphics[scale=0.16,trim=0 0 0 0,clip]{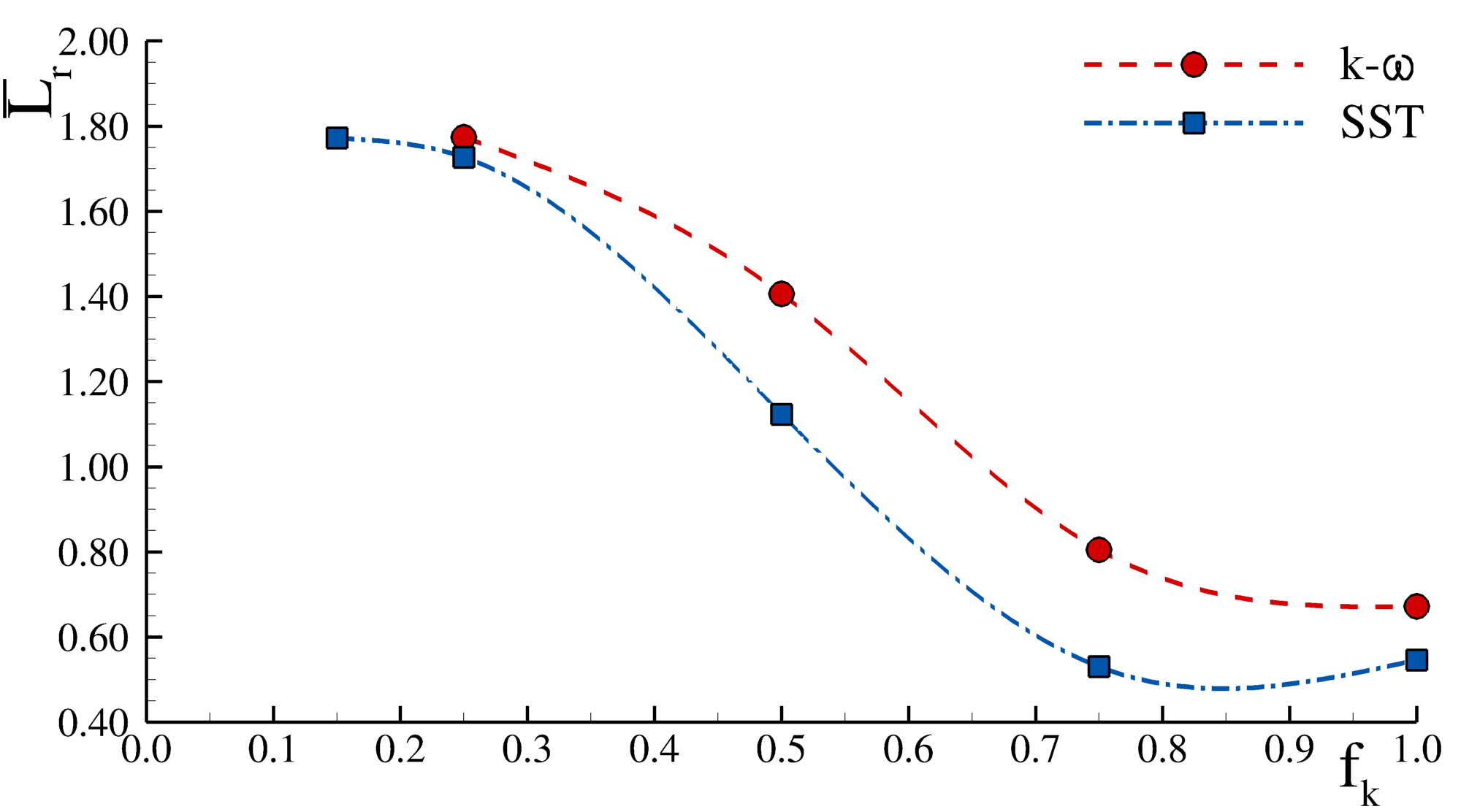}}
\caption{Convergence of the root-mean-square lift coefficient, $C_L'$, time-averaged drag coefficient, $\overline{C}_D$, pressure base coefficient, $\overline{C}_{pb}$, and recirculation length, $\overline{L}_r$, with the physical resolution for different PANS closures  (data connected through splines).}
\label{fig:5.1_2}
\end{figure}

\begin{figure}
\centering
\subfloat[ $\overline{C}_p(\theta)$.]{\label{fig:5.1_3a}
\includegraphics[scale=0.16,trim=0 0 0 0,clip]{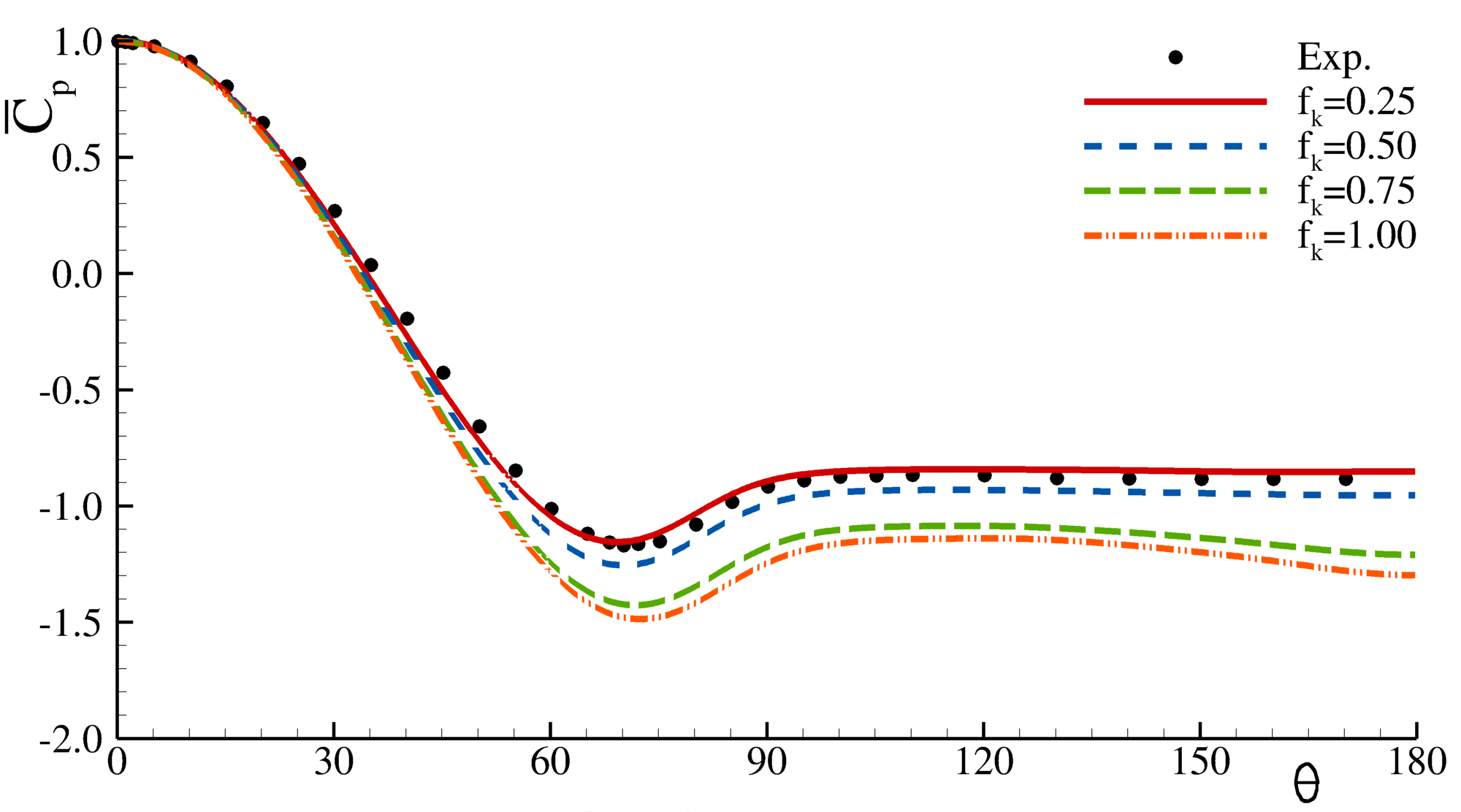}}
~
\subfloat[$\langle \overline{V_1} \rangle$.]{\label{fig:5.1_3b}
\includegraphics[scale=0.16,trim=0 0 0 0,clip]{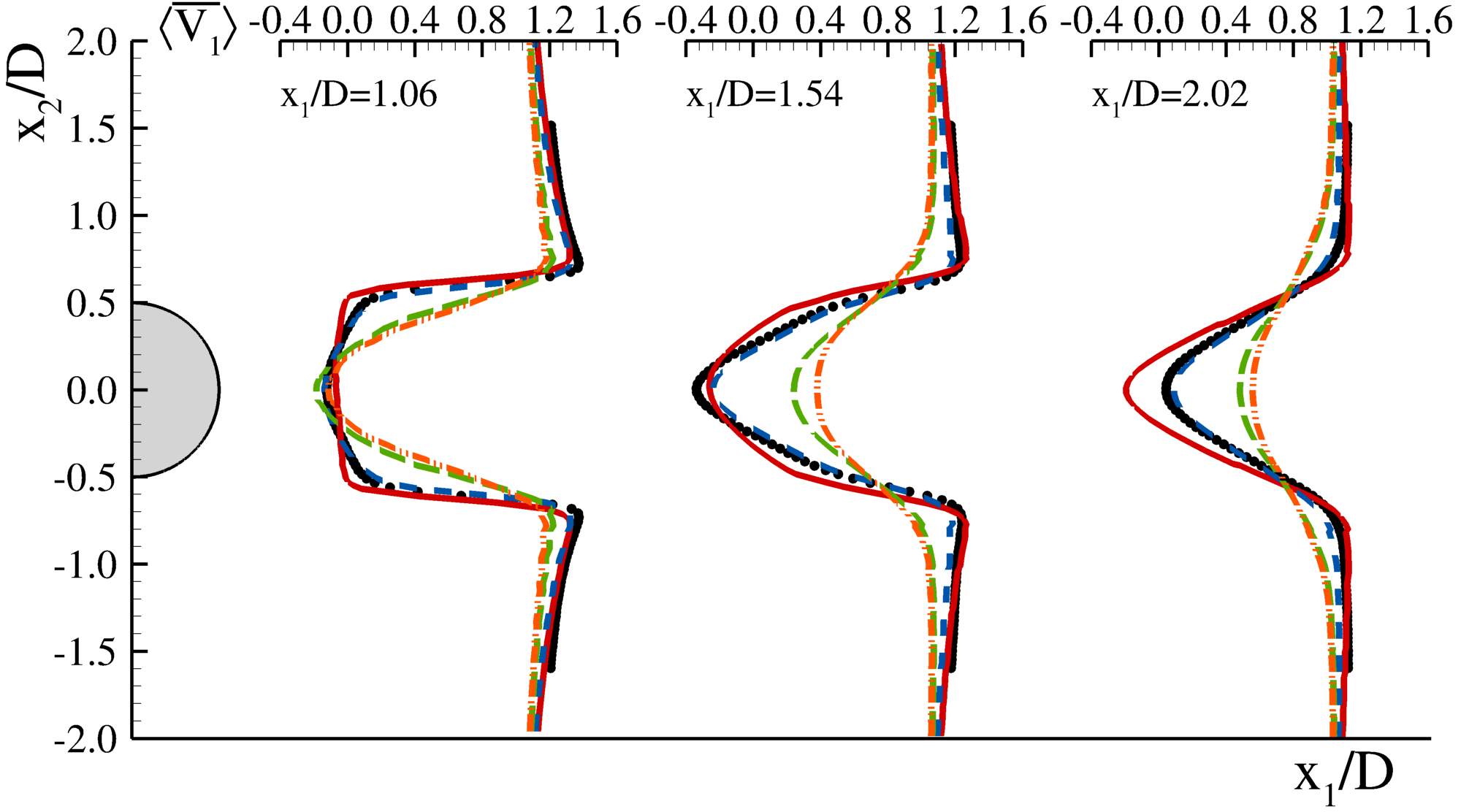}}
\\
\subfloat[ $\overline{v_1v_1}$.]{\label{fig:5.1_3c}
\includegraphics[scale=0.16,trim=0 0 0 0,clip]{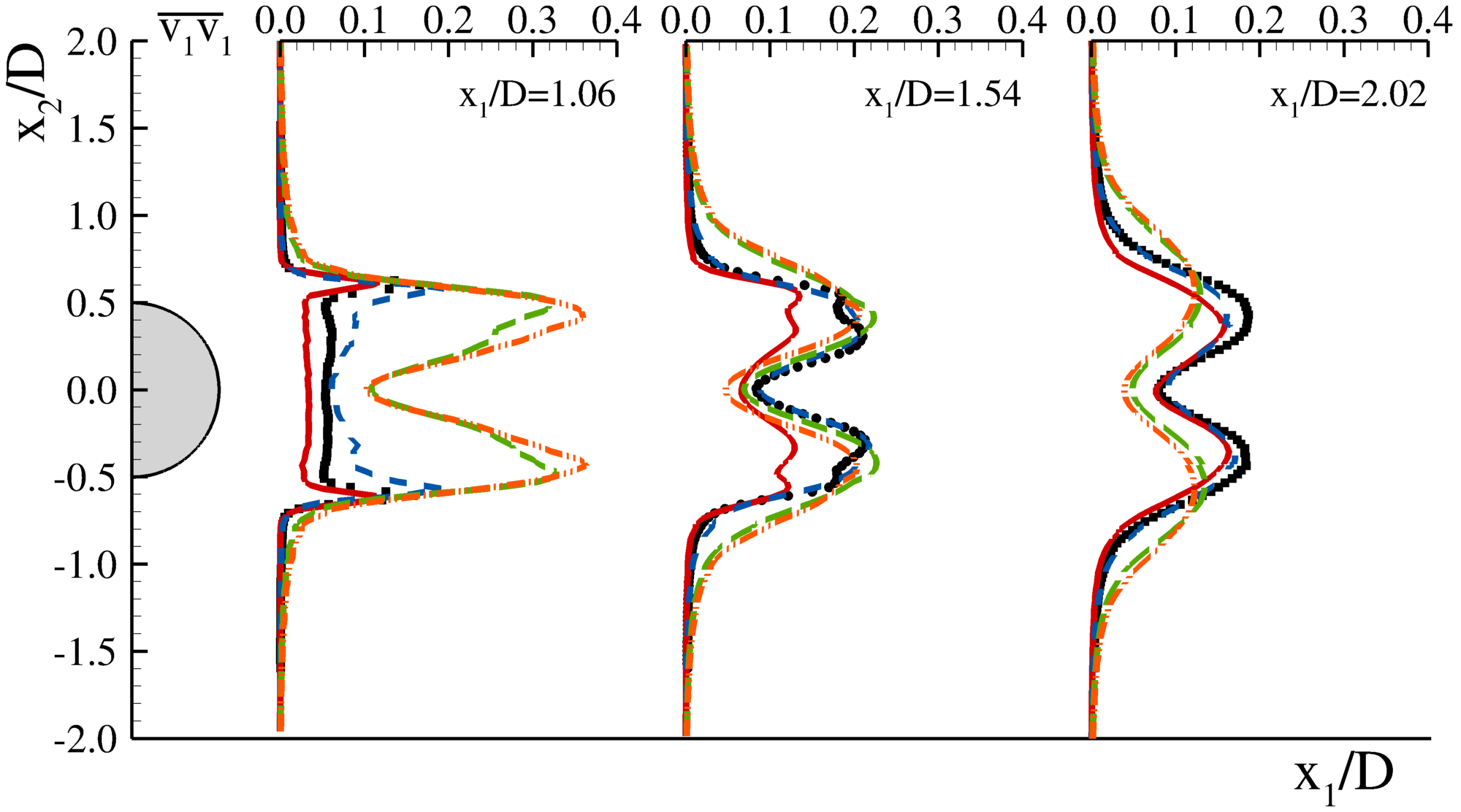}}
~
\subfloat[$\overline{v_1v_2}$.]{\label{fig:5.1_3d}
\includegraphics[scale=0.16,trim=0 0 0 0,clip]{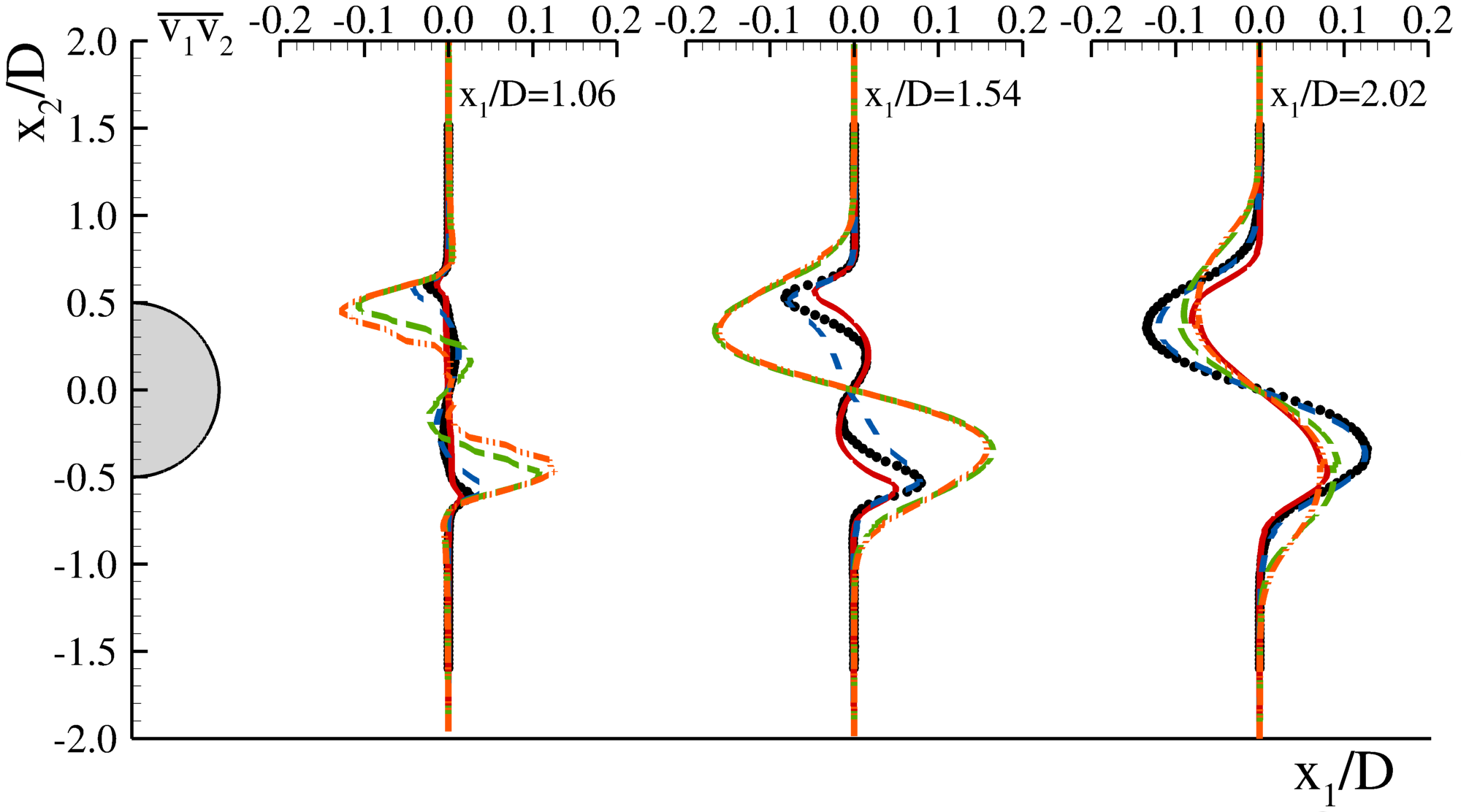}}
\caption{Time-averaged pressure distribution on the cylinder's surface, $\overline{C}_p(\theta)$, and stream-wise velocity magnitude, $\langle \overline{V_1} \rangle$, variance, $\overline{v_1v_1}$, and covariance, $\overline{v_1 v_2}$, in the near-wake for different physical resolutions. Experiments taken from \cite{NORBERG_BBVIV3_2002} ($\overline{C}_p(\theta)$) and \cite{PARNAUDEAU_PF_2008} ($\langle\overline{V}_1\rangle$, $\overline{v_1v_1}$, and $\overline{v_1v_2}$). Results for $k-\omega$ PANS closure.}
\label{fig:5.1_3}
\end{figure}

\begin{figure}
\centering
\subfloat[$f_k=1.00$.]{\label{fig:5.1_4a}
\includegraphics[scale=0.16,trim=0 0 0 0,clip]{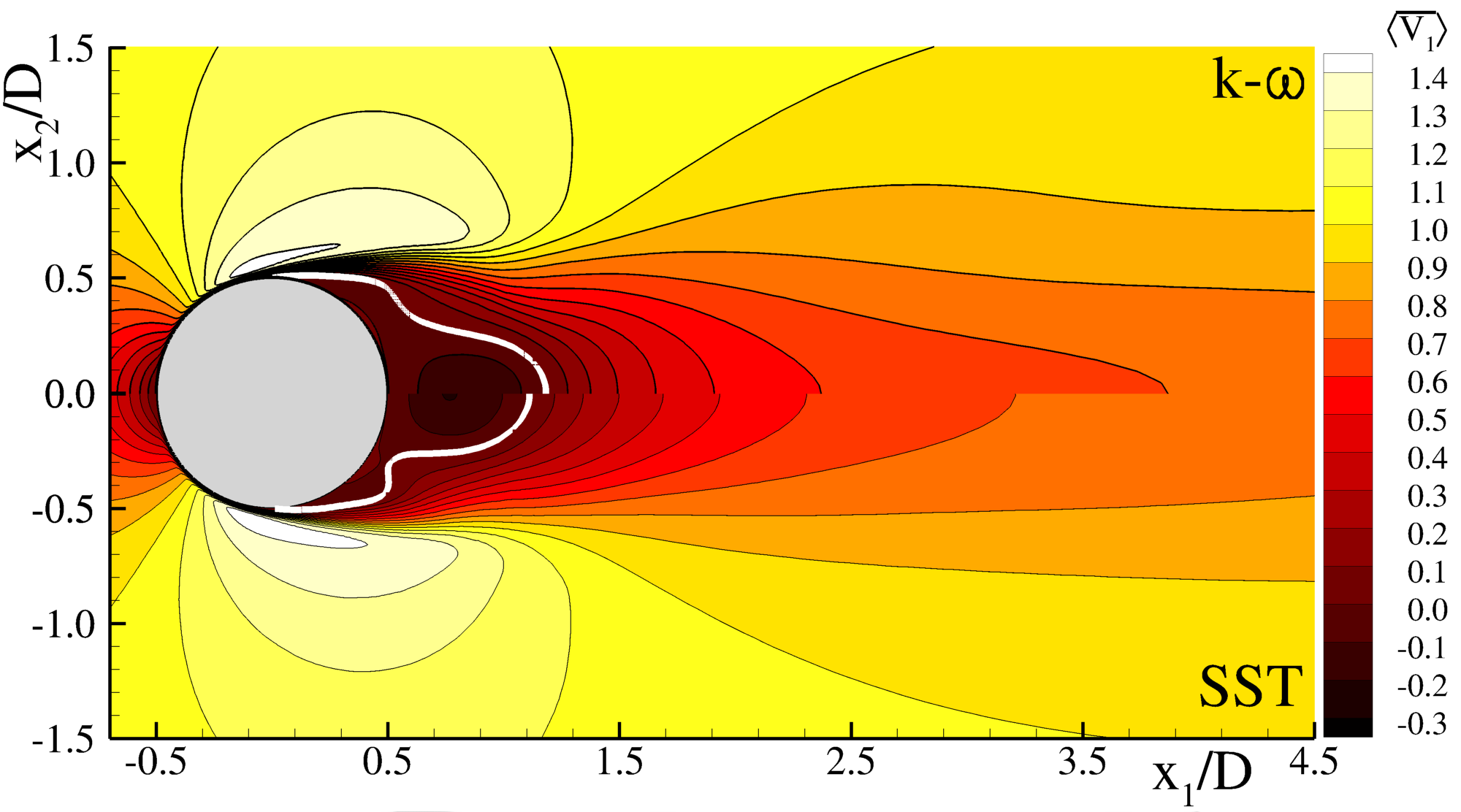}}
~
\subfloat[$f_k=0.75$.]{\label{fig:5.1_4b}
\includegraphics[scale=0.16,trim=0 0 0 0,clip]{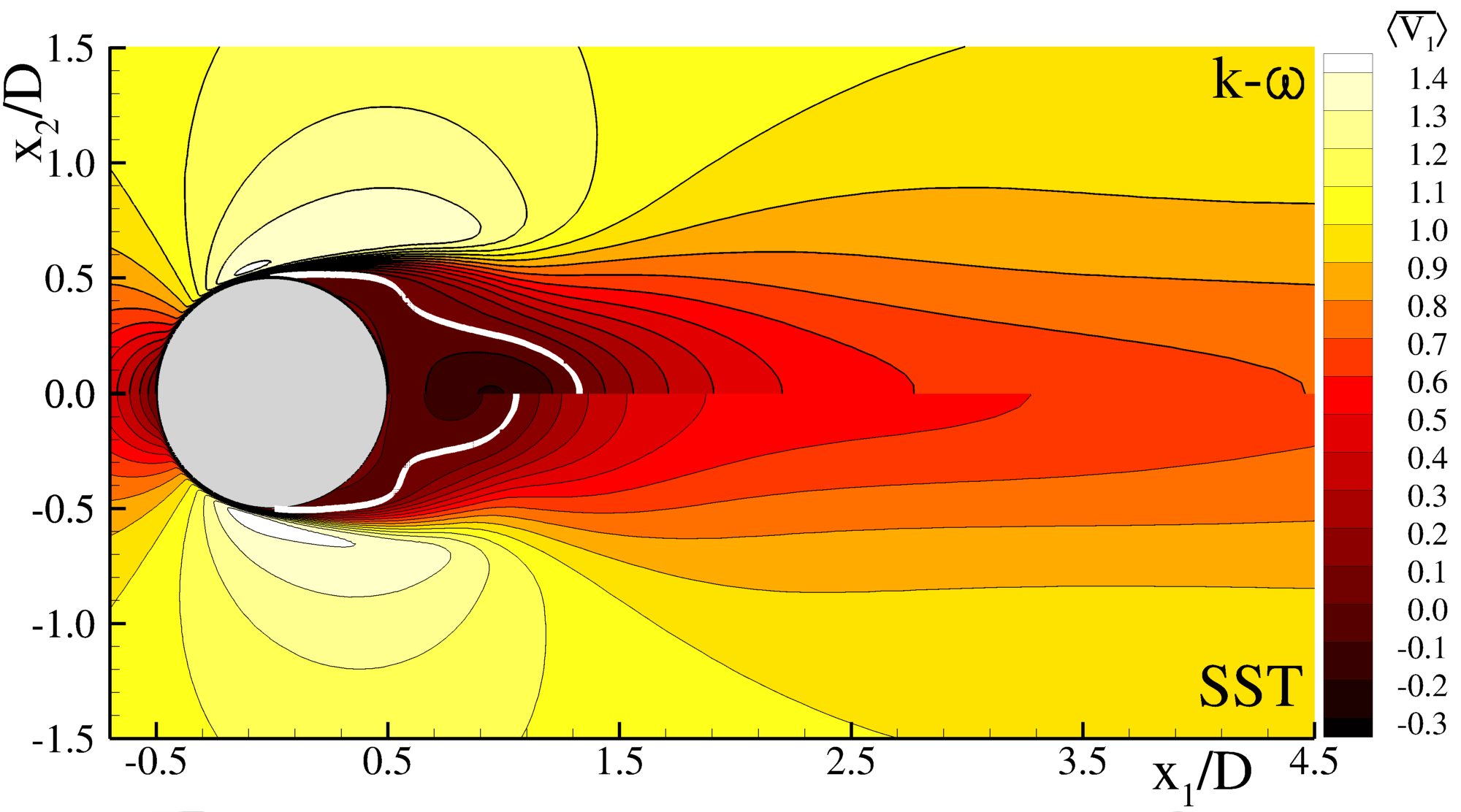}}
\\
\subfloat[$f_k=0.50$.]{\label{fig:5.1_4c}
\includegraphics[scale=0.16,trim=0 0 0 0,clip]{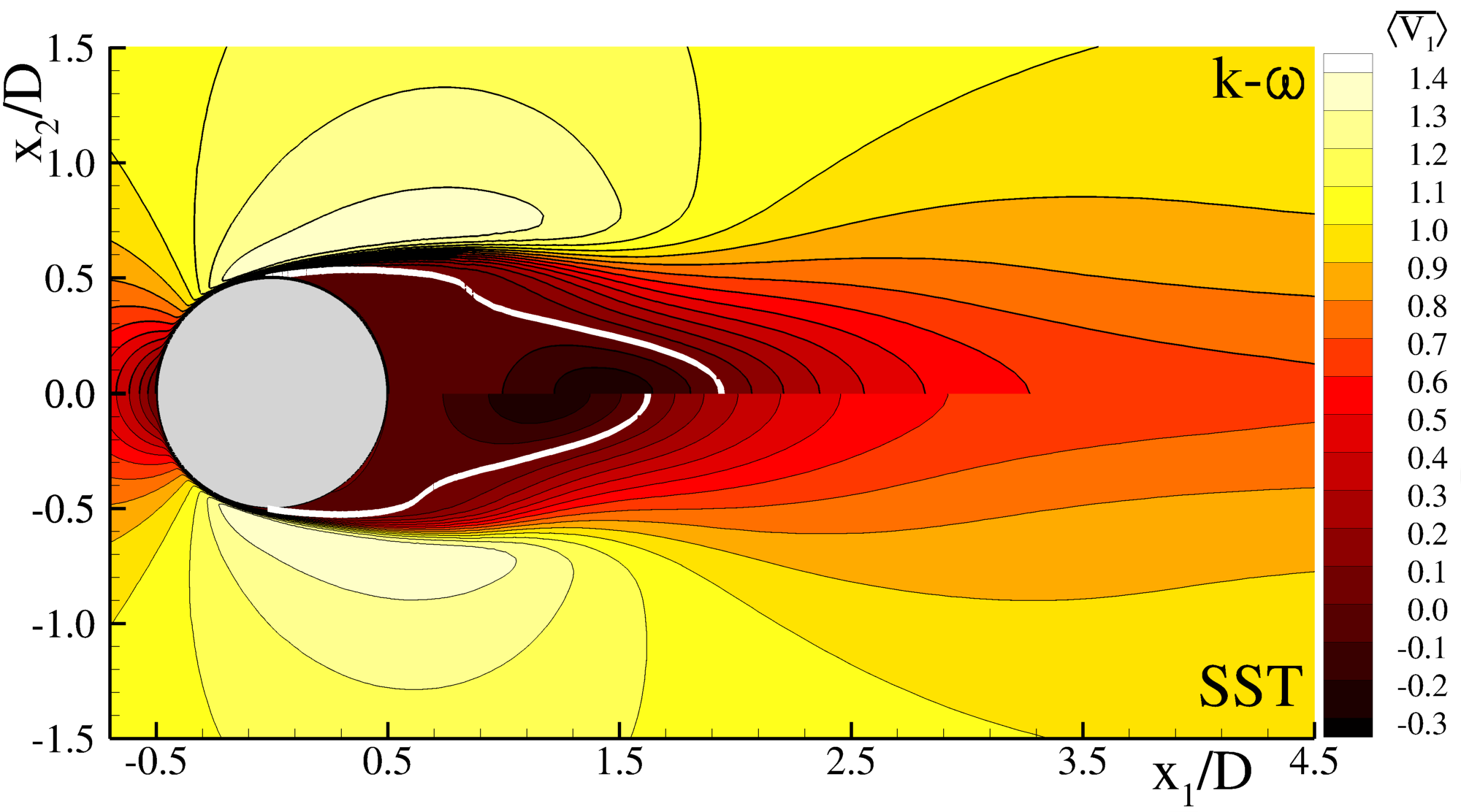}}
~
\subfloat[$f_k=0.25$.]{\label{fig:5.1_4d}
\includegraphics[scale=0.16,trim=0 0 0 0,clip]{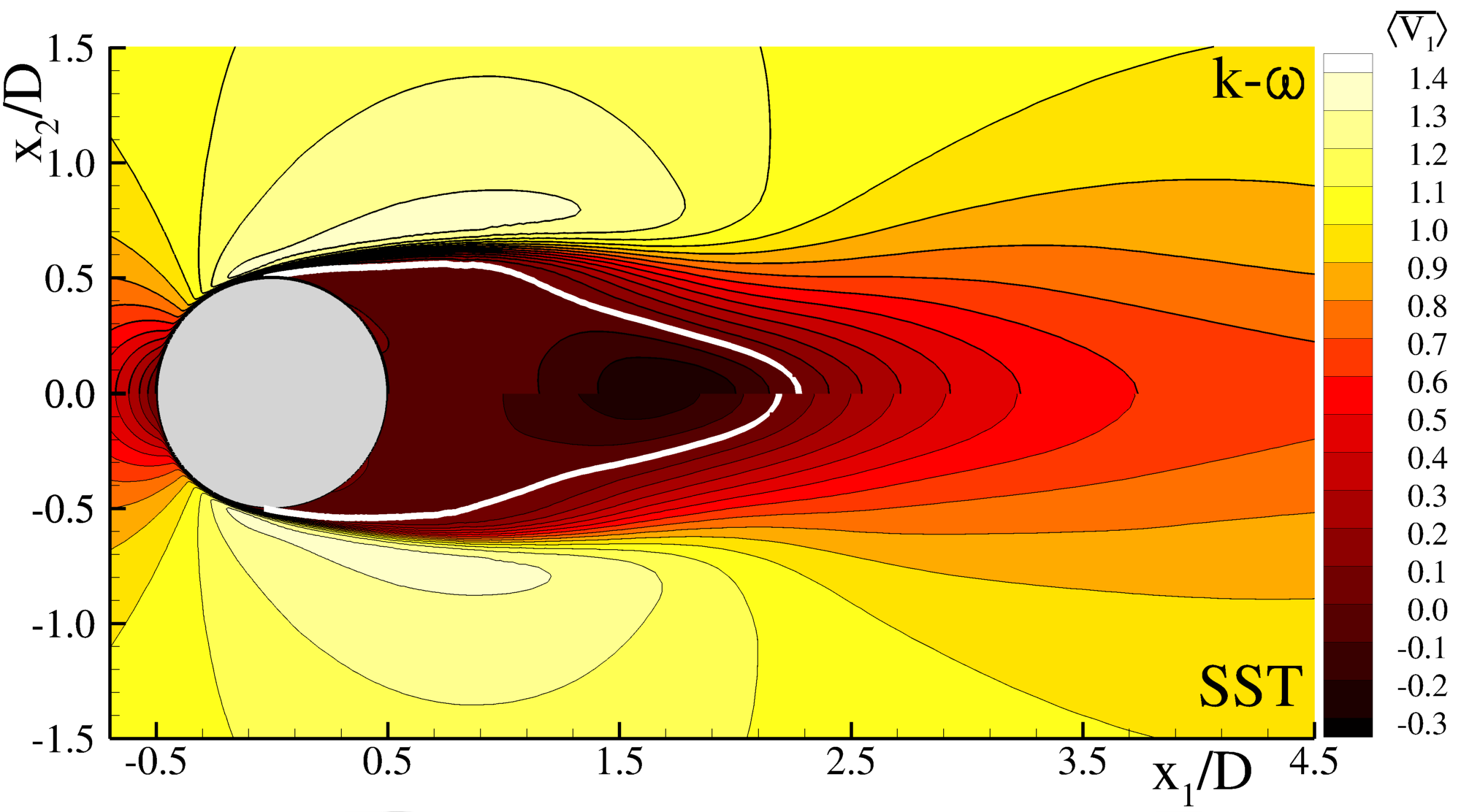}}
\\
\subfloat[Experiments of \cite{PARNAUDEAU_PF_2008}.]{\label{fig:5.1_4e}
\includegraphics[scale=0.16,trim=0 0 0 0,clip]{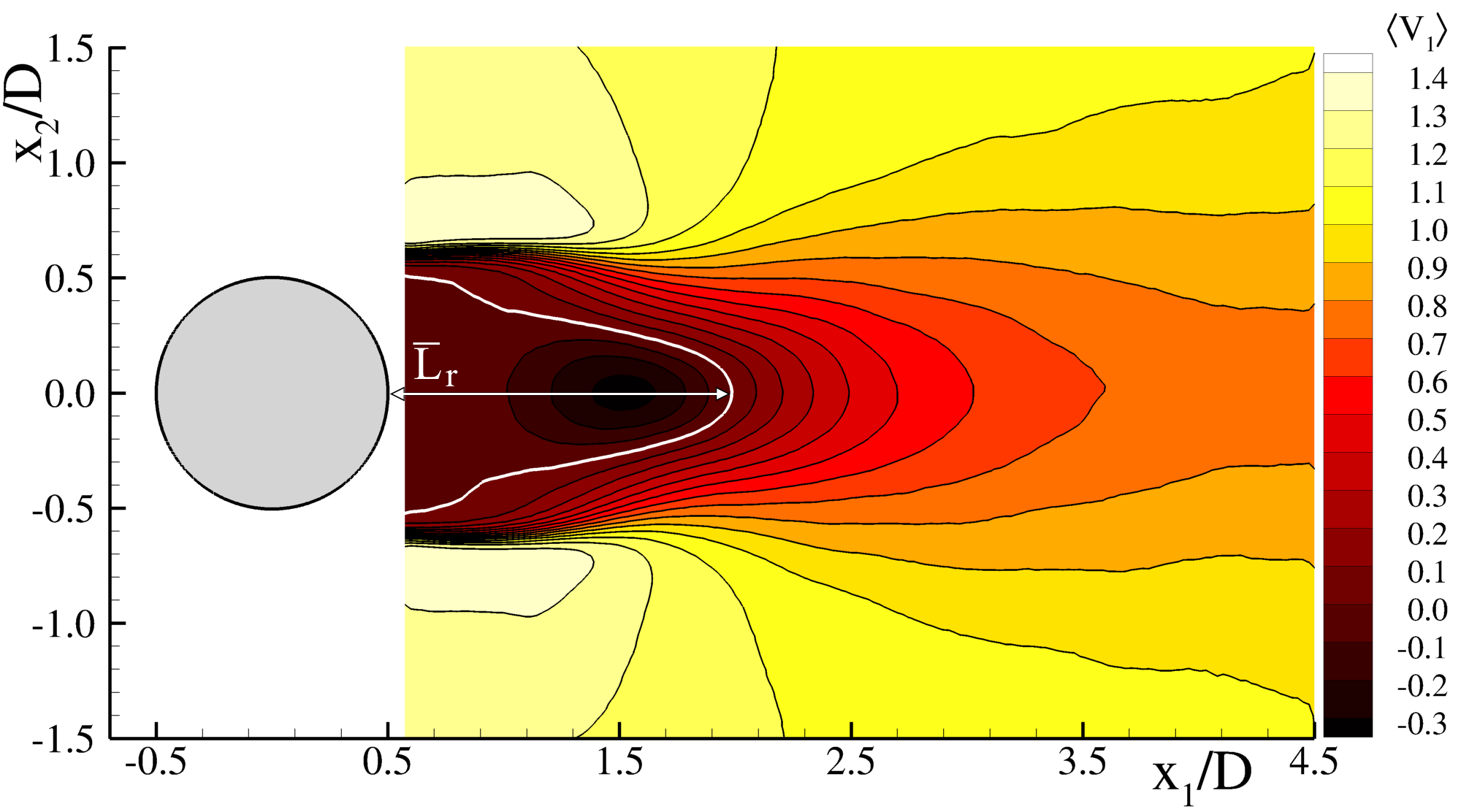}}
\caption{Time-averaged stream-wise velocity magnitude, $\langle \overline{V_1} \rangle$, for different physical resolutions and PANS closures. White line delimits recirculation region, $\langle \overline{V_1} \rangle=0$. Experiments taken from \cite{PARNAUDEAU_PF_2008}.}
\label{fig:5.1_4}
\end{figure}

The results from different $f_k$ simulations of various integral and local flow quantities are shown in figures \ref{fig:5.1_1} to \ref{fig:5.1_4} and table \ref{tab:5.1_1}. As expected, the data show that as the physical resolution increases ($f_k \rightarrow 0$), a wider range of scales are resolved - figure \ref{fig:5.1_1}. 

Referring to table \ref{tab:5.1_1}, it is seen that the characteristic frequency of vortex-shedding - the Strouhal number - is only weakly dependent on the physical resolution. The PANS simulations at different $f_k$ values capture this quantity quite accurately. On the other hand, the data demonstrate a clear overprediction of the resistance coefficients ($\overline{C}_D$, $C_L'$, and $\overline{C}_{pb}$) in low resolution simulations (here denoted as $f_k>0.50)$.  
For the case of $C_L'$, it is observed that the prediction for this quantity resulting from simulations at $f_k=1.00$ is nearly five times higher than the experimental observation. Yet, as $f_k$ decreases, the results approach the experiments. Similar trends are obtained for the time-averaged lengths of recirculation and formation of the vortex-shedding instability, $\overline{L}_r$ and $\overline{L}_F$. The refinement of resolution leads to slightly larger values of $\overline{L}_r$ and $\overline{L}_F$ than those measured experimentally. This result is addressed later. Overall, the data of table \ref{tab:5.1_1} indicates that when the resolution is improved beyond $f_k=0.50$, two important outcomes are observed: \textit{i)} both $k-\omega$ and SST results begin to converge - figure \ref{fig:5.1_2}; and \textit{ii)} the converged values are clearly within the experimental range of the respective quantities.

Next we consider the profiles of the time-averaged pressure coefficient on the cylinder's surface, $\overline{C}_p(\theta)$, stream-wise velocity, $\langle \overline{V_1} \rangle$, and stresses, $\overline{v_iv_j}$, at different locations in figure \ref{fig:5.1_3}. Whereas high resolution (here denoted as $f_k \leq 0.50$) simulations closely match the experimental measurements of the pressure coefficient distribution, low resolutions exhibit large discrepancies. Similar conclusions can be inferred for the velocity and stresses profiles. For instance, considering the profiles of $\langle \overline{V_1} \rangle$ at $x_1/D=1.06$, these move from a "V" to an "U" shape matching the experimental observations. The stresses are clearly overpredicted in low resolution simulations.

\begin{table}
  \begin{center}
\def~{\hphantom{0}}
\begin{tabular}{L{1.2cm}C{1.0cm}C{1.6cm}C{1.6cm}C{1.6cm}C{1.6cm}C{1.6cm}C{1.6cm}C{1.6cm}}
Model & $f_k$ & $St$& $C_L'$ & $\overline{C}_D$  & $-\overline{C}_{pb}$  & $\overline{L}_r$& $\overline{L}_F$\\[3pt] \hline
\multirow{ 4}{*}{$k-\omega$}	& 1.00    	& 0.214	& 0.56	&1.18  	& 1.30  	& 0.67 & 0.48\\
 						& 0.75  	& 0.214	& 0.46	& 1.13  	& 1.21  	& 0.81 & 0.60\\
 						& 0.50  	& 0.214	& 0.18	& 0.98  	& 0.96  	& 1.41 & 1.31\\
						& 0.25  	& 0.211	& 0.07 	& 0.92  	& 0.85  	& 1.77 & 1.64\\ \hline
\multirow{ 5}{*}{SST}		& 1.00 	& 0.211	& 0.66	& 1.25 	& 1.41 	 & 0.55 & 0.37\\
 						& 0.75	& 0.211	& 0.67	& 1.25 	& 1.43 	 & 0.53 & 0.45\\
 						& 0.50	& 0.214	& 0.28	& 1.04 	& 1.05 	 & 1.12 & 1.10\\
						& 0.25 	& 0.208	& 0.10	& 0.93 	& 0.86	 & 1.73 & 1.73\\ 
						& 0.15 	& 0.205	& 0.08	& 0.92 	& 0.85 	 & 1.77 & 1.79\\ \hline
Exp.1  					&-		& 0.208	& 0.10 	& 0.98  		& 0.88  		& 1.51 	& 1.31\\ 
Exp.2					&-		& 0.194-0.209	& - 	& 0.98-1.03  	& 0.68-0.89  	& - 		& 1.44-1.65\\ \hline 
\end{tabular}
\caption{Strouhal number, $St$, root-mean-square lift coefficient, $C_L'$, time-averaged drag coefficient, $\overline{C}_D$, pressure base coefficient, $\overline{C}_{pb}$, recirculation length, $\overline{L}_r$, and formation length (second peak in $\overline{v_1v_1}$ at $x_2=0$), $\overline{L}_F$, for different physical resolutions and PANS closures. Experiments 1 taken from \cite{PARNAUDEAU_PF_2008}, \cite{NORBERG_BBVIV3_2002}, and \cite{NORBERG_JFS_2003}; while experiments 2 denote a data range collected from table \ref{tab:2_1_1} at $Re=3000$ and $3900$.}
\label{tab:5.1_1}
  \end{center}
\end{table}

The time-averaged stream-wise velocity fields on the $x_1-x_2$ plane are shown in figure \ref{fig:5.1_4} for the two closure models at different resolutions. These fields are compared against the experimental data of \cite{PARNAUDEAU_PF_2008} presented in figure \ref{fig:5.1_4e}. The recirculation zone and reattachment length can be inferred from the figures. It is once again evident that the results generally improve by decreasing $f_k$. The computed reattachment length moves a little further downstream from the experimental location at the finest resolution. Many experimental works \citep{FAGE_ARC_1929,SADEH_NASA_1982,NORBERG_JFS_1987,NORBERG_TREP_1987} have noted that the reattachment length is a strong function of the upstream (inflow boundary condition) turbulence intensity, $I$. \cite{NORBERG_TREP_1987} has demonstrated that reducing $I$ from $1.4\%$ to $0.1\%$ leads to an increase of $11\%$ in the recirculation length - see table \ref{tab:2_1_1}. In the experiment considered here \citep{PARNAUDEAU_PF_2008}, the authors do not report a precise value of $I$, but indicate that it is below $0.2\%$. We believe that this is the main reason (the aspect ratio may also be an influence - see table \ref{tab:2_1_1}) for the observed differences between PANS simulations at $f_k\leq 0.25$ and the experiments of \cite{PARNAUDEAU_PF_2008}.

In summary, it is evident that the simulation fidelity varies substantially with improving resolution in the range $f_k>0.50$. For $f_k \leq 0.50$, however, the results are only weakly dependent on resolution, and converge towards the experiments with improving resolution. We now examine the underlying physics to explain the convergence beyond $f_k$ equal to $0.50$.

%
%
\subsection{Physics of Scale-Resolving Simulation Flow}
\label{sec:5.2}
Experiments (discussed in Section \ref{sec:2}) clearly indicate that the flow comprises four critical stages: \textit{i)} onset of the Kelvin-Helmholtz (free shear-layer) instability; \textit{ii)} spatial development of the Kelvin-Helmholtz rollers; \textit{iii)} breakdown to high-intensity turbulence; and \textit{iv)} vortex-shedding. All these stages must be reasonably replicated in a simulation in order to achieve an adequate agreement with experiments. Furthermore, experimental observations indicate that this process is strongly related with the magnitude of $Re$. Most notably, the Kelvin-Helmholtz rollers that dictate the flow dynamics in the free shear-layer are only observed beyond a Reynolds number of $1200$. We will now examine how the closure model and the range of resolved scales influence these flow features, and ultimately lead the simulation data to be in agreement with the experiments. In the remainder of this sub-section we will examine $i)$ how the instability is modelled at different resolutions, and $ii)$ the manner in which spatially-developing turbulence which is far from equilibrium is represented.

\begin{figure}
\centering
\subfloat[$k-\omega$.]{\label{fig:5.2_1a}
\includegraphics[scale=0.27,trim=0 0 0 0,clip]{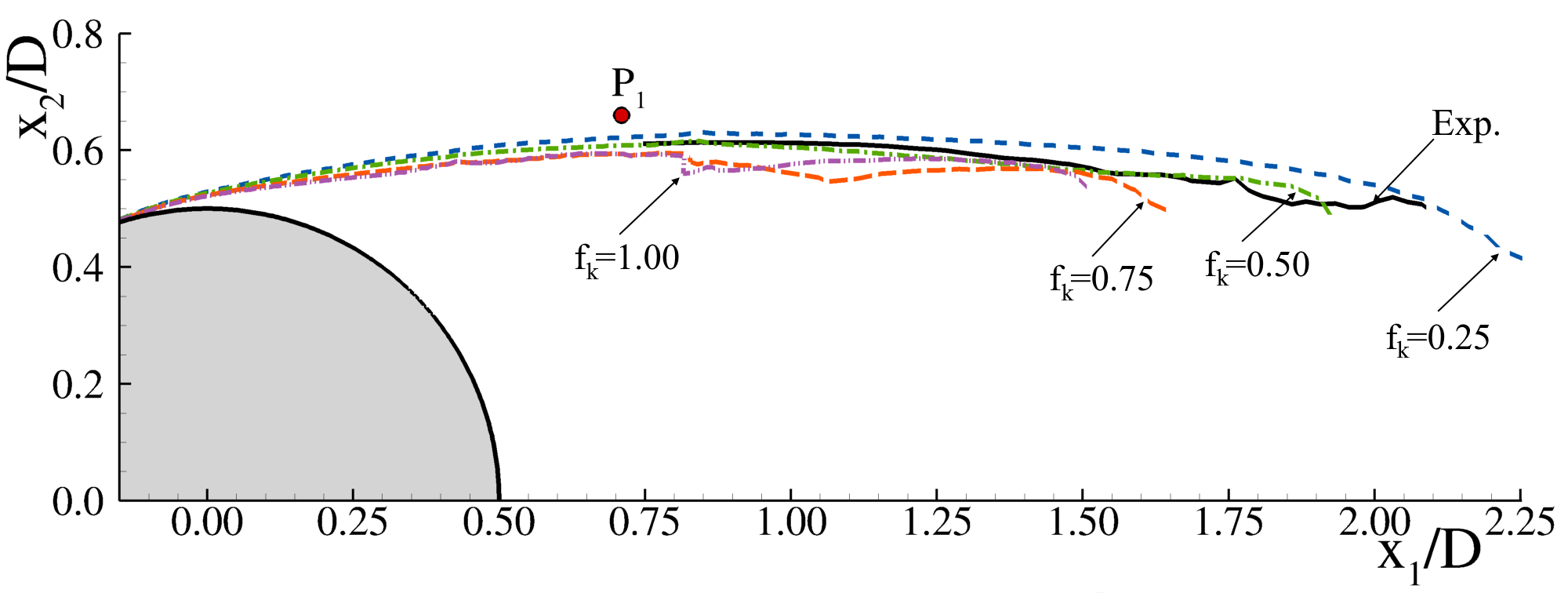}}
\\
\subfloat[SST.]{\label{fig:5.2_1b}
\includegraphics[scale=0.27,trim=0 0 0 0,clip]{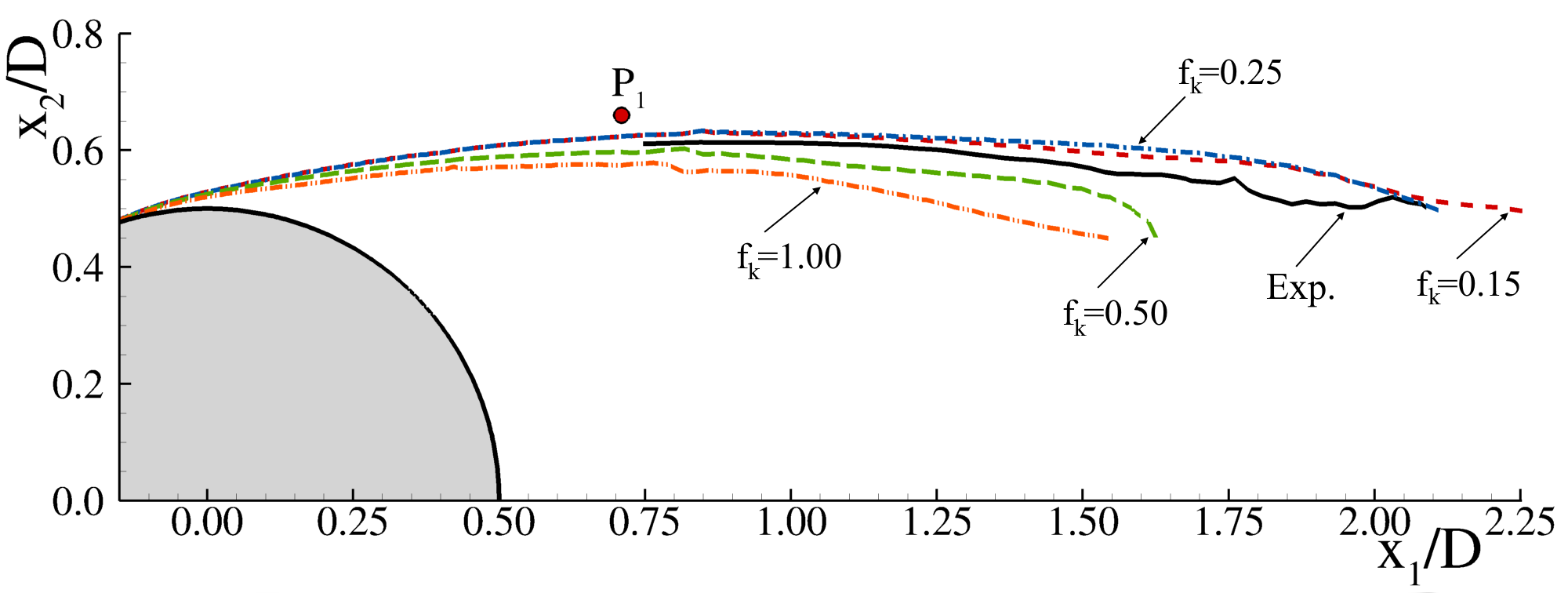}}
\caption{Time-averaged inflection line of the free shear-layer, $\partial^2 \langle \overline{V_1} \rangle / \partial x_2^2=0$, for different physical resolutions and PANS closures. Experiments taken from \cite{PARNAUDEAU_PF_2008}. $P_1$ ($x_1/D=0.71;x_2/D=0.66$) denotes the location of the probe used in figures \ref{fig:5.2_5} and \ref{fig:5.2_6}.}
\label{fig:5.2_1}
\end{figure}

It has been well known for a long time \citep{RAYLEIGH_LMS_1883} that the Kelvin-Helmholtz instability originates and develops along the locus of the background velocity field inflection point - the so-called inflection line. It is therefore of primary importance in a numerical simulation to accurately replicate the location and development of the inflection line. In figure \ref{fig:5.2_1}, the inflection lines computed from simulations at different resolutions are compared against that from the experimental measurements of \cite{PARNAUDEAU_PF_2008}. All simulations accurately capture the initial location of the inflection line. However, the subsequent development changes as a function of the resolution $f_k$. For the $f_k = 1.00$ and $0.75$ cases, the inflection lines curve inward in comparison to the experimental data. On the other hand, the $f_k= 0.50$ and $0.25$ cases follow the experimental trajectory quite accurately. This observation holds for both $k-\omega$ and SST closures. Yet, the deviation is more pronounced in the SST case. As a result of the distinct change in the behaviour of simulations at $f_k>0.50$ and $f_k\leq 0.50$, we now analyse them individually. 

\subsubsection{Low Resolution Simulations}
\label{sec:5.2.1}

\begin{figure}
\centering
\subfloat[$f_k=1.00$.]{\label{fig:5.2_2a}
\includegraphics[scale=0.16,trim=0 0 0 0,clip]{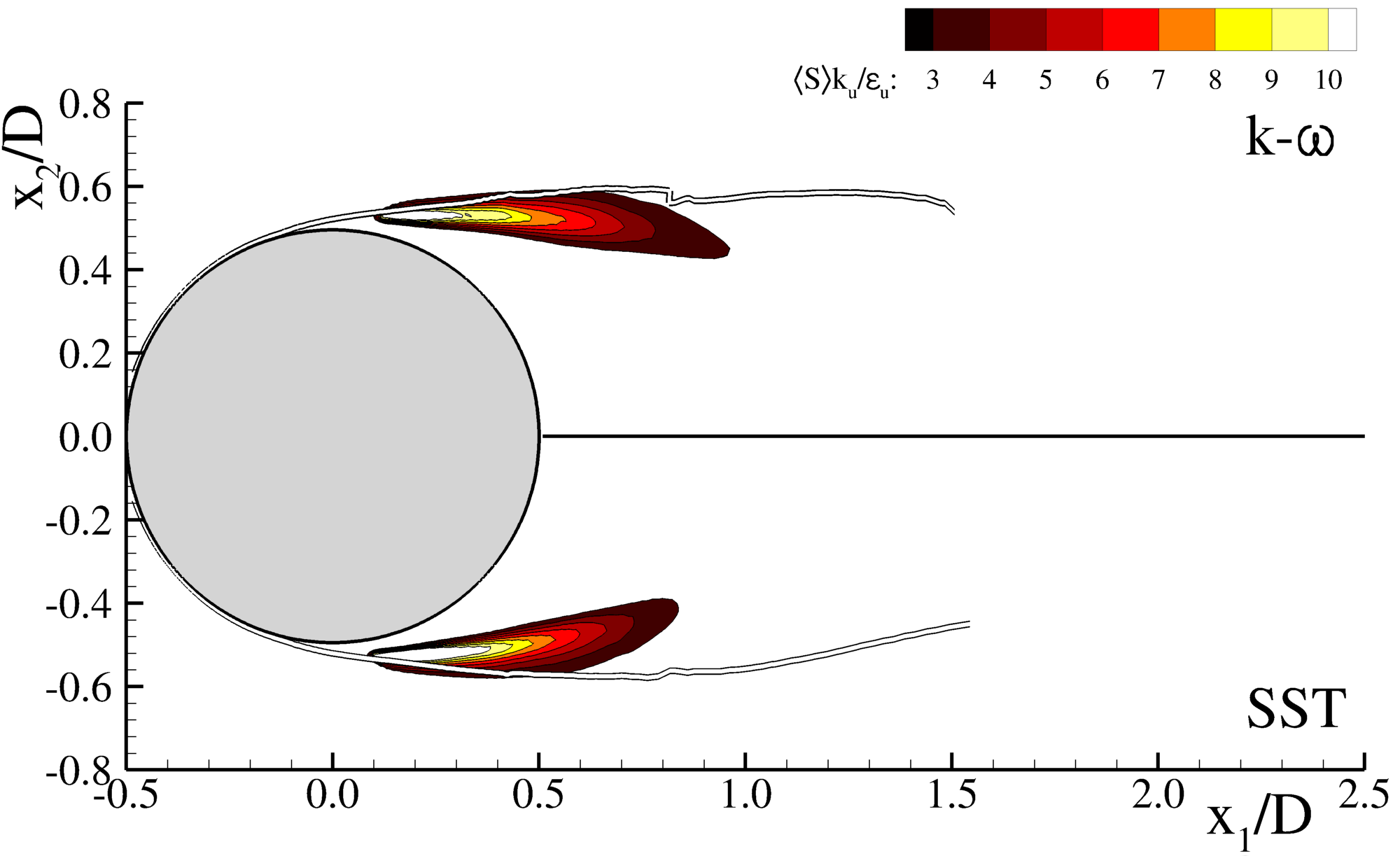}}
~
\subfloat[$f_k=0.75$.]{\label{fig:5.2_2b}
\includegraphics[scale=0.16,trim=0 0 0 0,clip]{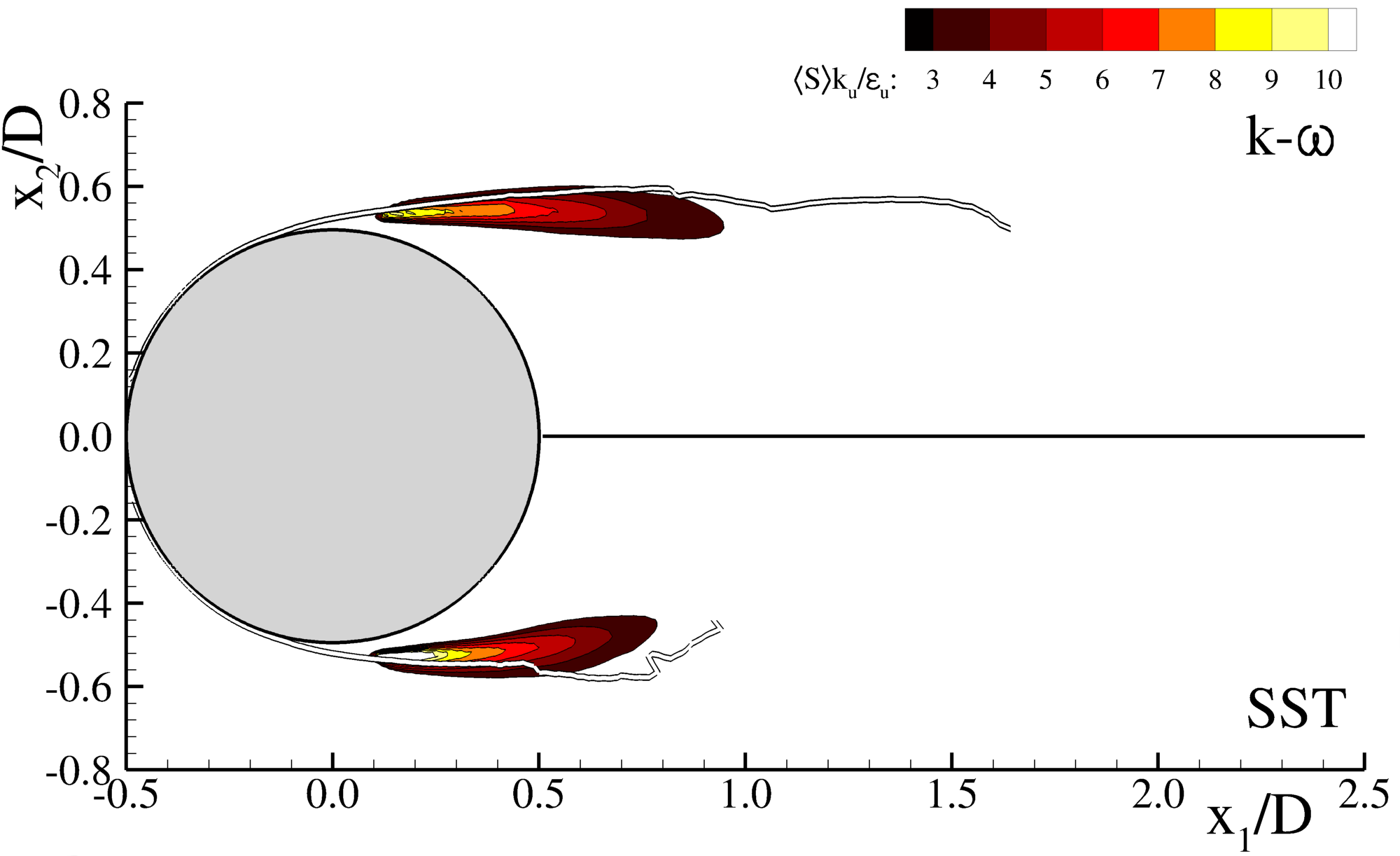}}
\\
\subfloat[$f_k=0.50$.]{\label{fig:5.2_2c}
\includegraphics[scale=0.16,trim=0 0 0 0,clip]{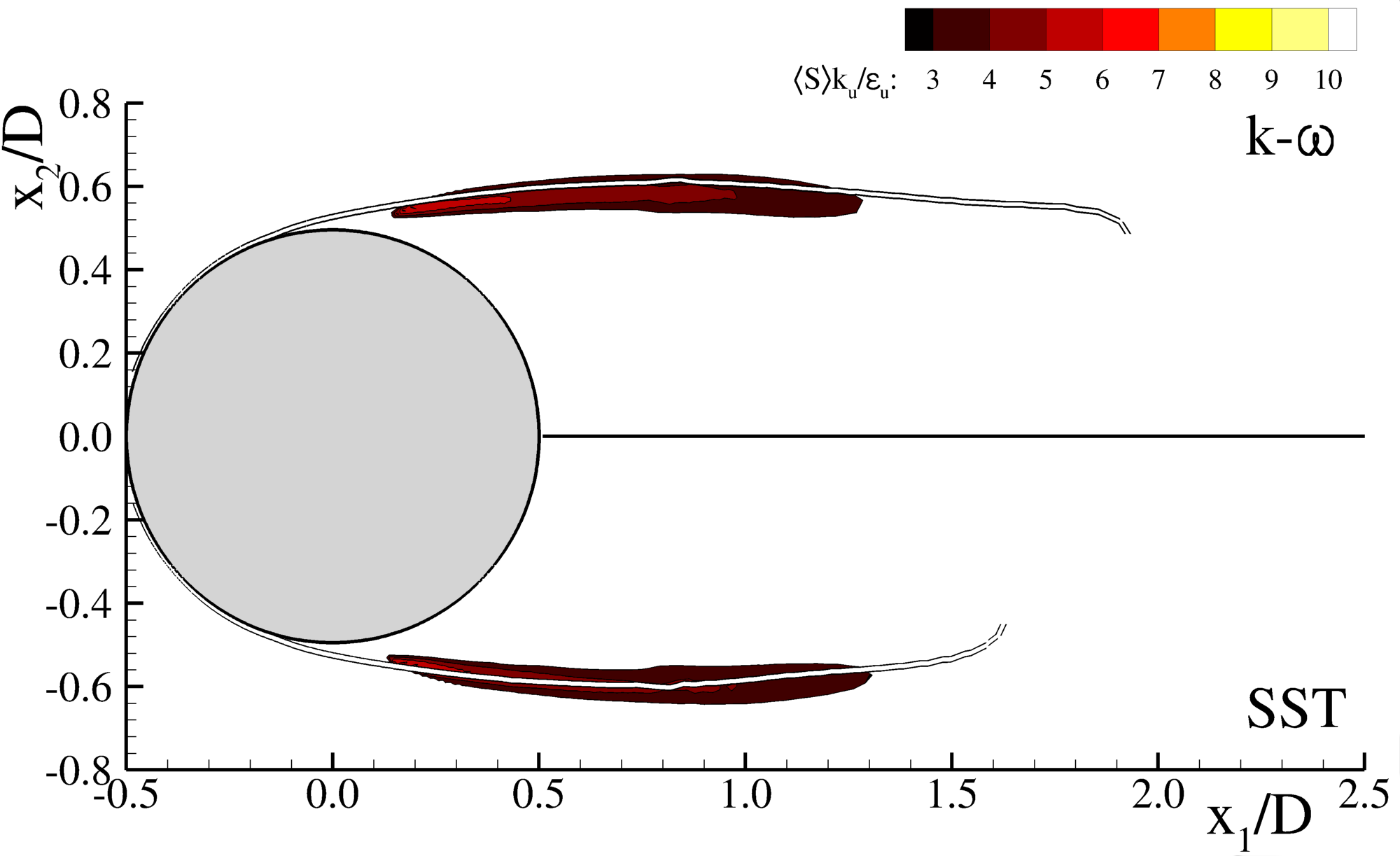}}
~
\subfloat[$f_k=0.25$.]{\label{fig:5.2_2d}
\includegraphics[scale=0.16,trim=0 0 0 0,clip]{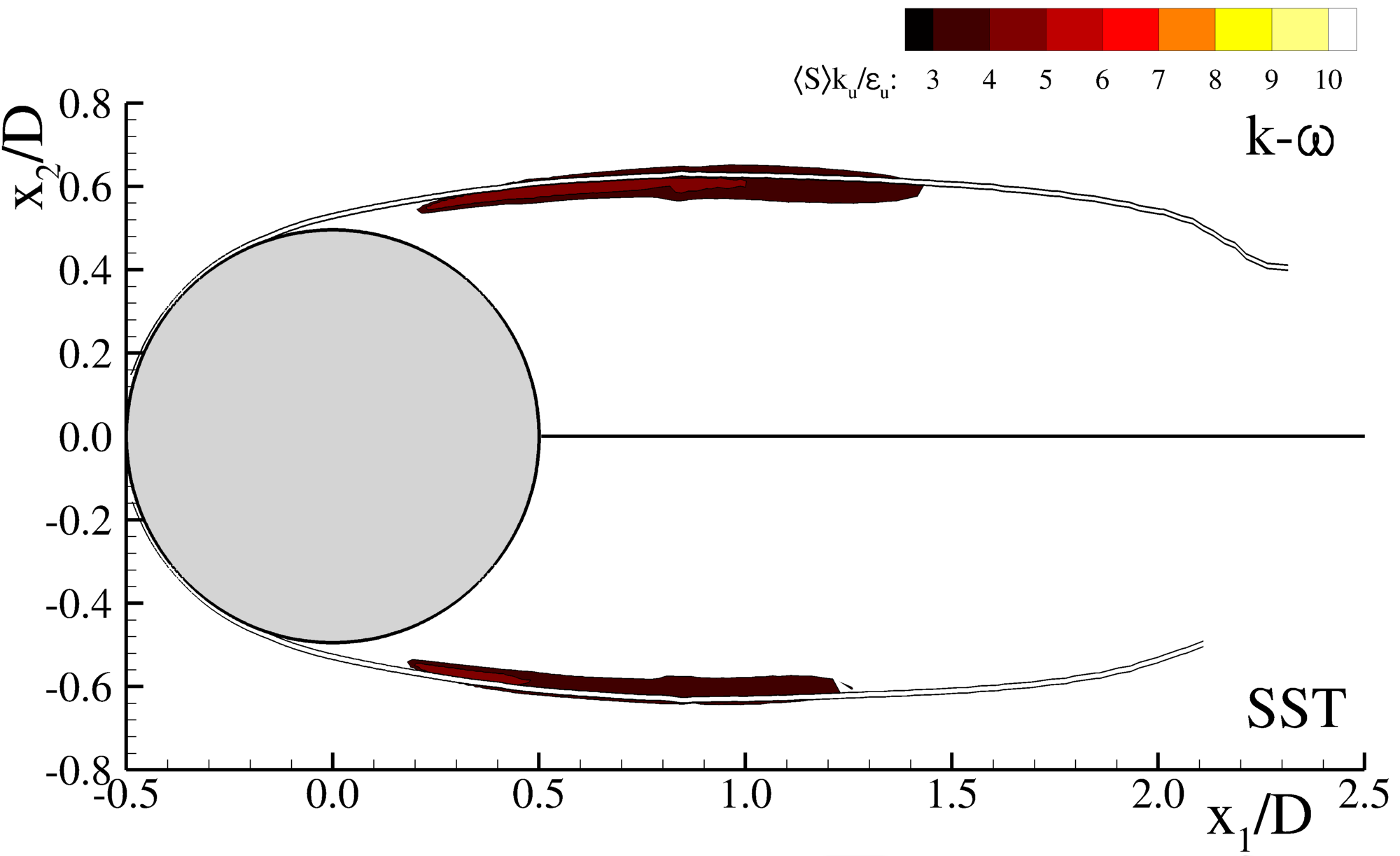}}
\caption{Time-averaged mean-to-unresolved field strain-rate ratio, $\langle S\rangle k_u/\epsilon_u$ for different physical resolutions and PANS closures. White/black lines represent the stream-wise velocity inflection point, $\partial^2 \langle \overline{V_1} \rangle/\partial x_2^2=0$.}
\label{fig:5.2_2}
\end{figure}

The evolution of the inflection line for high $f_k$ (including RANS) cases clearly indicates that the closure model is inadequate in this region. As mentioned in Section 3, the RANS models, by design, are best suited for regions of fully-developed turbulence. To examine the state of turbulence in the inflection-line region, we present the contours of the time-averaged strain-rate ratio of mean and unresolved fields, $\langle S \rangle k_u/\epsilon_u$, in figure \ref{fig:5.2_2}. High values of this quantity (typically $\langle S \rangle k_u/\epsilon_u>8$) indicate that turbulence is dominated by linear processes and the instabilities are still growing. Therefore, turbulence constitutive relations are elastic rather than viscous in the linear regime, leading to the overprediction of turbulent viscosity in these regions. On the other hand, low magnitudes of $\langle S \rangle k_u/\epsilon_u$ (between 3 and 6) denote fully-developed turbulence. Boussinesq relations typically provide reasonable closure only in these regions.

Both $f_k=1.00$ and $0.75$ simulations exhibit large areas where $\langle S \rangle k_u/\epsilon_u>8$. This means that these regions where turbulence is developing and far from equilibrium are being represented by closures that are only appropriate for fully-developed turbulence. Consequently, figure \ref{fig:5.2_2} demonstrates that RANS and coarse resolution SRS tend to significantly overestimate the turbulent viscosity in the vicinity of the inflection line. This evidently stems from the unphysically large levels of production and turbulence kinetic energy predicted by the closure model in these regions. 


\begin{figure}
\centering
\includegraphics[scale=0.25,trim=0 0 0 0,clip]{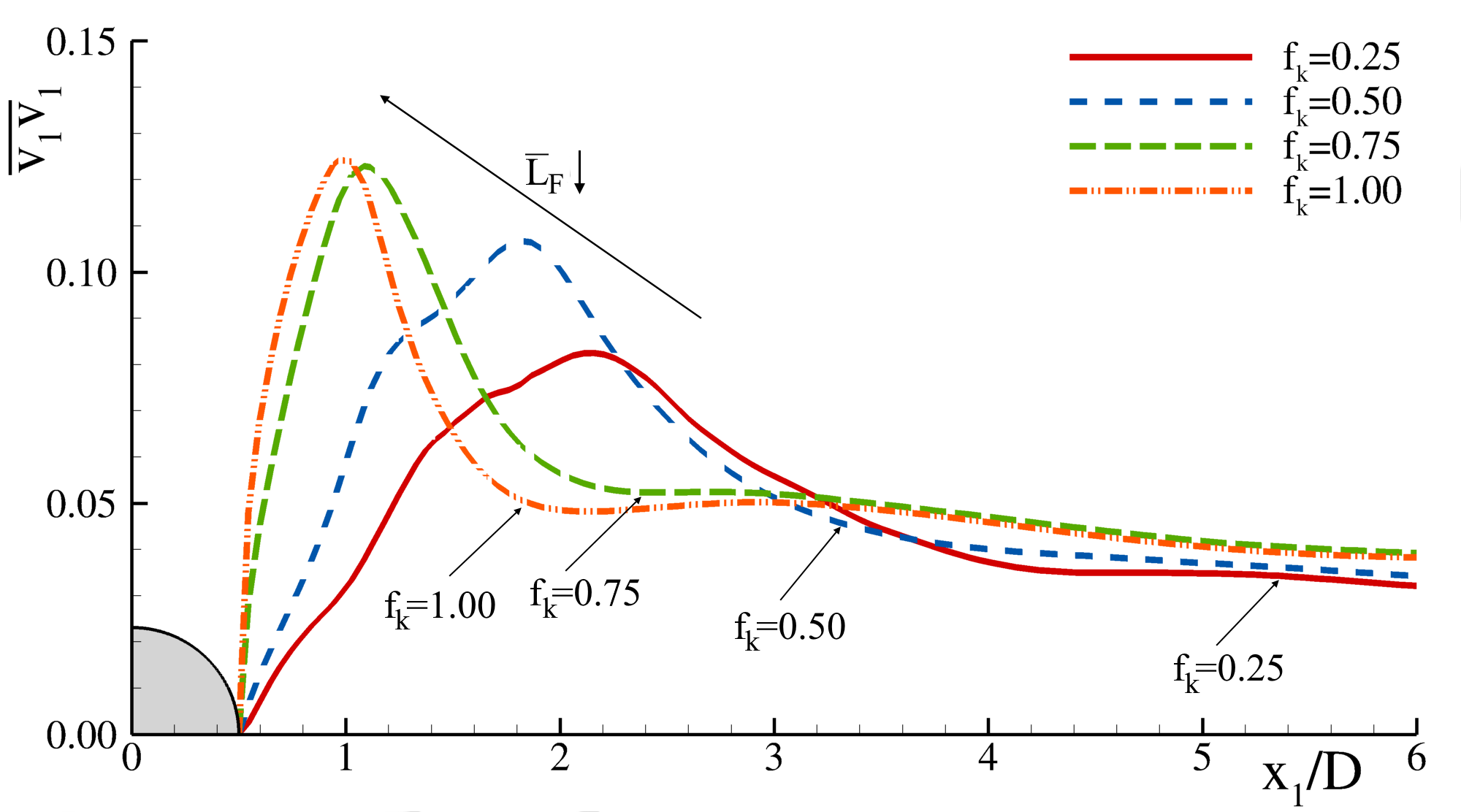}
\caption{Time-averaged stream-wise velocity variance, $\overline{v_1v_1}$, at $x_2/D=0$ for different physical resolutions. Results for $k-\omega$ PANS closure.}
\label{fig:5.2_3}
\end{figure}

The first and most significant consequence of the high turbulent viscosity level is that the breakdown to high-intensity turbulence occurs prematurely, well ahead of the experimental location. To illustrate this point, figure \ref{fig:5.2_3} presents the time-averaged variance (or normal stress) of the stream-wise velocity field in the near-wake at $x_2/D=0.0$. The location of the second and highest peak marks the beginning of vortex-shedding, whereas the second the onset of instabilities (in \cite{NORBERG_AUBBWVIV_1998}, it is compared to the location of the breakdown to high intensity turbulence). The data shown in these figures exhibit a single peak for the low resolution simulations which moves upstream with the increase of $f_k$. This is a direct consequence of the high level of turbulent viscosity. As a result, RANS and low resolution SRS do not distinguish between the onset of the Kelvin-Helmholtz instability and the breakdown to turbulence.

\begin{figure}
\centering
\subfloat[$f_k=1.00$.]{\label{fig:5.2_4a}
\includegraphics[scale=0.16,trim=0 0 0 0,clip]{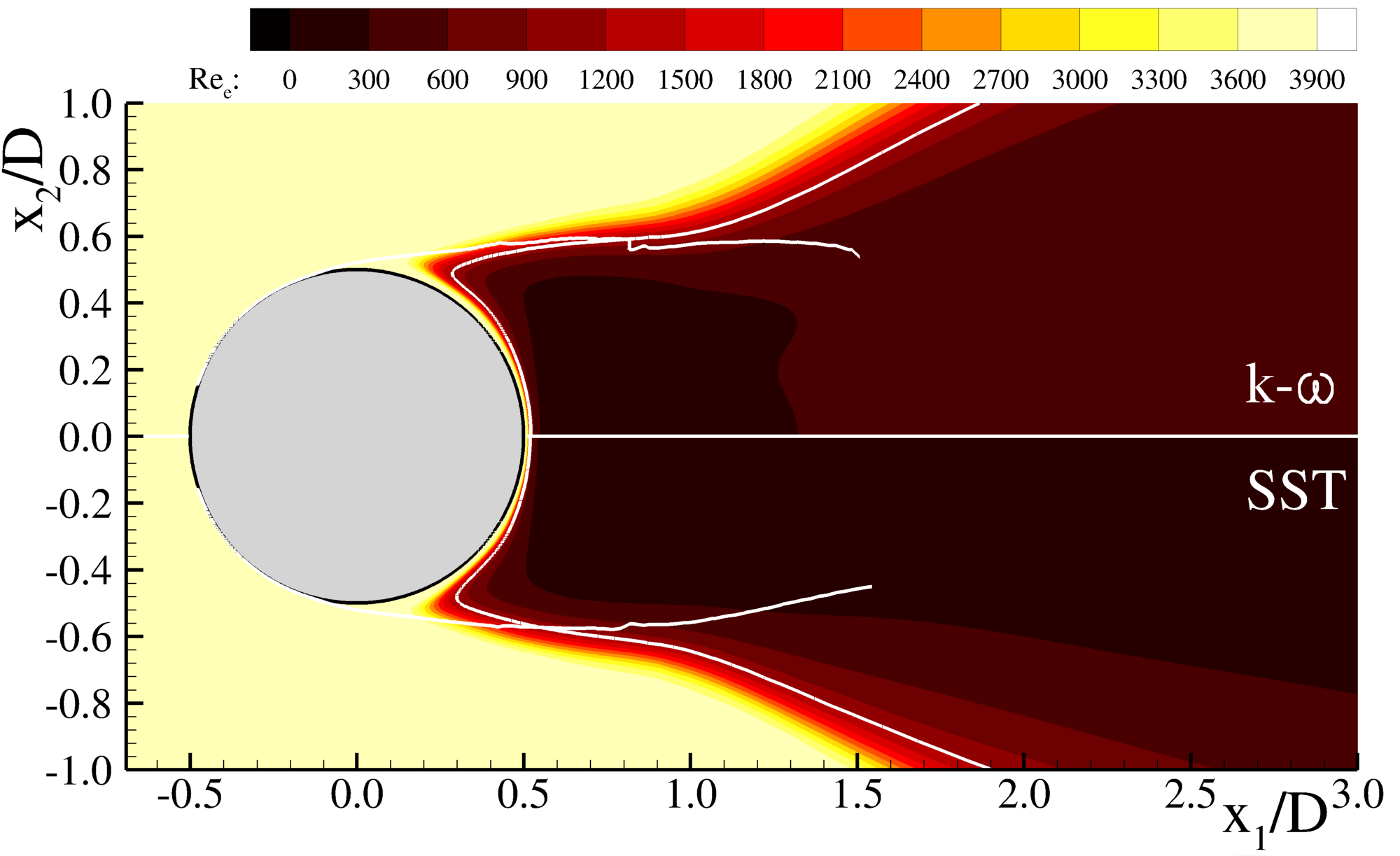}}
~
\subfloat[$f_k=0.75$.]{\label{fig:5.2_4b}
\includegraphics[scale=0.16,trim=0 0 0 0,clip]{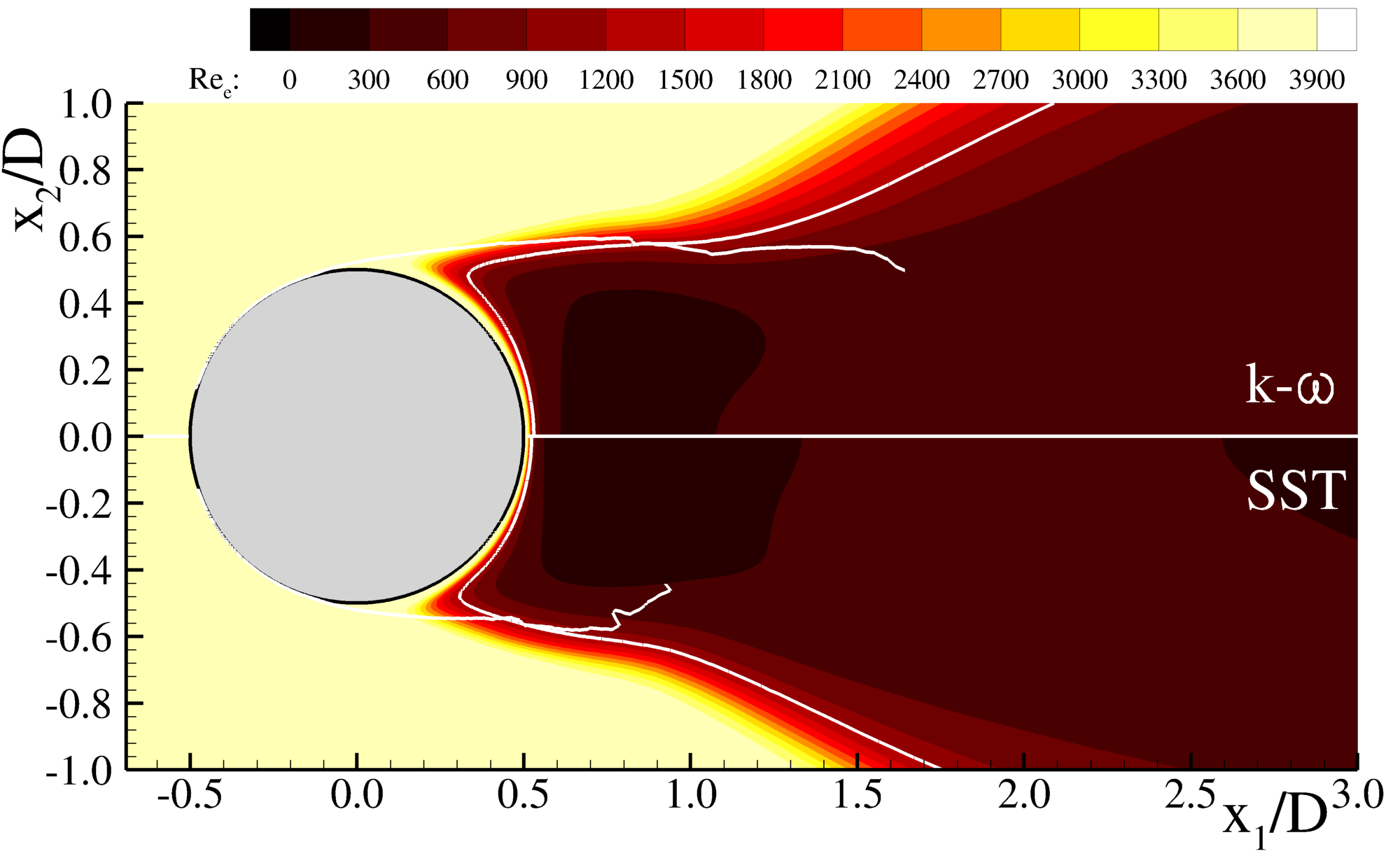}}
\\
\subfloat[$f_k=0.50$.]{\label{fig:5.2_4c}
\includegraphics[scale=0.16,trim=0 0 0 0,clip]{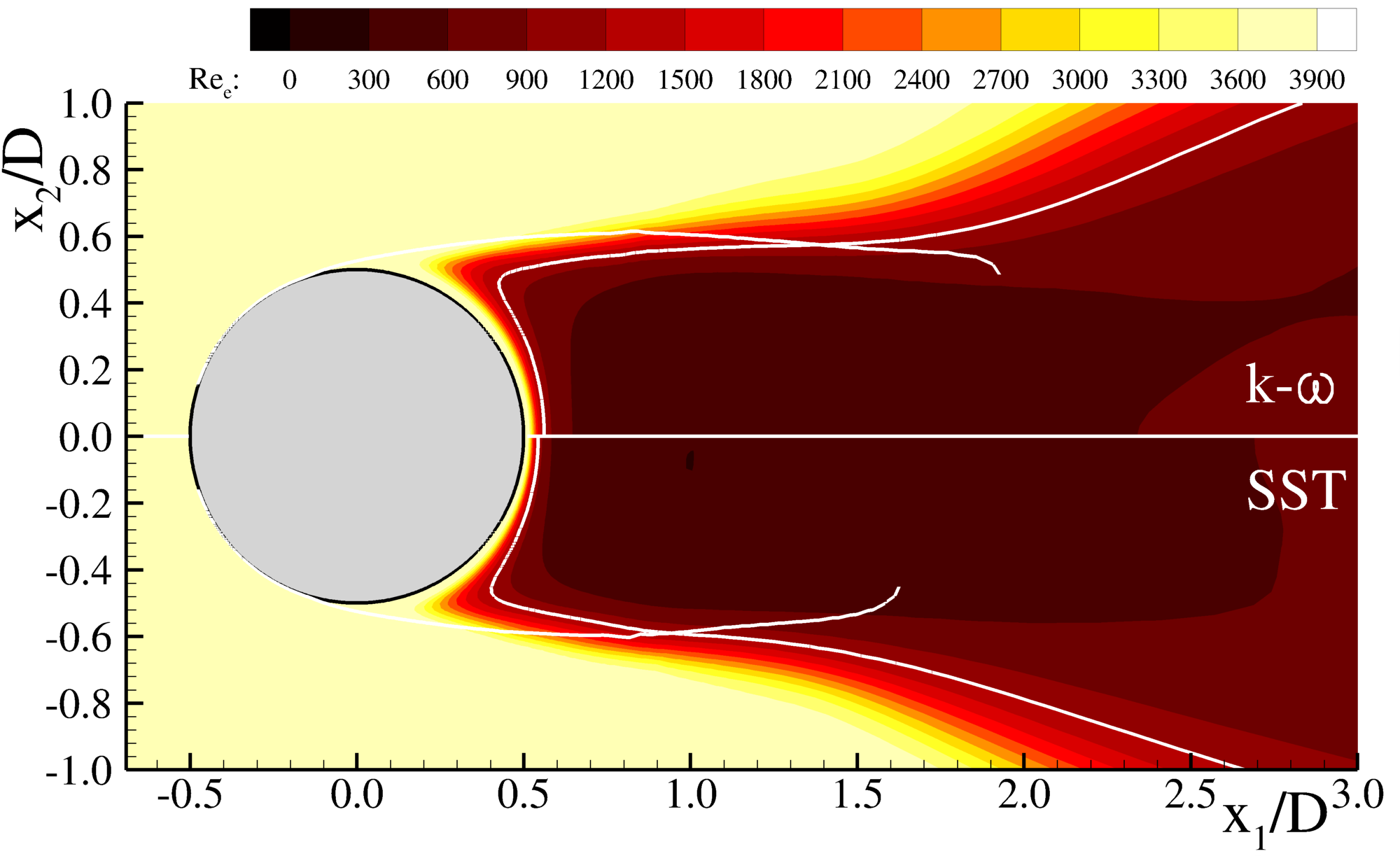}}
~
\subfloat[$f_k=0.25$.]{\label{fig:5.2_4d}
\includegraphics[scale=0.16,trim=0 0 0 0,clip]{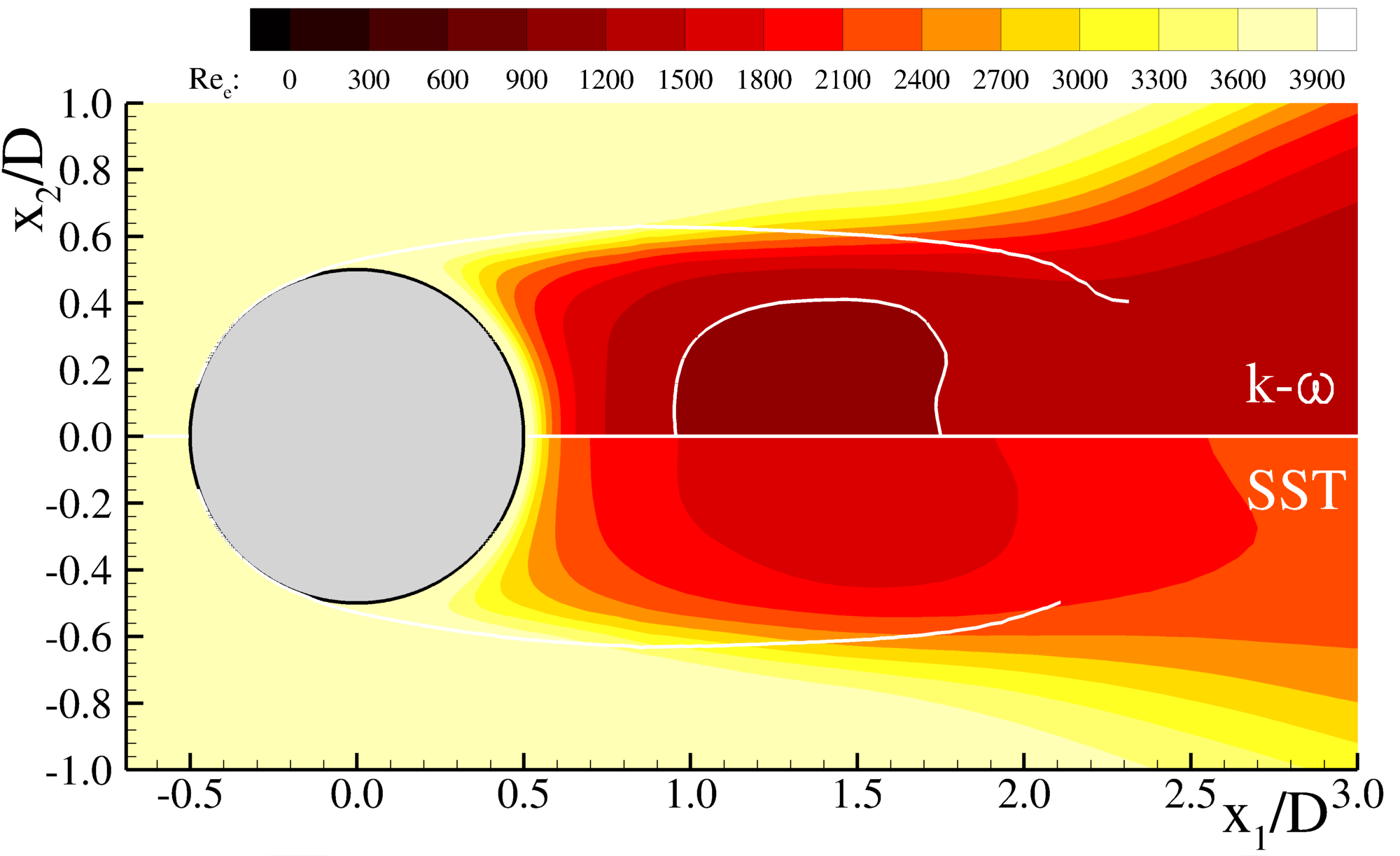}}
\caption{Time-averaged computational Reynolds number, $Re_e$, in the near-wake for different physical resolutions and PANS closures. White/black lines represent the stream-wise velocity inflection point, $\partial^2 \langle \overline{V_1} \rangle/\partial x_2^2=0$, while white line delimits $Re_e=1200$.}
\label{fig:5.2_4}
\end{figure}

The overproduction of turbulence kinetic energy and turbulent viscosity causes the reduction of the effective computational Reynolds number shown in figure \ref{fig:5.2_4}. For the two lowest resolutions tested, it is evident that the effective computational Reynolds number $Re_e$ is lower than $1200$ (the lower limit to experimentally observe the Kelvin-Helmholtz rollers). In fact, figures \ref{fig:5.2_4a} and \ref{fig:5.2_4b} show that the start of the inward curving of the inflection lines is related with the point where the inflection lines experience $Re \le  1200$.

\begin{figure}
\centering
\subfloat[$\langle \omega_3 \rangle$ at $f_k=1.00$ (SST).]{\label{fig:5.2_5a}
\includegraphics[scale=0.16,trim=0 0 0 0,clip]{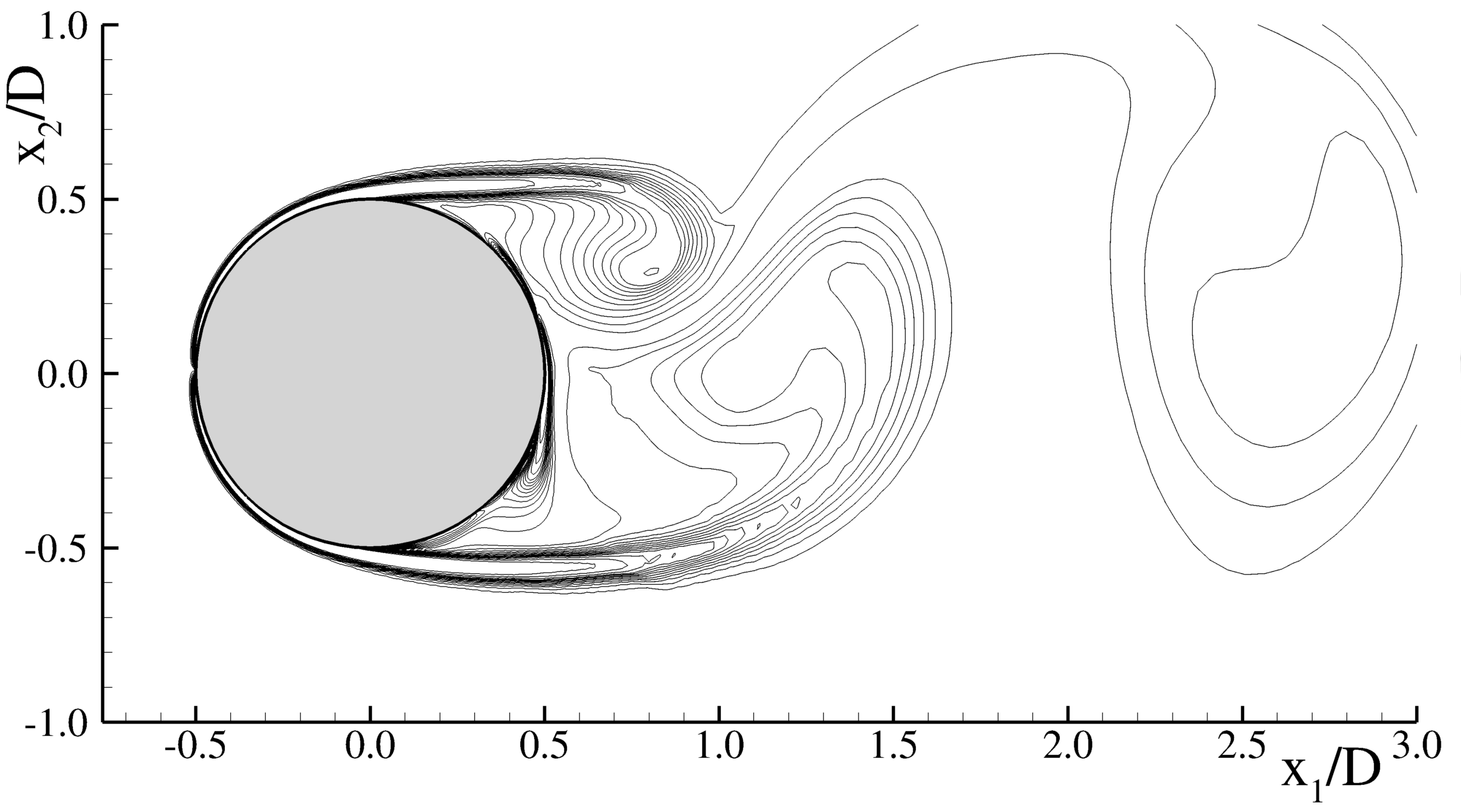}}
~
\subfloat[$\langle \omega_3 \rangle$ at $f_k=0.75$ (SST).]{\label{fig:5.2_5b}
\includegraphics[scale=0.16,trim=0 0 0 0,clip]{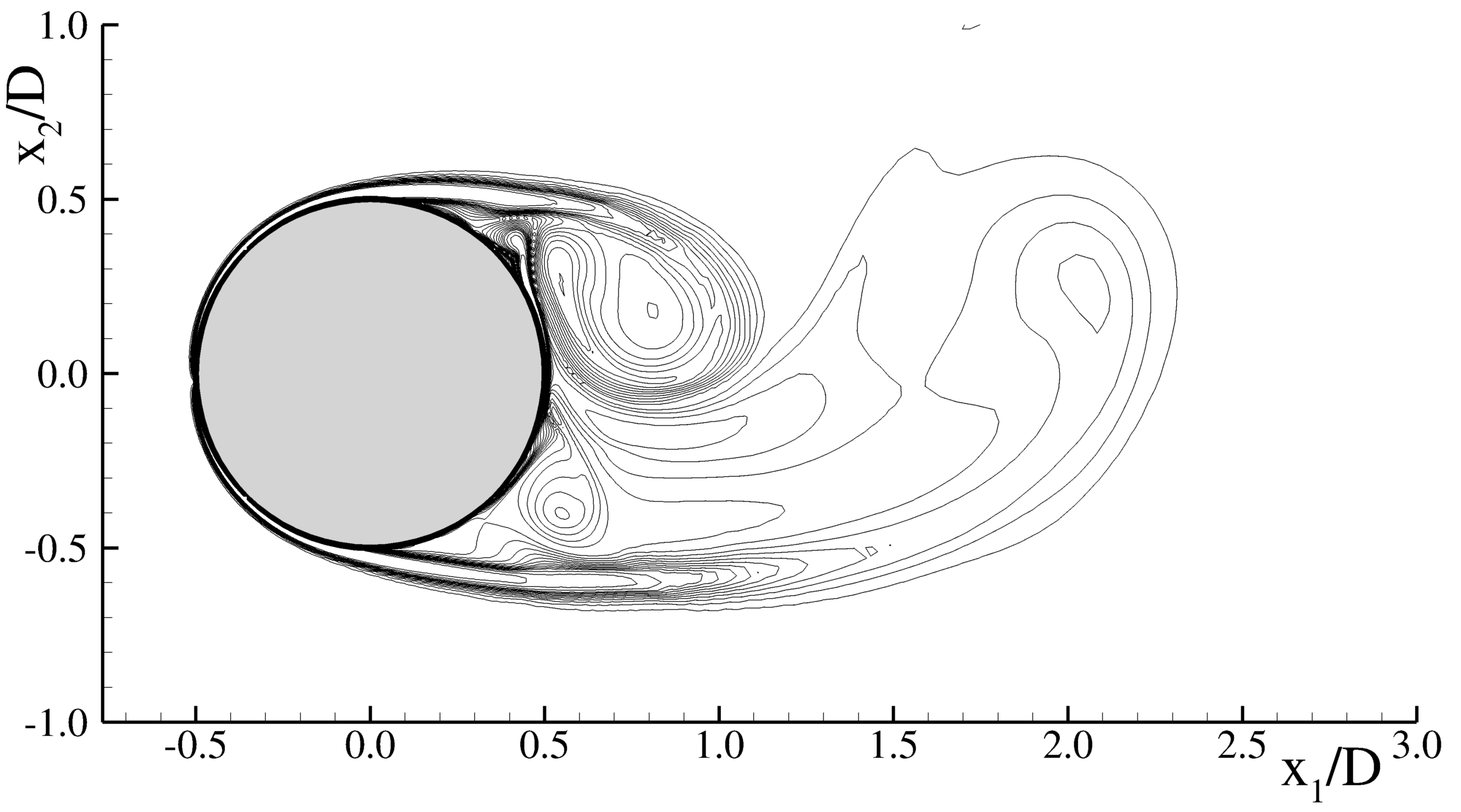}}
\\
\subfloat[$\langle V_1\rangle (t)$ at $f_k=1.00$ (SST).]{\label{fig:5.2_5c}
\includegraphics[scale=0.16,trim=0 0 0 0,clip]{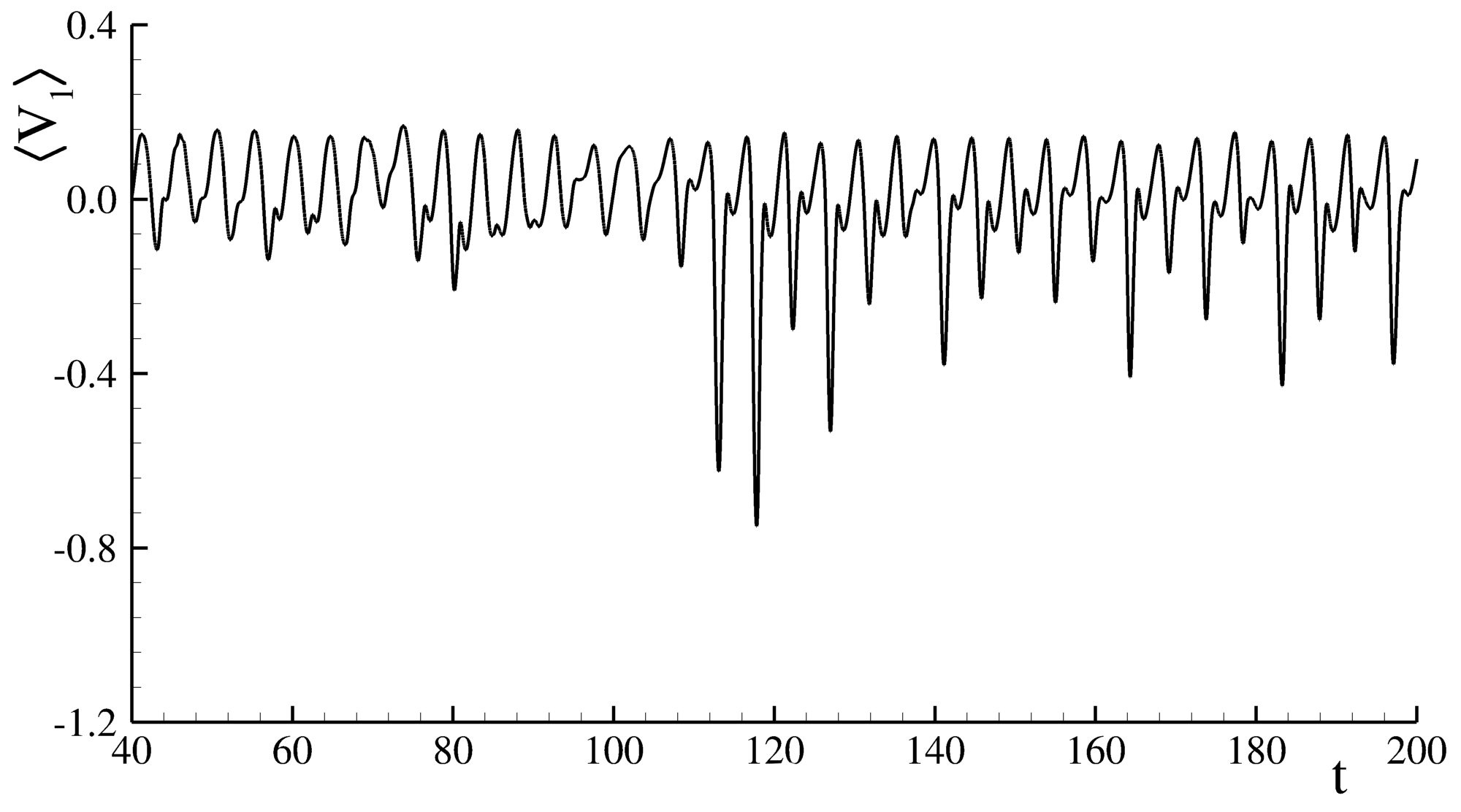}}
~
\subfloat[$\langle V_1\rangle (t)$ at $f_k=0.75$ (SST).]{\label{fig:5.2_5d}
\includegraphics[scale=0.16,trim=0 0 0 0,clip]{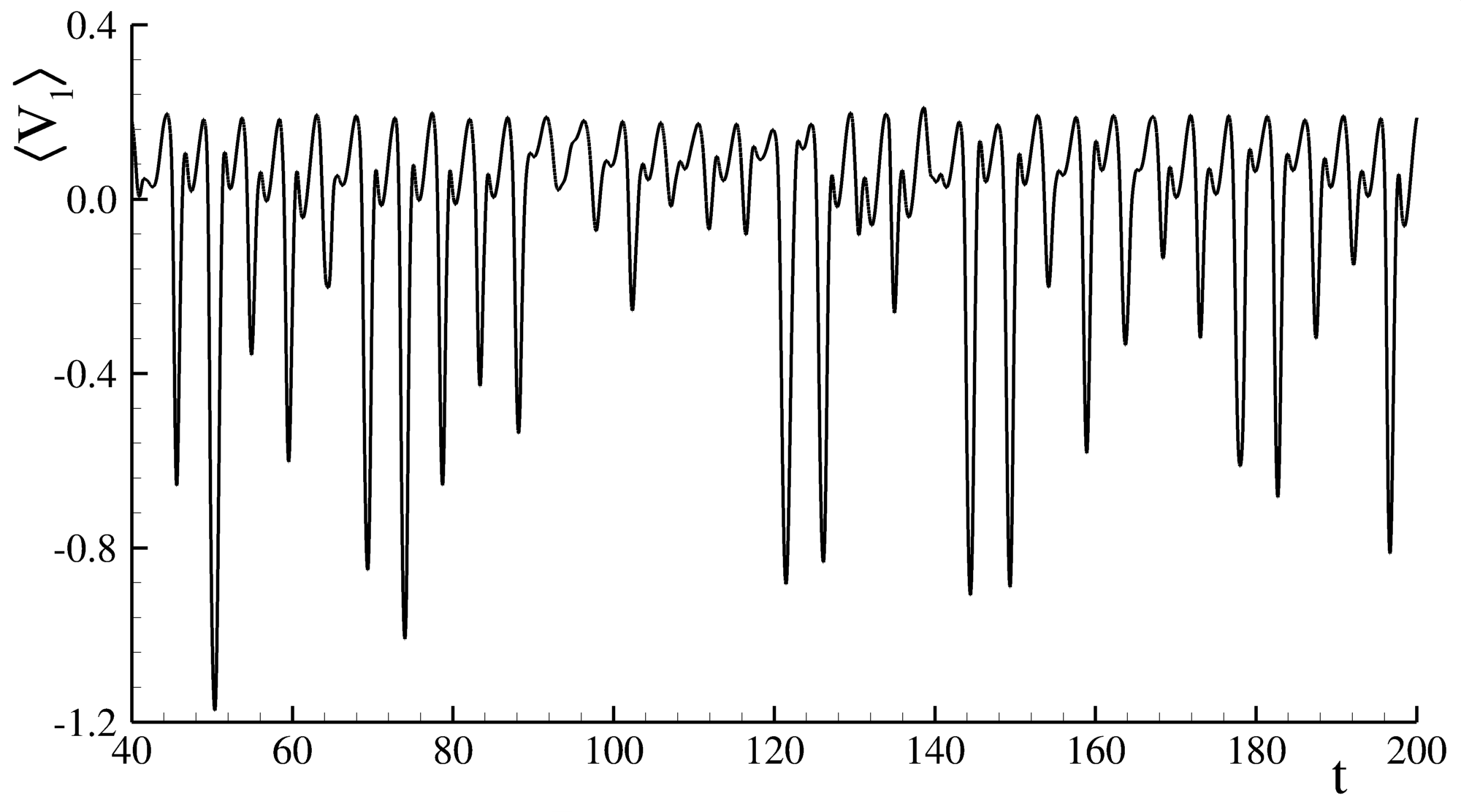}}
\\
\subfloat[$E(\langle V_1 \rangle)$ at $f_k=1.00$.]{\label{fig:5.2_5e}
\includegraphics[scale=0.16,trim=0 0 0 0,clip]{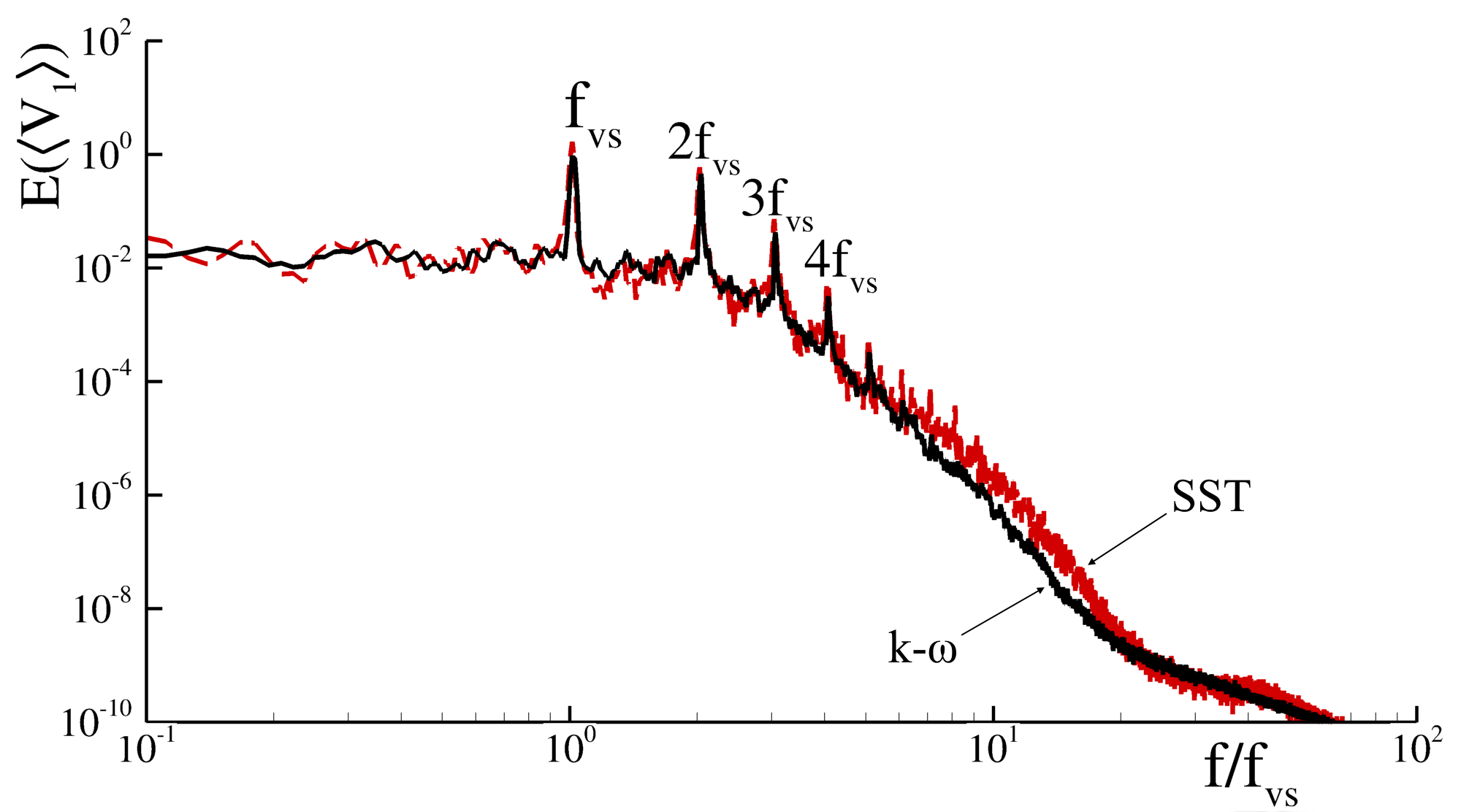}}
~
\subfloat[$E(\langle V_1 \rangle)$ at $f_k=0.75$.]{\label{fig:5.2_5f}
\includegraphics[scale=0.16,trim=0 0 0 0,clip]{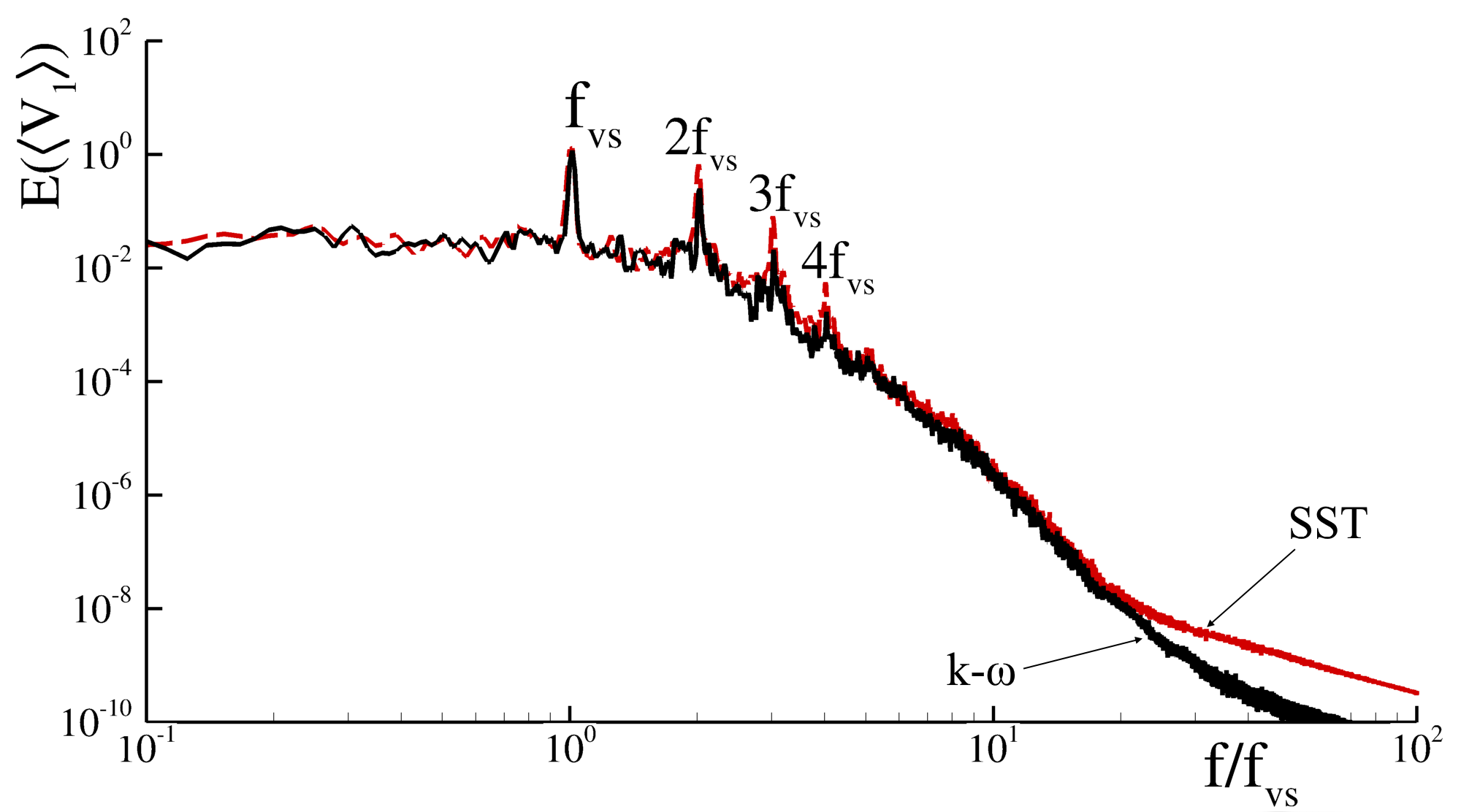}}
\caption{Schematic of the instantaneous span-wise vorticity, $\langle \omega_3 \rangle$, field, time-trace of the stream-wise velocity, $\langle V_1\rangle (t)$, field and respective frequency spectrum, $E(\langle V_1 \rangle)$, at $P_1$ ($x_1/D=0.71$; $x_2/D=0.66$) for $f_k=1.00$ and $0.75$ and different PANS closures.}
\label{fig:5.2_5}
\end{figure}

These results indicate that the poor quality of the low resolution simulations is related to their inability to generate the instability and capture the coherent structures developing in the free shear-layer. Figure \ref{fig:5.2_5} confirms this point by depicting the contour of the instantaneous resolved span-wise vorticity, $\langle \omega_3 \rangle$, field in the near-wake, the temporal evolution of the resolved stream-wise velocity at $P_1$= ($x_1/D=0.71$; $x_2/D=0.66$), and the resulting frequency spectrum. Compared to figure \ref{fig:2.1_2}, it is evident that the vortex-shedding inception moves upstream with the increase of $f_k$, and that the development of the Kelvin-Helmholtz rollers is suppressed since neither the vorticity contours or the spectra evidence the presence of such instability. 

\begin{figure}
\centering
\subfloat[$\langle \omega_3 \rangle$ at $f_k=0.50$ (SST).]{\label{fig:5.2_6a}
\includegraphics[scale=0.16,trim=0 0 0 0,clip]{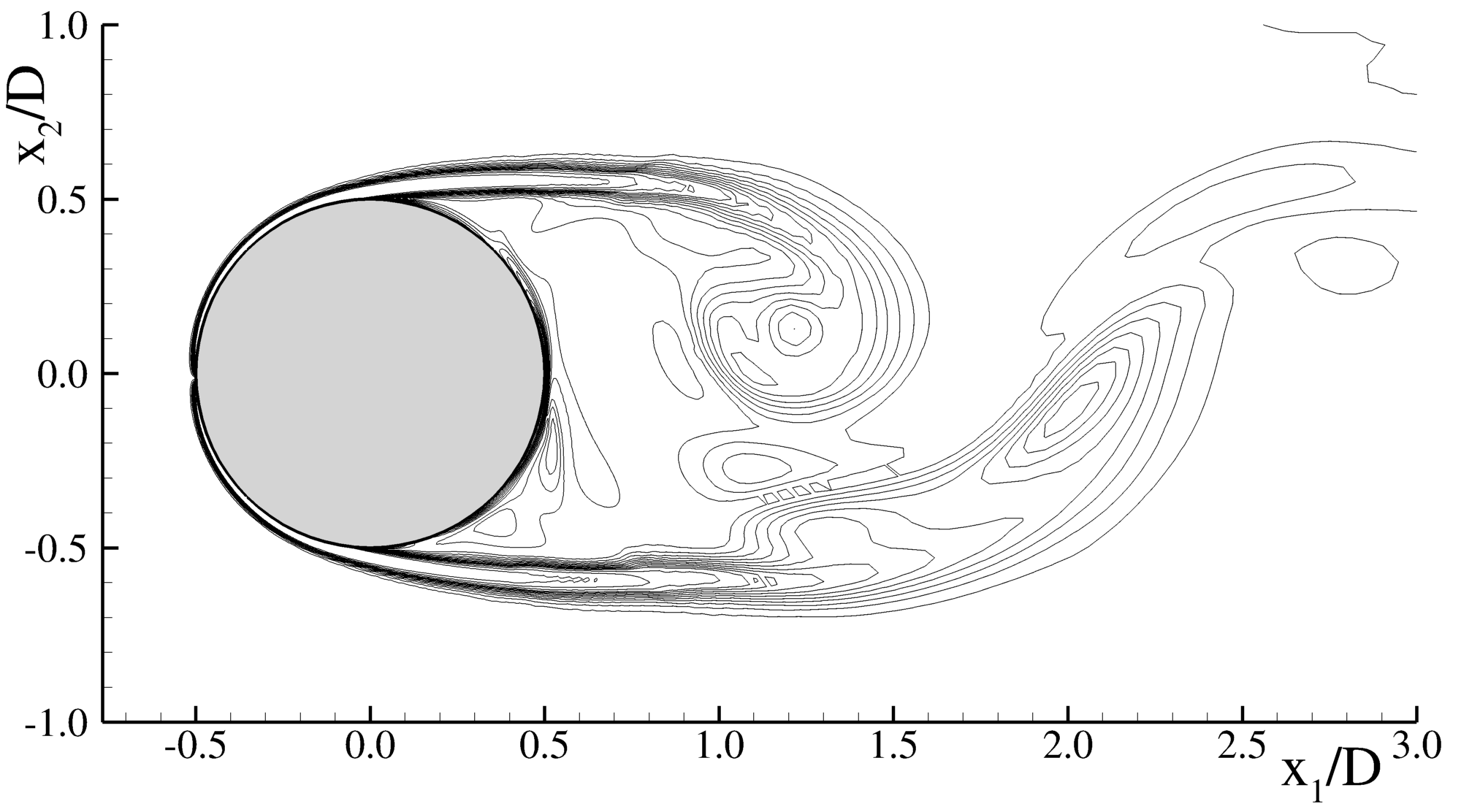}}
~
\subfloat[$\langle \omega_3 \rangle$ at $f_k=0.25$ (SST).]{\label{fig:5.2_6b}
\includegraphics[scale=0.16,trim=0 0 0 0,clip]{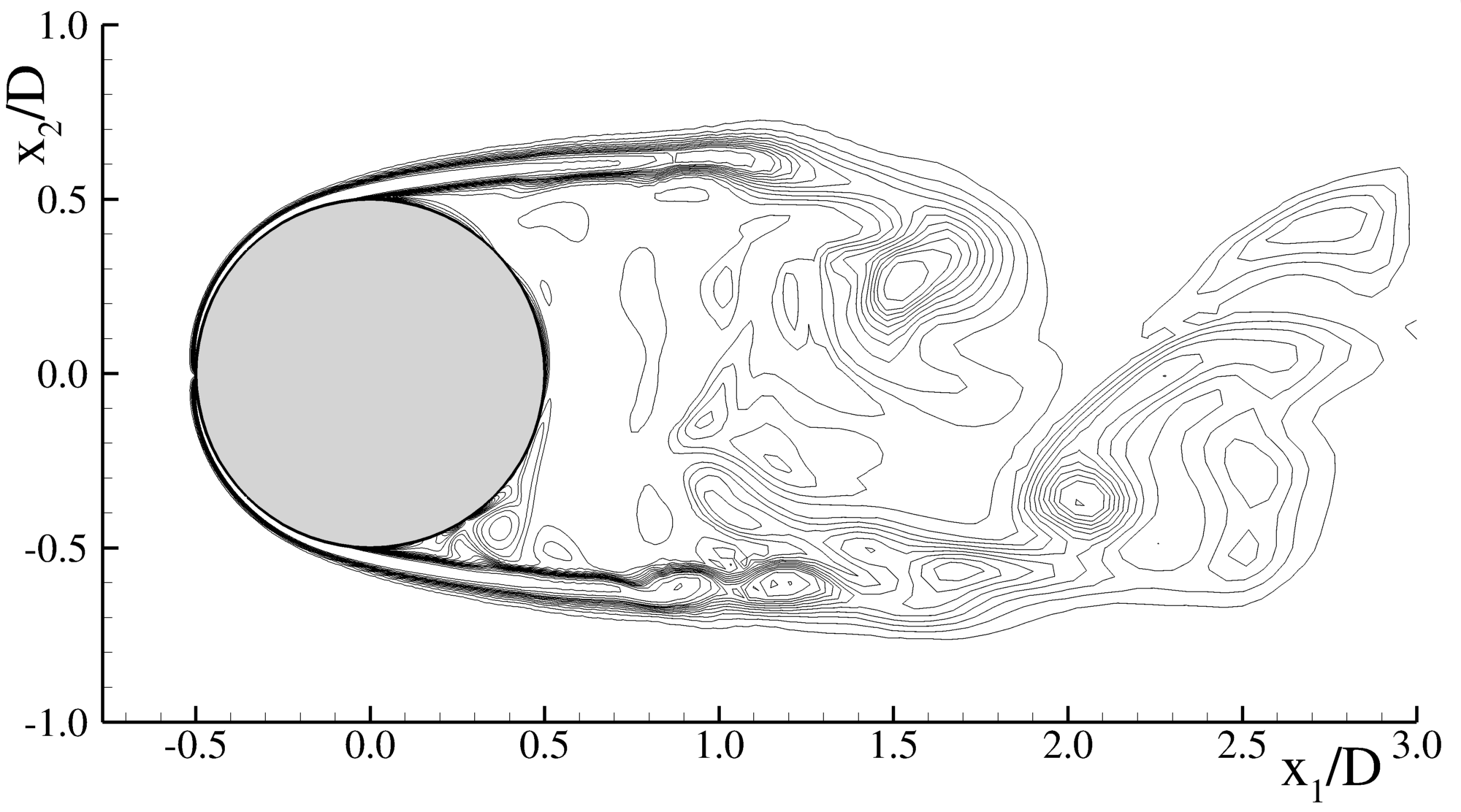}}
\\
\subfloat[$\langle V_1\rangle (t)$ at $f_k=0.50$ (SST).]{\label{fig:5.2_6c}
\includegraphics[scale=0.16,trim=0 0 0 0,clip]{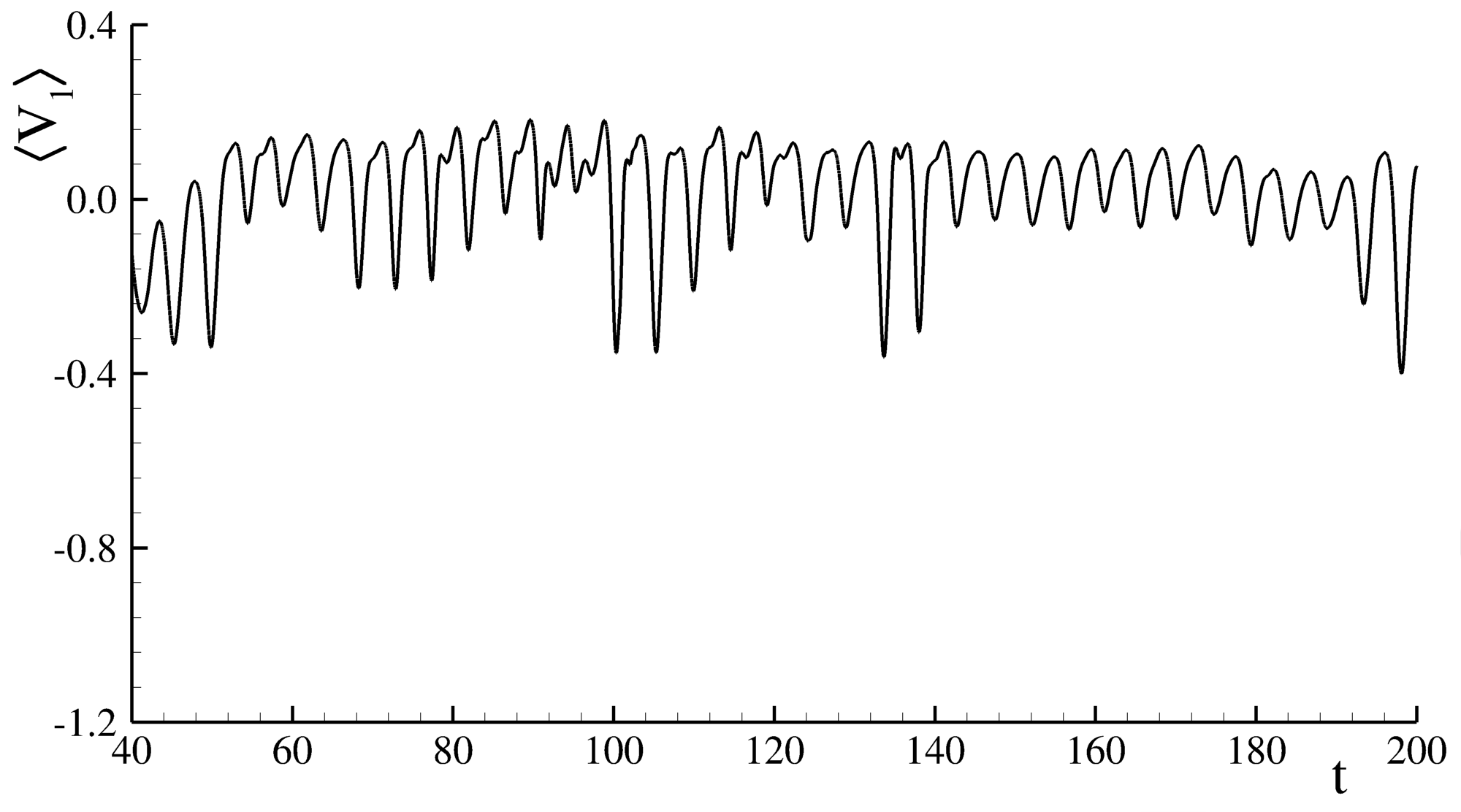}}%
~
\subfloat[$\langle V_1\rangle (t)$ at $f_k=0.25$ (SST).]{\label{fig:5.2_6d}
\includegraphics[scale=0.16,trim=0 0 0 0,clip]{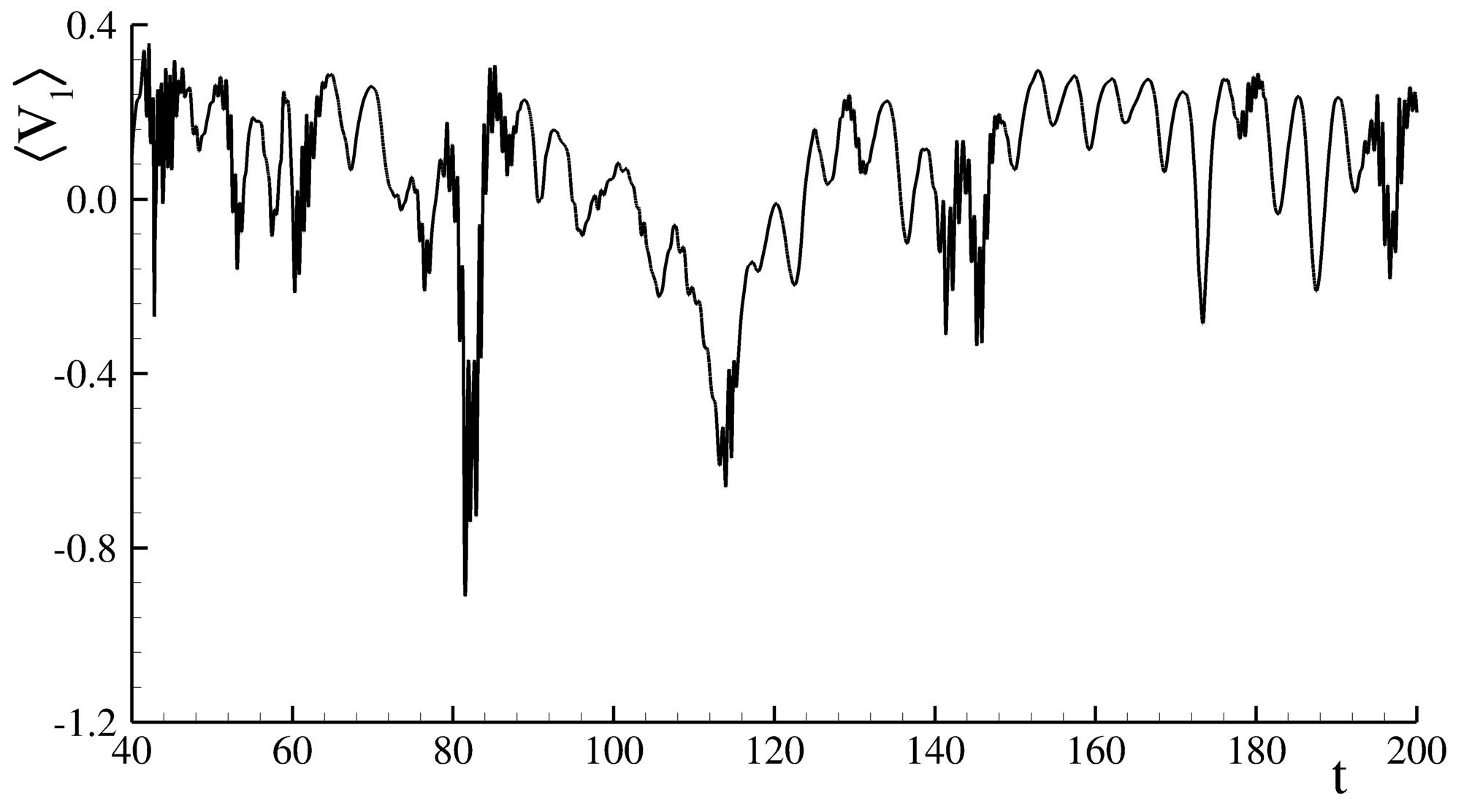}}
\\
\subfloat[$E(\langle V_1 \rangle)$ at $f_k=0.50$.]{\label{fig:5.2_6e}
\includegraphics[scale=0.16,trim=0 0 0 0,clip]{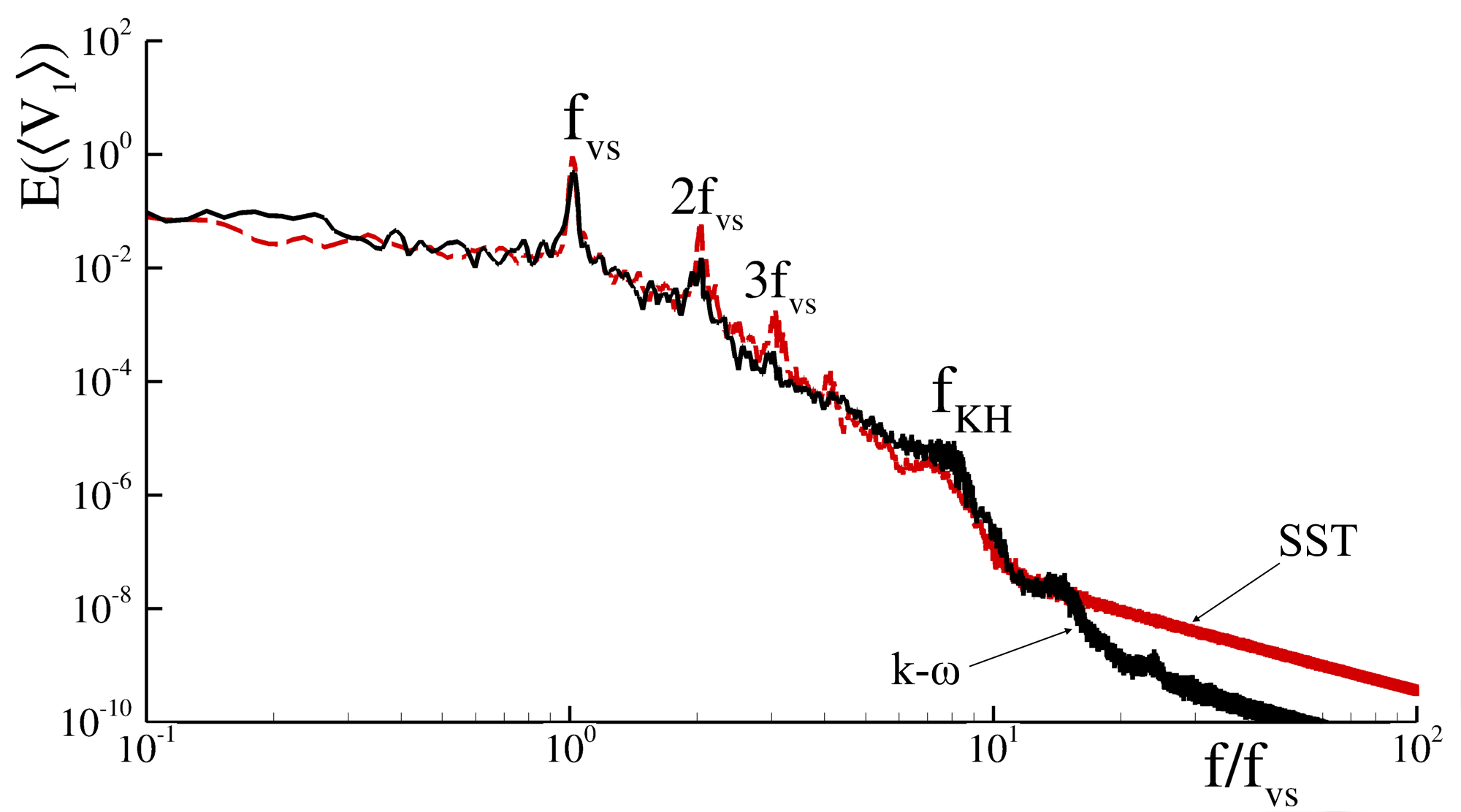}}
~
\subfloat[$E(\langle V_1 \rangle)$ at $f_k=0.25$.]{\label{fig:5.2_6f}
\includegraphics[scale=0.16,trim=0 0 0 0,clip]{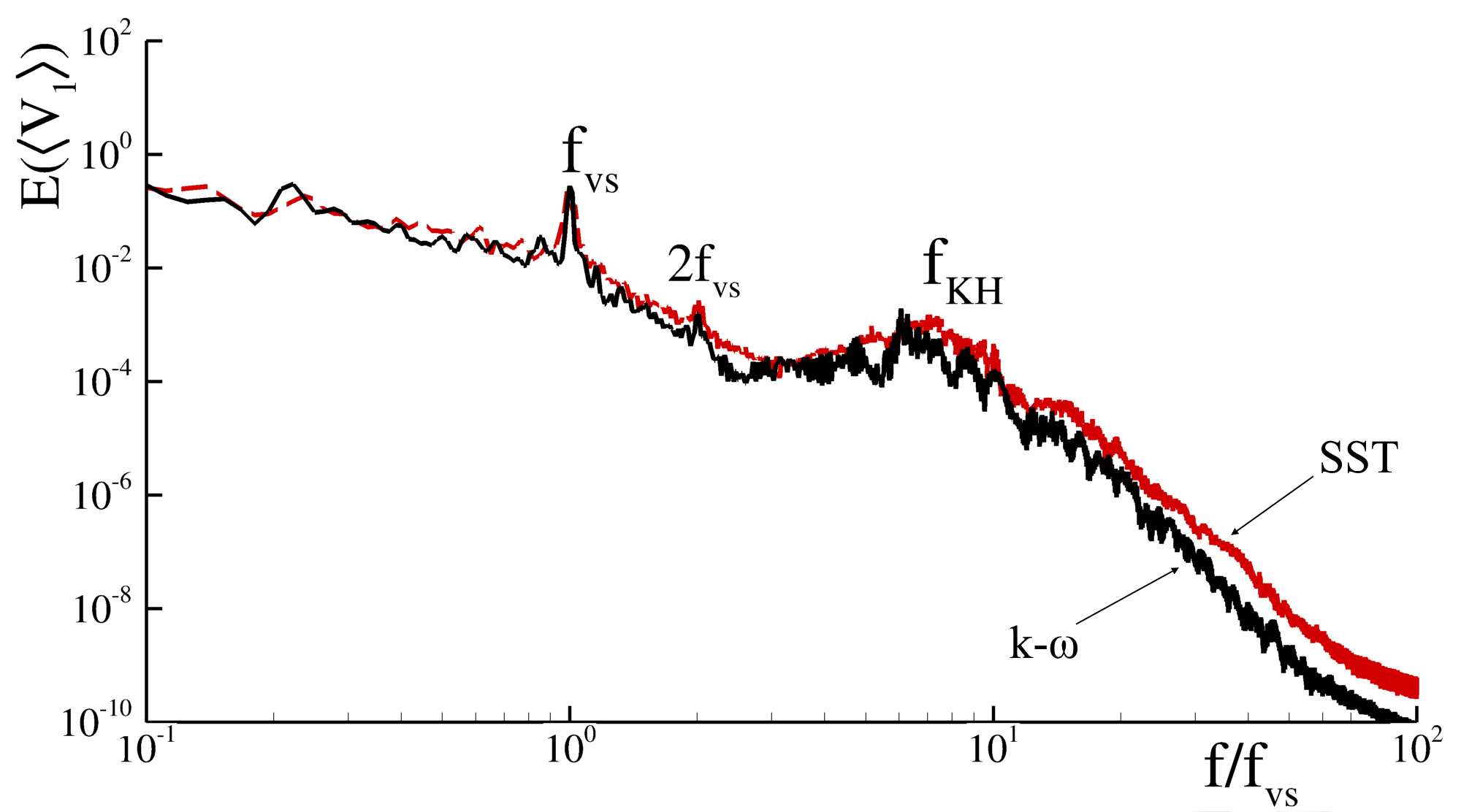}}
\caption{Schematic of the instantaneous span-wise vorticity, $\langle \omega_3 \rangle$, field, time-trace of the stream-wise velocity, $\langle V_1\rangle (t)$, field and respective frequency spectrum, $E(\langle V_1)$, at $P_1$ ($x_1/D=0.71$; $x_2/D=0.66$) for $f_k=0.50$ and $0.25$ and different PANS closures.}
\label{fig:5.2_6}
\end{figure}

\subsubsection{High Resolution Simulations}
\label{sec:5.2.2}

Although the complex nature of turbulence along the inflection line region limits the application of the PANS closures, the increase of resolution reduces the model influence on the simulations. This permits capturing the appropriate physics in the resolved scales. For this reason, simulations employing $f_k\leq 0.50$ yield $\langle S \rangle k_u/\epsilon_u \leq 6$ nearly throughout the flow domain - figure \ref{fig:5.2_2}. The closure model is used only to represent appropriate physics. Figure \ref{fig:5.2_2} also shows that contrary to the low resolution simulations, the region where the maximum $\langle S \rangle k_u/\epsilon_u$ occurs coincides with the inflection line.

The increase in the extent of the linear-physics region results in a more accurate representation of the turbulence kinetic energy production. The formation length increases with the reduction of $f_k$ and gets closer to the experimental measurements - figure \ref{fig:5.2_3}. The stream-wise normal stresses in the centreline now exhibit two peaks. This indicates that the aforementioned four stages process is better represented, especially for the $f_k=0.25$ case. Recall that the second peak marks the vortex-shedding formation length, while the first indicates the onset of instability.

Referring to the $Re_e$ contours of figure \ref{fig:5.2_4}, the results confirm that the extension of the free shear-layer is closely related to the magnitude $Re_e$. Whereas for the $f_k=0.25$ case the inflection line never experiences $Re_e \leq 1200$, for $f_k=0.50$ that occurs much further downstream than for the low resolution simulations. The high values of $Re_e$ guarantee the adequate replication of the four stages of the vortex-shedding structure development. 

The consequences of all these results are evident in figure \ref{fig:5.2_6}. At the $f_k=0.25$ case it is now possible to visualize the Kelvin-Helmholtz rollers, and evidently detect them in the velocity field and correspondent spectrum. Note that the similarities between figures \ref{fig:2.1_2}, \ref{fig:5.2_6a} and \ref{fig:5.2_6e} are remarkable. For the $f_k=0.50$ case, however, this instability is less pronounced, but still visible in the spectrum. Overall, although $f_k=0.50$ and $f_k=0.25$ show some differences, most flow statistics are quite similar. Thus, $f_k=0.50$ may be suffice for most engineering applications.

\subsection{Towards Criteria for Optimal SRS Resolution}
\label{sec:5.3}

Our study demonstrates that successful SRS computations of engineering flows with spatially-developing coherent structures depend upon three key factors: $i)$ identification of the physical mechanisms responsible for generating coherent structures; $ii)$ optimal resolution of the key features of the coherent flow field that are not amenable to modelling; and $iii)$ modelling only the fully-developed (stochastic) portion of the turbulence flow field which lends itself to tractable closures. Based on the findings of the current study, we propose the following criteria for achieving optimal SRS computations to provide a reasonable balance between accuracy and computational effort. 

The {\bf first task} is to identify the region in the flow field wherein the coherent structures are generated. In general, the origin of the coherent structures can be traced back to an underlying instability in the flow field. In the present case, the Kelvin-Helmholtz instability is the mechanism responsible for generating coherent structures in the flow. It is well-known that this instability occurs along the inflection-point locus of the velocity field. Thus, we search the resolved field to identify the region wherein the second derivative of the time-averaged stream-wise velocity vanishes - figure \ref{fig:5.2_1}. For other types of instabilities, the corresponding instability criteria must be used to locate the region of coherent-structure generation.

The {\bf second step} is to ensure that the instability is fully manifested, leading to the natural spatial development of coherent structures. It is evident from the results presented in this paper that the instability is fully established if, and only if, the effective computational Reynolds number exceeds the critical Reynolds number of that instability ($Re > 1200$ in the present case) in the coherence region. For smaller effective Reynolds numbers ($f_k > 0.50$), the
 development of the coherent structure is highly compromised leading to very short separation bubbles and large errors in the flow statistics - figures \ref{fig:5.1_1} to \ref{fig:5.1_4} and table \ref{tab:5.1_1}. For finer resolutions ($f_k \leq 0.50$), the computational Reynolds number becomes progressively larger ($Re > 1200$) and the development of the coherent structures is in good agreement with the experimental observations - figures \ref{fig:5.2_1}, \ref{fig:5.2_4} and \ref{fig:5.2_6}. The advantage to be gained from finer-resolution (beyond $f_k \le 0.50$) is not significant enough to justify the added computational burden - figures \ref{fig:5.1_1} to \ref{fig:5.1_4} and table \ref{tab:5.1_1}. Thus, the optimal resolution in the region of coherent structure development  is such that the computational Reynolds number is marginally higher than the critical Reynolds number. It must also be noted that the resolution outside the coherent flow region does not appear to play a significant role. These findings also support the paradigm that the coherent structures can be considered as instabilities in a non-Newtonian fluid whose viscosity is given by the effective computational viscosity.

The {\bf third task} is to confirm that the modelled flow field component is comprised only of fully-developed (stochastic) turbulence. This can be adjudged by examining the strain-rate ratio of the mean and unresolved fields ($\langle S \rangle k_u/ \epsilon_u$ or $\langle S \rangle/\omega_u$) throughout the flow field. The SRS can be considered physically reasonable if, and only if, the strain-rate ratio is below $6$ which is an indication that the modelled field is fully-turbulent or stochastic. If the computed ratio exceeds a value of about $6$, that indicates that the turbulence in that region is experiencing linear effects and the closure model is likely to be inadequate. In low-resolution simulations ($f_k > 0.50$), it is seen that the strain-rate ratio is very high ($\langle S \rangle k_u/ \epsilon_u>8$) over significant regions of coherent structure development - figure \ref{fig:5.2_2}. In these cases, the SRS computation attempts to model the coherent flow leading to a poor agreement with the experimental observations. In fine resolution simulations ($f_k \leq 0.50$), the strain-rate ratio is below 5 everywhere - figure \ref{fig:5.2_2}. This indicates that the closure model is only being applied to the stochastic (fully-turbulent) component of the flow.

\section{Conclusions}
\label{sec:6}

We investigate the effectiveness of the Scale-Resolving Simulation approach in computing flows with spatially-developing coherent structures. SRS methods of practical interest seek to resolve only the key features of coherent structures and represent the residual stochastic turbulent field with closure models. The objective of this study is to identify the closure modelling/simulation challenges and develop criteria for a reasonable resolution that provides an optimal balance between accuracy and computational effort. The selected canonical case is the flow around a circular cylinder at $Re=3900$ which comprises several complex flow features. The SRS approach employed is the PANS method of \cite{GIRIMAJI_JAM_2005}. 

SRS computations are performed at four levels of physical resolution (cut-off length scale) designated according to the fraction of unresolved turbulence kinetic energy $f_k = 1.00$, $0.75$, $0.50$ and $0.25$. It is important to note that $f_k = 1.00$ corresponds to a RANS (Reynolds-Averaged Navier-Stokes) computation and decreasing $f_k$ implies increasing resolution. It is exhibited that the quality of the results improves with increasing resolution (decreasing $f_k$) but begins to approach the experimental observation only beyond $f_k \leq 0.50$. It is shown that a good agreement with the experimental measurements can be achieved only if all the stages of the coherent structure development are adequately resolved. Four key stages are identified: Kelvin-Helmholtz instability onset in the shear-layer behind the cylinder; gradual spatial development of the Kelvin-Helmholtz rollers; breakdown to high intensity turbulence; and vortex-shedding.

The behaviour of the SRS flow fields at different resolutions is examined to propose general guidelines for optimal SRS of flows with coherent structures:
\begin{itemize}
\item[$i)$] The critical instability region of coherent structure generation and development must be first identified. Here, the resolved field is investigated to locate the region wherein the second derivative of the time-averaged stream-wise velocity vanishes. This region is designated as the ``critical'' zone in which the resolution must be sufficient in order to reasonably replicate the flow structures;
\item[$ii)$]The effective computational Reynolds number in the coherence region must exceed the critical Reynolds number needed for the onset of the responsible instability. In the present case, the responsible mechanism is the Kelvin-Helmholtz instability  and the critical Reynolds number is $1200$. It is shown that coarser resolutions cause a truncated development of the instability leading to a poor agreement of the flow statistics with the experiments. On the other hand, finer resolutions with larger effective computational Reynolds number ($f_k \le 0.50$; $Re > 1200$) do not lead to much improvement in the quality of the simulation results. Thus, the optimal resolution is established to be that which produces a computational Reynolds number of about $1200$ in the free shear-layer region; 
\item[$iii)$] Finally, it must be confirmed using the strain-rate ratio of mean and unresolved fields analysis that the modelled field is comprised mostly of fully-developed (stochastic) turbulence for acceptable accuracy. Here, it is demonstrated that for coarse resolution cases ($f_k > 0.50$; $Re< 1200$), a significant amount of the coherent flow field is modelled resulting in a poor agreement with the experiments. 
\end{itemize}

We propose the above ``good practice" guidelines to achieve reasonable SRS predictions of practical engineering flows for which LES is not viable and one-point closures are not adequate. It is important to note that the proposed criteria are necessary, but may not be sufficient, conditions for ensuring a reasonable SRS computation.

\section*{Acknowledgements}

The authors would like to express their sincere gratitude to J. Carlier, R. Govardhan, E. Lamballais, C. Norberg, S. Rajagopalan, and C.H.K. Williamson for providing the experimental measurements used in this study. The authors also gratefully acknowledge J. de Wilde for the interesting discussions, as well as the Maritime Research Institute of Netherlands (MARIN) for providing the necessary HPC resources.

\bibliographystyle{jfm}
\bibliography{SRC/SR/references}

\begin{thebibliography}{64}
\expandafter\ifx\csname natexlab\endcsname\relax\def\natexlab#1{#1}\fi
\def\au#1{#1} \def\ed#1{#1} \def\yr#1{#1}\def\at#1{#1}\def\jt#1{\textit{#1}}
  \def\bt#1{#1}\def\bvol#1{\textbf{#1}} \def\vol#1{#1} \def\pg#1{#1}
  \def\publ#1{#1}\def\arxiv#1{#1}\def\org#1{#1}\def\st#1{\textit{#1}}

\bibitem[Achenbach(1971)]{ACHENBACH_JFM_1971}
{\sc \au{Achenbach, E.}} \yr{1971}  \at{{Influence of Surface Roughness on the
  Cross-Flow Around a Circular Cylinder}}.  \jt{Journal of Fluid Mechanics}
  \bvol{46}~(2),  \pg{321--335}.

\bibitem[Basara {\em et~al.\/}(2011)Basara, Krajnovi\'c, Girimaji \&
  Pavlovic]{BASARA_AIAA_2011}
{\sc \au{Basara, B.}, \au{Krajnovi\'c, S.}, \au{Girimaji, S.} \& \au{Pavlovic,
  Z.}} \yr{2011}  \at{{Near-Wall Formulation of the Partially Averaged
  Navier-Stokes Turbulence Model}}.  \jt{American Institute of Aeronautics and
  Astronautics Journal}  \bvol{49}~(12),  \pg{2627--2636}.

\bibitem[Bloor(1964)]{BLOOR_JFM_1964}
{\sc \au{Bloor, M.S.}} \yr{1964}  \at{{The Transition to Turbulence in the Wake
  of a Circular Cylinder}}.  \jt{Journal of Fluid Mechanics}  \bvol{19}~(2),
  \pg{290--304}.

\bibitem[Chaouat \& Schiestel(2005)]{CHAOUAT_PF_2005}
{\sc \au{Chaouat, B.} \& \au{Schiestel, R.}} \yr{2005}  \at{{A New Partially
  Integrated Transport Model for Subgrid-Scale Stresses and Dissipation Rate
  for Turbulent Developing Flows}}.  \jt{Physics of Fluids}  \bvol{17}, 065106.

\bibitem[Fage \& Warsap(1929)]{FAGE_ARC_1929}
{\sc \au{Fage, A.} \& \au{Warsap, J.H.}} \yr{1929}  \bt{{The Effects of
  Turbulence and Surface Roughness on the Drag of a Circular Cylinder}}.
  Reports and Memoranda 12083.  \org{Aeronautical Research Committee}.

\bibitem[Fasel {\em et~al.\/}(2002)Fasel, Seidel \& Wernz]{FASEL_JFE_2002}
{\sc \au{Fasel, H.F.}, \au{Seidel, J.} \& \au{Wernz, S.}} \yr{2002}  \at{{A
  Methodology for Simulations for Complex Turbulent Flows}}.  \jt{Journal of
  Fluids Engineering}  \bvol{124}~(4),  \pg{933--942}.

\bibitem[Gatski \& Speziale(1993)]{GATSKI_JFM_1993}
{\sc \au{Gatski, T.B.} \& \au{Speziale, C.G.}} \yr{1993}  \at{{On Explicit
  Algebraic Stress Models for Complex Turbulent Flows}}.  \jt{Journal of Fluid
  Mechanics}  \bvol{254},  \pg{59--78}.

\bibitem[Gerich \& Eckelmann(1982)]{GERICH_JFM_1982}
{\sc \au{Gerich, D.} \& \au{Eckelmann, H.}} \yr{1982}  \at{{Influence of End
  Plates and Free Ends on the Shedding Frequency of Circular Cylinders}}.
  \jt{Journal of Fluid Mechanics}  \bvol{122},  \pg{109--121}.

\bibitem[Germano(1992)]{GERMANO_JFM_1992}
{\sc \au{Germano, M.}} \yr{1992}  \at{{Turbulence: the Filtering Approach}}.
  \jt{Journal of Fluid Mechanics}  \bvol{238},  \pg{325--336}.

\bibitem[Girimaji(1996)]{GIRIMAJI_TCFD_1996}
{\sc \au{Girimaji, S.S.}} \yr{1996}  \at{{Fully Explicit and Self-Consistent
  Algebraic Reynolds Stress Model}}.  \jt{Theoretical and Computational Fluid
  Dynamics}  \bvol{8}~(6),  \pg{387--402}.

\bibitem[Girimaji(2000)]{GIRIMAJI_JFM_2000}
{\sc \au{Girimaji, S.S.}} \yr{2000}  \at{{Pressure-Strain Correlation Modelling
  of Complex Turbulent Flows}}.  \jt{Journal of Fluid Mechanics}  \bvol{422},
  \pg{91--123}.

\bibitem[Girimaji(2001)]{GIRIMAJI_TCFD_2001}
{\sc \au{Girimaji, S.S.}} \yr{2001}  \at{{Lower-Dimensional Manifold
  (Algebraic) Representation of Reynolds Stress Closure Equations}}.
  \jt{Theoretical and Computational Fluid Dynamics}  \bvol{14}~(4),
  \pg{259--281}.

\bibitem[Girimaji(2005)]{GIRIMAJI_JAM_2005}
{\sc \au{Girimaji, S.}} \yr{2005}  \at{{Partially-Averaged Navier-Stokes Model
  for Turbulence: A Reynolds-Averaged Navier-Stokes to Direct Numerical
  Simulation Bridging Method}}.  \jt{Journal of Applied Mechanics}
  \bvol{73}~(3),  \pg{413--421}.

\bibitem[Girimaji \& Abdol-Hamid(2005)]{GIRIMAJI_AIAA43_2005}
{\sc \au{Girimaji, S.S.} \& \au{Abdol-Hamid, K.S.}} \yr{2005}
  {Partially-Averaged Navier Stokes Model for Turbulence: Implementation and
  Validation}.  \bt{In {\em 43\textsuperscript{rd} American Institute of
  Aeronautics and Astronautics (AIAA) Aerospace Sciences Meeting and
  Exhibit\/}}. Reno, Nevada.

\bibitem[Girimaji {\em et~al.\/}(2005)Girimaji, Jeong \&
  Srinivasan]{GIRIMAJI_JAM2_2005}
{\sc \au{Girimaji, S.S.}, \au{Jeong, E.} \& \au{Srinivasan, R.}} \yr{2005}
  \at{{Partially Averaged Navier-Stokes Method for Turbulence: Fixed Point
  Analysis and Comparison With Unsteady Partially Averaged Navier-Stokes}}.
  \jt{Journal of Applied Mechanics}  \bvol{73}~(3),  \pg{422--429}.

\bibitem[G\"{u}ven {\em et~al.\/}(1980)G\"{u}ven, Farrel \&
  Patel]{GUVEN_JFM_1980}
{\sc \au{G\"{u}ven, O.}, \au{Farrel, C.} \& \au{Patel, V.C.}} \yr{1980}
  \at{{Surface-Roughness Effects on the Mean Flow Past Circular Cylinder}}.
  \jt{Journal of Fluid Mechanics}  \bvol{98}~(4),  \pg{673--701}.

\bibitem[Huang \& Rajagopal(1996)]{HUANG_TCFD_1996}
{\sc \au{Huang, Y.-N.} \& \au{Rajagopal, K.R.}} \yr{1996}  \at{{On a
  Generalized Nonlinear $k-\epsilon$ Model for Turbulence that Models
  Relaxation Effects}}.  \jt{Theoretical and Computational Fluid Dynamics}
  \bvol{8}~(4),  \pg{275--288}.

\bibitem[Hussain \& Reynolds(1970)]{REYNOLDS_JFM_1970}
{\sc \au{Hussain, A.K.M.F.} \& \au{Reynolds, W.C.}} \yr{1970}  \at{{The
  Mechanics of an Organized Wave in Turbulent Shear Flow}}.  \jt{Journal of
  Fluid Mechanics}  \bvol{41}~(2),  \pg{241--258}.

\bibitem[Kok(2017)]{KOK_FTC_2017}
{\sc \au{Kok, J.C.}} \yr{2017}  \at{{A Stochastic Backscatter Model for
  Grey-Area Mitigation in Detached Eddy Simulations}}.  \jt{Flow, Turbulence
  and Combustion}  \bvol{99}~(1),  \pg{119--150}.

\bibitem[Kok {\em et~al.\/}(2004)Kok, Dol, Oskam \& der Ven]{KOK_AIAA42_2004}
{\sc \au{Kok, J.C.}, \au{Dol, H.S.}, \au{Oskam, B.} \& \au{der Ven, H.V.}}
  \yr{2004} {Extra-Large Eddy Simulation of Massively Separated Flows}.  \bt{In
  {\em 42\textsuperscript{nd} American Institute of Aeronautics and
  Astronautics (AIAA) Aerospace Sciences Meeting and Exhibit\/}}. Reno, United
  States of America.

\bibitem[Koopmann(1967)]{KOOPMANN_JFM_1967}
{\sc \au{Koopmann, G.H.}} \yr{1967}  \at{{The Vortex Wake of Vibrating
  Cylinders at Low Reynolds Numbers}}.  \jt{Journal of Fluid Mechanics}
  \bvol{28}~(3),  \pg{501--512}.

\bibitem[Lakshmipathy \& Girimaji(2006)]{LAKSHMIPATHY_AIAA_2006}
{\sc \au{Lakshmipathy, S.} \& \au{Girimaji, S.S.}} \yr{2006}
  {Partially-Averaged Navier-Stokes Method for Turbulent Flows: $k-\omega$
  Model Implementation}.  \bt{In {\em 44\textsuperscript{th} American Institute
  of Aeronautics and Astronautics (AIAA) Aerospace Sciences Meeting and
  Exhibit\/}}. Reno, United States of America.

\bibitem[Launder {\em et~al.\/}(1975)Launder, Reece \& Rodi]{LAUNDER_JFM_1975}
{\sc \au{Launder, B.E.}, \au{Reece, G.J.} \& \au{Rodi, W.}} \yr{1975}
  \at{{Progress in the Development of a Reynolds-Stress Turbulence Closure}}.
  \jt{Journal of Fluid Mechanics}  \bvol{68}~(3),  \pg{537--566}.

\bibitem[Lehmkuhl {\em et~al.\/}(2013)Lehmkuhl, Rodriguez, Borrel \&
  Oliva]{LEHMKUHL_PF_2013}
{\sc \au{Lehmkuhl, O.}, \au{Rodriguez, I.}, \au{Borrel, R.} \& \au{Oliva, A.}}
  \yr{2013}  \at{{Low-Frequency Unsteadiness in the Vortex Formation of a
  Circular Cylinder}}.  \jt{Physics of Fluids}  \bvol{25}, 085109.

\bibitem[Lin {\em et~al.\/}(1995)Lin, Vorobieff \& Rockwell]{LIN_JFS_1995}
{\sc \au{Lin, J.-C.}, \au{Vorobieff, P.} \& \au{Rockwell, D.}} \yr{1995}
  \at{{Three-Dimensional Patterns of Streamwise Vorticity in the Near-Wake of a
  Cylinder}}.  \jt{Journal of Fluids and Structures}  \bvol{9}~(2),
  \pg{231--234}.

\bibitem[Mansy {\em et~al.\/}(1994)Mansy, Yang \& Williams]{MANSY_JFM_1994}
{\sc \au{Mansy, H.}, \au{Yang, P.-M.} \& \au{Williams, D.R.}} \yr{1994}
  \at{{Quantitative Measurements of Three-Dimensional Structures in the Wake of
  a Circular Cylinder}}.  \jt{Journal of Fluid Mechanics}  \bvol{270},
  \pg{277--296}.

\bibitem[Menter {\em et~al.\/}(2003)Menter, Kuntz \& Langtry]{MENTER_THMT_2003}
{\sc \au{Menter, F.R.}, \au{Kuntz, M.} \& \au{Langtry, R.}} \yr{2003} {Ten
  Years of Industrial Experience with the SST Turbulence Model}.  \bt{In {\em
  Turbulence, Heat and Mass Transfer 4\/} (ed. \ed{K.~Hanjali\'c, Y.~Nagano \&
  M.~Tummers})}.  \publ{Antalya, Turkey: Begell House, Inc}.

\bibitem[Mishra \& Girimaji(2010)]{MISHRA_FTC_2010}
{\sc \au{Mishra, A.A.} \& \au{Girimaji, S.S.}} \yr{2010}  \at{{Pressure-Strain
  Correlation Modeling: Towards Achieving Consistency with Rapid Distortion
  Theory}}.  \jt{Flow, Turbulence and Combustion}  \bvol{85}~(3-4),
  \pg{593--619}.

\bibitem[Mishra \& Girimaji(2013)]{MISHRA_JFM_2013}
{\sc \au{Mishra, A.A.} \& \au{Girimaji, S.S.}} \yr{2013}  \at{{Intercomponent
  Energy Transfer in Incompressible Homogeneous Turbulence: Multi-Point Physics
  and Amenability to One-Point Closures}}.  \jt{Journal of Fluid Mechanics}
  \bvol{731},  \pg{639--681}.

\bibitem[Mishra \& Girimaji(2017)]{MISHRA_JFM_2017}
{\sc \au{Mishra, A.A.} \& \au{Girimaji, S.S.}} \yr{2017}  \at{{Toward
  Approximating Non-Local Dynamics in Single-Point Pressure-Strain Correlation
  Closures}}.  \jt{Journal of Fluid Mechanics}  \bvol{811},  \pg{168--188}.

\bibitem[Muschinski(1996)]{MUSCHINSKI_JFM_1996}
{\sc \au{Muschinski, A.}} \yr{1996}  \at{{A Similarity Theory of Locally
  Homogeneous and Isotropic Turbulence Generated by a Smagorinsky-Type LES}}.
  \jt{Journal of Fluid Mechanics}  \bvol{325},  \pg{239--260}.

\bibitem[Norberg(1987)]{NORBERG_TREP_1987}
{\sc \au{Norberg, C.}} \yr{1987}  \bt{{Effects of Reynolds Numbers and a
  Low-Intensity Freestream Turbulence on the Flow Around a Circular Cylinder}}.
  {\em Tech. Rep.\/} 87/2.  \org{Chalmers University of Technology},
  G\"oteborg, Sweden.

\bibitem[Norberg(1994)]{NORBERG_JFM_1994}
{\sc \au{Norberg, C.}} \yr{1994}  \at{{An Experimental Investigation of the
  Flow Around a Circular Cylinder: Influence of Aspect Ratio}}.  \jt{Journal of
  Fluid Mechanics}  \bvol{258},  \pg{287--319}.

\bibitem[Norberg(1998)]{NORBERG_AUBBWVIV_1998}
{\sc \au{Norberg, C.}} \yr{1998} {LDV-Measurements in the Near Wake of a
  Circular Cylinder}.  \bt{In {\em Advances in Understanding of Bluff Body
  Wakes and Vortex-Induced Vibration\/}}. Washington DC, United States of
  America.

\bibitem[Norberg(2002)]{NORBERG_BBVIV3_2002}
{\sc \au{Norberg, C.}} \yr{2002} {Pressure Distributions around a Circular
  Cylinder in Cross-Flow}.  \bt{In {\em Symposium on Bluff Body Wakes and
  Vortex-Induced Vibrations (BBVIV3)\/}}. Port Arthur, Australia.

\bibitem[Norberg(2003)]{NORBERG_JFS_2003}
{\sc \au{Norberg, C.}} \yr{2003}  \at{{Fluctuating Lift on a Circular Cylinder:
  Review and New Measurements}}.  \jt{Journal of Fluids and Structures}
  \bvol{17}~(1),  \pg{59--96}.

\bibitem[Norberg \& Sund\'{e}n(1987)]{NORBERG_JFS_1987}
{\sc \au{Norberg, C.} \& \au{Sund\'{e}n, B.}} \yr{1987}  \at{{Turbulence and
  Reynolds Number Effects on the Flow and Fluid Forces on a Single Cylinder in
  Cross Flow}}.  \jt{Journal of Fluids and Structures}  \bvol{1}~(3),
  \pg{337--357}.

\bibitem[Parnaudeau {\em et~al.\/}(2008)Parnaudeau, Carlier, Heitz \&
  Lamballais]{PARNAUDEAU_PF_2008}
{\sc \au{Parnaudeau, P.}, \au{Carlier, J.}, \au{Heitz, D.} \& \au{Lamballais,
  E.}} \yr{2008}  \at{{Experimental and Numerical Studies of the Flow over a
  Circular Cylinder at Reynolds Number 3900}}.  \jt{Physics of Fluids}
  \bvol{20}, 085101.

\bibitem[Pereira {\em et~al.\/}(2015)Pereira, Vaz \&
  E\c{c}a]{PEREIRA_THMT_2015}
{\sc \au{Pereira, F.S.}, \au{Vaz, G.} \& \au{E\c{c}a, L.}} \yr{2015} {An
  Assessment of Scale-Resolving Simulation Models for the Flow Around a
  Circular Cylinder}.  \bt{In {\em Turbulence, Heat and Mass Transfer (THMT)
  8\/}}. Sarajevo, Bosnia and Herzegovina.

\bibitem[Pereira {\em et~al.\/}(2017)Pereira, Vaz, E\c{c}a \&
  Girimaji]{PEREIRA_IJHFF_2017}
{\sc \au{Pereira, F.S.}, \au{Vaz, G.}, \au{E\c{c}a, L.} \& \au{Girimaji, S.S.}}
  \yr{2017}  \at{{Simulation of the Flow Around a Circular Cylinder at
  $Re=3900$ with Partially-Averaged Navier-Stokes Equations}}.
  \jt{(Submitted)} .

\bibitem[Pope(1975)]{POPE_JFM_1975}
{\sc \au{Pope, S.B.}} \yr{1975}  \at{{A More General Effective-Viscosity
  Hypothesis}}.  \jt{Journal of Fluid Mechanics}  \bvol{72}~(2),
  \pg{331--340}.

\bibitem[Prasad \& Williamson(1997{\natexlab{{\em a\/}}})]{PRADAD_JFM_1997A}
{\sc \au{Prasad, A.} \& \au{Williamson, C.H.K.}} \yr{1997{\natexlab{{\em
  a\/}}}}  \at{{The Instability of the Shear Layer Separating from a Bluff
  Body}}.  \jt{Journal of Fluid Mechanics}  \bvol{333},  \pg{375--402}.

\bibitem[Prasad \& Williamson(1997{\natexlab{{\em b\/}}})]{PRASAD_JFM_1997B}
{\sc \au{Prasad, A.} \& \au{Williamson, C.H.K.}} \yr{1997{\natexlab{{\em
  b\/}}}}  \at{{Three-Dimensional Effects in Turbulent Bluff-Body Wake}}.
  \jt{Journal of Fluid Mechanics}  \bvol{343},  \pg{235--265}.

\bibitem[Rajagopalan \& Antonia(2005)]{RAJAGOPALAN_EF_2005}
{\sc \au{Rajagopalan, S.} \& \au{Antonia, R.A.}} \yr{2005}  \at{{Flow Around a
  Circular Cylinder - Structure of the Near Wake Shear Layer}}.
  \jt{Experiments in Fluids}  \bvol{38}~(4),  \pg{393--402}.

\bibitem[Rayleigh(1883)]{RAYLEIGH_LMS_1883}
{\sc \au{Rayleigh, L.}} \yr{1883}  \at{{Investigation of the Character of the
  Equilibrium of an Incompressible Heavy Fluid of Variable Density}}.
  \jt{London Mathematical Society}  \bvol{14},  \pg{170--177}.

\bibitem[ReFRESCO(2017)]{ReFRESCO}
{\sc \au{ReFRESCO}} \yr{2017} www.refresco.org.

\bibitem[Reyes {\em et~al.\/}(2014)Reyes, Cooper \& Girimaji]{DASIA_PF_2014}
{\sc \au{Reyes, D.A.}, \au{Cooper, J.M.} \& \au{Girimaji, S.S.}} \yr{2014}
  \at{{Characterizing Velocity Fluctuations in Partially Resolved Turbulence
  Simulations}}.  \jt{Physics of Fluids}  \bvol{26}, 085106.

\bibitem[Rivlin(1957)]{RIVLIN_QAM_1957}
{\sc \au{Rivlin, R.S.}} \yr{1957}  \at{{The Relation Between the Flow of
  Non-Newtonian Fluids and Turbulence Newtonian Fluids}}.  \jt{Quarterly of
  Applied Mathematics}  \bvol{XV}~(2),  \pg{212--215}.

\bibitem[Rodi(1976)]{RODI_ZAMM_1976}
{\sc \au{Rodi, W.}} \yr{1976}  \at{{A New Algebraic Relation for Calculating
  the Reynolds Stresses}}.  \jt{Zeitschrift fuer Angewandte Mathematik und
  Mechanik (ZAMM)}  \bvol{56},  \pg{219--221}.

\bibitem[Roshko(1954)]{ROSHKO_TECH_1954}
{\sc \au{Roshko, A.}} \yr{1954}  \bt{{On the Development of Turbulent Wake from
  Vortex Streets}}. {\em Tech. Rep.\/} 1191.  \org{National Advisory Committee
  for Aeronautics}.

\bibitem[Sadeh \& Saharon(1982)]{SADEH_NASA_1982}
{\sc \au{Sadeh, W.Z.} \& \au{Saharon, D.B.}} \yr{1982}  \bt{{Turbulence Effect
  on Crossflow Around a Circular Cylinder at Subcritical Reynolds Numbers}}.
  NASA Contractor Report 3622.  \org{National Aeronautics and Space
  Administration (NASA)}.

\bibitem[Schiestel(1987)]{SCHIESTEL_PF_1987}
{\sc \au{Schiestel, R.}} \yr{1987}  \at{{Multiple-Time-Scale Modeling of
  Turbulent Flows in One-Point Closures}}.  \jt{Physics of Fluids}
  \bvol{30}~(3),  \pg{722--731}.

\bibitem[Schiestel \& Dejoan(2005)]{SCHIESTEL_TCFD_2005}
{\sc \au{Schiestel, R.} \& \au{Dejoan, A.}} \yr{2005}  \at{{Towards a New
  Partially Integrated Transport Model for Coarse Grid and Unsteady Flow
  Simulation}}.  \jt{Theoretical and Computational Fluid Dynamics}
  \bvol{18}~(6),  \pg{443--468}.

\bibitem[Shur {\em et~al.\/}(2008)Shur, Spalart, Strelets \&
  Travin]{SHUR_IJHFF_2008}
{\sc \au{Shur, M.L.}, \au{Spalart, P.R.}, \au{Strelets, M.Kh.} \& \au{Travin,
  A.K.}} \yr{2008}  \at{{A Hybrid RANS-LES Approach with Delayed-DES and
  Wall-Modelled LES Capabilities}}.  \jt{International Journal of Heat and
  Fluid Flow}  \bvol{29}~(6),  \pg{1639--1649}.

\bibitem[Spalart {\em et~al.\/}(1997)Spalart, Jou, Strelets \&
  Allmaras]{SPALART_AFOSRIC_1997}
{\sc \au{Spalart, P.R.}, \au{Jou, W.-H.}, \au{Strelets, M.} \& \au{Allmaras,
  S.}} \yr{1997} {Comments on the Feasibility of LES for Wings and on the
  Hybrid RANS/LES Approach}.  \bt{In {\em 1\textsuperscript{st} Air Force
  Office of Scientific Research (AFOSR) International Conference on DNS/LES -
  Advances in DNS/LES\/}}. Ruston, United Statues of America.

\bibitem[Speziale(1996)]{SPEZIALE_ICNMFD_1996}
{\sc \au{Speziale, C.}} \yr{1996} {Computing Non-Equilibrium Turbulent Flow
  with Time-Dependent RANS and VLES}.  \bt{In {\em 15\textsuperscript{th}
  International Conference on Numerical Methods in Fluid Dynamics\/}}.
  Monterey, United States of America.

\bibitem[Speziale {\em et~al.\/}(1991)Speziale, Sarkar \&
  Gatski]{SPEZIALE_JFM_1991}
{\sc \au{Speziale, C.G.}, \au{Sarkar, S.} \& \au{Gatski, T.B}} \yr{1991}
  \at{{Modelling the Pressure Strain-Rate Correlation of Turbulence: An
  Invariant Dynamical Systems Approach}}.  \jt{Journal of Fluid Mechanics}
  \bvol{227},  \pg{245--272}.

\bibitem[Szepessy \& Bearman(1992)]{SZEPESSY_JFM_1992}
{\sc \au{Szepessy, S.} \& \au{Bearman, P.W.}} \yr{1992}  \at{{Aspect Ratio and
  End Plate Effects on Vortex Shedding From a Circular Cylinder}}.  \jt{Journal
  of Fluid Mechanics}  \bvol{234},  \pg{191--217}.

\bibitem[Tokumaru \& Dimotakis(1991)]{TOKUMARU_JFM_1991}
{\sc \au{Tokumaru, P.T.} \& \au{Dimotakis, P.E.}} \yr{1991}  \at{{Rotary
  Oscillation Control of a Cylinder Wake}}.  \jt{Journal of Fluid Mechanics}
  \bvol{224},  \pg{77--90}.

\bibitem[Unal \& Rockwell(1988)]{UNAL_JFM_1988}
{\sc \au{Unal, M.F.} \& \au{Rockwell, D.}} \yr{1988}  \at{{On Vortex Shedding
  from a Cylinder. Part 1. The Initial Instability}}.  \jt{Journal of Fluid
  Mechanics}  \bvol{190},  \pg{419--512}.

\bibitem[West \& Apelt(1982)]{WEST_JFM_1982}
{\sc \au{West, G.S.} \& \au{Apelt, C.J.}} \yr{1982}  \at{{The Effects of Tunnel
  Blockage and Aspect Ratio on the Mean Flow Past a Circular Cylinder with
  Reynolds Numbers Between $10^4$ and $10^5$}}.  \jt{Journal of Fluid
  Mechanics}  \bvol{114},  \pg{361--377}.

\bibitem[Wilcox(1988)]{WILCOX_AIAA_1988}
{\sc \au{Wilcox, D.C.}} \yr{1988}  \at{{Reassessment of the Scale-Determining
  Equation for Advanced Turbulence Models}}.  \jt{American Institute of
  Aeronautics and Astronautics Journal}  \bvol{26}~(11),  \pg{1299--1310}.

\bibitem[Williamson(1996)]{WILLIAMSON_ARFM_1996}
{\sc \au{Williamson, C.H.K.}} \yr{1996}  \at{{Vortex Dynamics in the Cylinder
  Wake}}.  \jt{Annual Review of Fluid Mechanics}  \bvol{28},  \pg{477--539}.

\bibitem[Zdravkovich(1997)]{ZDRAVKOVICH_BOOK_1997}
{\sc \au{Zdravkovich, M.M.}} \yr{1997} {\em Flow Around Circular Cylinder
  Volume 1: Fundamentals\/}.  \publ{Oxford Science Publications}.

\end{thebibliography}

\end{document}